\newif\ifsubmit
\submitfalse

\documentclass[acmsmall,authorversion,screen]{acmart}%

\usepackage[T1]{fontenc}
\usepackage[utf8]{inputenc}
\usepackage[english]{babel}

\usepackage{graphicx}
\usepackage[scaled=0.9]{DejaVuSansMono} % A decent mono font

\usepackage{amsmath}
\usepackage{amsfonts}
\usepackage{amssymb}
\usepackage{amsthm}
\usepackage{mathtools}
\usepackage{textcomp}

\theoremstyle{definition}
\newtheorem{theorem}{Theorem}
\newtheorem{lemma}{Lemma}
\newtheorem{corollary}{Corollary}
\newtheorem{definition}{Definition}
\newtheorem{hypothesis}{Hypothesis}

\usepackage{natbib}

\usepackage{wrapfig}
\usepackage{caption}
\usepackage{subcaption}
\usepackage{placeins}
\usepackage{paralist}
\usepackage{enumitem}% Compact lists
\usepackage{pifont}

\usepackage{syntax} % Grammar definitions
\usepackage{verbatim}
\usepackage{listings} % Code
\usepackage{xspace} % Useful for macros
\usepackage{mathpartir} % Syntax trees
\usepackage{placeins} % for \FloatBarrier
\usepackage{afterpage}

\usepackage{textcomp}

\newcommand{\ie}[0]{\emph{i.e.}, }

\newcommand\TODO[1]{{\ \\\color{red}\large\textbf{TODO} #1}\\}

\newcommand\mysc[1]{{\textsc{#1}}\xspace}
\newcommand\htag[1]{\shortintertext{\textbf{#1}}}

\newcommand\links{\mysc{Links}}
\newcommand\hop{\mysc{Hop}}
\newcommand\opa{\mysc{Opa}}
\newcommand\urweb{\mysc{Ur/Web}}

\newcommand\http{\mysc{HTTP}}
\newcommand\js{\mysc{JavaScript}}
\newcommand\php{\mysc{php}}
\newcommand\html{\mysc{HTML}}

\newcommand\ml{\mysc{ML}}
\newcommand\ocaml{\mysc{OCaml}}
\newcommand\eliom{\mysc{Eliom}}
\newcommand\ocsigen{\mysc{Ocsigen}}
\newcommand\jsocaml{\mysc{js\_of\_ocaml}}
\newcommand\lwt{\mysc{Lwt}}

\newcommand\code[1]{\texttt{#1}}

\lstdefinelanguage{javascript}{
  keywords={typeof, new, true, false, catch, function, return, null, catch, switch, var, if, in, while, do, else, case, break,class, export, boolean, throw, implements, import, this},
  sensitive=false,
  comment=[l]{//},
  morecomment=[s]{/*}{*/},
  morestring=[b]',
  morestring=[b]"
}

\lstdefinelanguage{eliom}[Objective]{Caml}{
  keywordstyle=[2]\bfseries\color{colorserver},
  keywordstyle=[3]\bfseries\color{colorclient},
  keywordstyle=[4]\bfseries\color{colorshared},
  morekeywords=[1]{base,implicit},
  morekeywords=[2]{server},
  morekeywords=[3]{client,fragment},
  morekeywords=[4]{shared,mixed},
}

\newcommand\ocamlc[1]{\mbox{\lstinline[language={[Objective]Caml},basicstyle=\ttfamily\normalsize]{#1}}}
\newcommand\eliomc[1]{\mbox{\lstinline[language={eliom},basicstyle=\ttfamily\normalsize]{#1}}}

\newcommand\inputeliom[2][]{\lstinputlisting[language=eliom,#1]{#2}}

\definecolor{colorshared}{HTML}{006E05}
\definecolor{colorserver}{HTML}{2E499E}
\definecolor{colorclient}{HTML}{838C00}
\newcommand\ddotop{\operatorname{:}}
\newcommand\rg[2]{\left[#1;#2\right]}

\newcommand\addedx[2][]{{\color{blue}#2}}
\newcommand\addedno[2][]{#2}
\newcommand\addedhide[2][]{{#1}}
\newcommand\added{\addedno}
\newcommand\addedrule{\addedno}

\newcommand\etiny{\eliom{}$_\varepsilon$\xspace}

\newcommand\emodule{\eliom{}$_m$\xspace}
\newcommand\ocsi{\mysc{ML$_\varepsilon$}}
\newcommand\ocsis{\mysc{ML\textsubscript{s}}}
\newcommand\ocsic{\mysc{ML\textsubscript{c}}}

\newcommand\rf[1]{\mathbf{#1}}

\newcommand\letm[3][\loc]{\mathtt{let}_{\added{#1}}\ #2 = #3}
\newcommand\valm[3][\loc]{\mathtt{val}_{\added{#1}}\ #2 : #3}
\newcommand\typem[3][\loc]{\mathtt{type}_{\added{#1}}\ #2 = #3}
\newcommand\typeabsm[2][\loc]{\mathtt{type}_{\added{#1}}\ #2}

\newcommand\letg{\mathtt{let}\ }

\newcommand\Y{\mathtt{Y}}
\newcommand\letin[3]{\mathtt{let}\ #1 = #2\ \mathtt{in}\ #3}
\newcommand\lam[2]{\lambda #1 . #2 }
\newcommand\closure[3]{\lambda #1.#2.#3}

\newcommand\injf[2]{#2\mathtt{\%}#1}
\newcommand\cv[1]{\{\{~#1~\}\}}

\newcommand\prog[1]{\mathtt{prog}\ #1\ \mathtt{end}}
\newcommand\ret{\mathtt{return}}

\newcommand\bindg{\mathtt{bind}\ }
\newcommand\pbind{\mathtt{bind}}
\newcommand\bind[2]{\pbind\ \rf{#1} = #2}
\newcommand\bindw[3]{\pbind\ \rf{#1} = #3\ \texttt{with}\ \rf{#2}}
\newcommand\bindenv[1]{\pbind\ \mathtt{env}\ \rf{#1}}

\newcommand\conv[2]{#1 \operatorname{\leadsto} #2}
\newcommand\serial{\mathtt{serial}}
\newcommand\fragment{\mathtt{frag}}

\newcommand\type[1]{\mathrm{#1}}
\newcommand\lamtype[2]{\type{#1} \operatorname{\rightarrow} \type{#2} }
\newcommand\app[2]{(#1\ #2)}
\newcommand\appty[2]{(#1)\type{#2}}
\newcommand\cvtype[1]{\{\type{#1}\}}

\newcommand\typicalt[1][*]{\type{t_i}}

\newcommand\mm{M}
\newcommand\vm{V}
\newcommand\sm{S}
\newcommand\dm{D}
\newcommand\Pm{P}
\newcommand\dstruct{\mathtt{struct}}
\newcommand\struct[1]{\dstruct\ #1\ \mathtt{end}}
\newcommand\appm[2]{#1(#2)}
\newcommand\constraint[2]{(#1 \ddotop #2)}
\newcommand\functor[4][]{\mathtt{functor}_{\added{#1}}\constraint{#2}{#3} #4}
\newcommand\functori[4][]
{\mathtt{functor}_{\added{#1}}\constraint{#2_i}{#3_i}_i #4}
\newcommand\functorenv[5][]
{\mathtt{functor}_{\added{#1}}(#2)\constraint{#3}{#4} #5}
\newcommand\functorenvi[5][]
{\mathtt{functor}_{\added{#1}}(#2)\constraint{#3_i}{#4_i}_i #5}
\newcommand\modulem[3][\mloc]{\mathtt{module}_{\added{#1}}\ #2 = #3}
\newcommand\signature[1]{\mathtt{sig}\ #1\ \mathtt{end}}
\newcommand\moduletym[3][\mloc]{\mathtt{module}_{\added{#1}}\ #2 : #3}
\newcommand\emptym\varepsilon

\newcommand\Mm{\mathcal M}
\newcommand\Sm{\mathcal S}
\newcommand\Dm{\mathcal D}
\newcommand\Xm{\mathcal X}

\newcommand\locs[1]{\operatorname{locations}(#1)}
\newcommand\sideX{m}
\newcommand\base{b}
\newcommand\cins{c/s}
\newcommand\loc{\ell}
\newcommand\loci{{\loc_i}}
\newcommand\csloc{\iota}
\newcommand\mloc{\varsigma}
\newcommand\subloc{\operatorname{\succ}}
\newcommand\notsubloc{\operatorname{\nsucc}}
\newcommand\canuseloc[2]{#1 \subloc #2}
\newcommand\cannotuseloc[2]{#1 \notsubloc #2}

\newcommand\inclloc{\operatorname{<:}}
\newcommand\canbedefloc[2]{#1 \inclloc #2}
\newcommand\withinlocs[3]{#1 \inclloc (#2 \subloc #3)}

\newcommand\substloc[3]{#3_{\left[#1 \mapsto #2\right]}}
\newcommand\substml[2][\loc]{\substloc{\ml}{#1}{#2}}

\newcommand\Env{\Gamma}

\newcommand\binding[3][\loc]{(\valm[#1]{#2}{#3})}
\newcommand\bindingty[3][\loc]{(\typem[#1]{#2}{#3})}
\newcommand\bindingabs[2][\loc]{(\typeabsm[#1]{#2})}
\newcommand\bindingm[3][\mloc]{(\moduletym[#1]{#2}{#3})}

\newcommand\iswt{\operatorname{\triangleright}}
\newcommand\iswtm{\operatorname{\blacktriangleright}}
\newcommand\wt[4][\loc]{#2 \iswt_{\added{#1}} #3 \ddotop \type{#4}}

\newcommand\wtm[4][\mloc]{#2 \iswtm_{\added{#1}} #3 \ddotop #4}

\newcommand\isequiv{\operatorname{\approx}}
\newcommand\equivty[4][\loc]{#2 \iswt_{\added{#1}} #3 \isequiv #4}

\newcommand\iswf{\operatorname{\vDash}}
\newcommand\wf[3][\loc]{#2 \iswf_{\added{#1}} #3}

\newcommand\wfm[3][\mloc]{#2 \iswf_{\added{#1}} #3}

\newcommand\bv[2][\loc]{\operatorname{BV_{\added{#1}}}(#2)}
\newcommand\ftv[1]{\operatorname{FreeTypeVar}(#1)}
\newcommand\injs[1]{\operatorname{INJS}(#1)}
\newcommand\frags[1]{\operatorname{FRAGS}(#1)}

\newcommand\specialize[3]{\added{\lfloor}#3\added{\rfloor_{#2}}}
\newcommand\strengthen[2]{#1/#2}

\newcommand\issubmod{<:}
\newcommand\submod[4][\mloc]{#2 \iswtm_{\added{#1}} #3 \issubmod #4}

\newcommand\instanceof[2]{#2 \operatorname{\succ} #1}
\newcommand\Close[2]{\text{\em Close}(#2,#1)}

\newcommand\Const{\text{\em Const}}

\newcommand\Typeof{\text{\em TypeOf}}

\newcommand\Cprim{E}
\renewcommand\C[1]{\Cprim[ #1 ]}
\newcommand\hole[1][\ ]{[#1]}

\newcommand\compile[3][]{\left\langle#3\right\rangle_{#2}^{#1}}

\newcommand\restrict[2]{#2\rvert_{#1}}

\newcommand\idclosure{\rf{f}}

\newcommand\injd{\mathtt{injection}}
\newcommand\pinj[2]{\mathtt{injection}\ \rf{#1}\ #2}
\newcommand\fragmentd{\mathtt{fragment}}
\newcommand\pfragment[2]{\mathtt{fragment}\ \rf{#1}\ #2}
\newcommand\dend{\mathtt{end}}
\newcommand\pend{\dend\ ()}
\newcommand\dexec{\mathtt{exec}}
\newcommand\pexec{\dexec\ ()}

\newcommand\pbinddyn{\mathtt{bind}_m}
\newcommand\binddyn[2]{\pbinddyn\ #1 = #2}
\newcommand\dyn{\mathtt{Dyn}}
\newcommand\fragm[2]{\mathtt{fragment}_m\ \rf{#1}\ #2}
\newcommand\getdyn[1]{#1.\dyn}

\newcommand\fragty{\mathtt{frag}}

\newcommand\subst[4][\loc]{#4\!\left[#2 \mapsto_{\added{#1}} #3\right]}
\newcommand\substi[4][\loc]{#4\!\left[#2 \mapsto_{\added{#1}} #3\right]_i}
\newcommand\substs[5][\loc]{#5\!\left[#2 \mapsto_{\added{#1}} #3\ \middle|\ #2\in#4\right]}
\newcommand\substenv[2]{#2\!\left[#1\right]}

\newcommand\envinj{\zeta}
\newcommand\envg{\gamma}

\newcommand\envfrag{\xi}
\newcommand\tokfrag[2]{\left\{\rf{#1}\operatorname{\!\mapsto\!}#2\right\}}
\newcommand\tokend{\dend}
\newcommand\cfrag{\operatorname{\!+\!\!+\!}}
\newcommand\emptyfrag{[\ ]}

\newcommand\clientprog{\mu}

\newcommand\trace{\theta}
\newcommand\singltr[1]{\langle#1\rangle}
\newcommand\niltr{\singltr{}}
\newcommand\ctr{\operatorname{@}}
\newcommand\print{\mathtt{print}}

\newcommand\envr{\rho}
\newcommand\bdr[2]{\left\{#1\operatorname{\!\mapsto\!}#2\right\}}
\newcommand\bdrin[3]{#3(#1) = #2}
\newcommand\emptyr{\{\}}

\newcommand\pred[2][]{\operatorname{\xRightarrow[#1]{\ #2\ }}}
\newcommand\redel[6][\mloc]
{#3\pred{#2}_{\added{#1}} #4\added{, #5}, #6}
\newcommand\reds[5]{\redel[s]{#1}{#2}{#3}{#4}{#5}}
\newcommand\redx[5]{\redel[\sideX]{#1}{#2}{#3}{#4}{#5}}
\newcommand\redxe[6]{\redel[\sideX]{#1}{#2}{#3}{#4,#5}{#6}}
\newcommand\redto[4]{\redel{#1}{#2}{#3}{\emptym}{#4}}
\newcommand\redp[4]{#2\pred{#1}#3, #4}
\newcommand\redc[6]
{#2\pred{#1\operatorname{|}#5\rightarrow#6}_{\added{c}} #3, #4}

\newcommand\redmls[6]
{#2\pred{#1}_{\added{\ocsis}} #3, #4, #5, #6}
\newcommand\redmlc[8]
{#2, #4\pred{#1\operatorname{|}#6\rightarrow#7}_{\added{\ocsic}} #3, #5, #8}
\newcommand\redmlce[6]
{#2\pred{#1\operatorname{|}#4\rightarrow#5}_{\added{\ocsic}} #3, #6}

\newcommand\appconst[3][\loc]{\delta_{\added{#1}}(#2,#3)}

\newcommand\noinj[1]{\overline{#1}}
\newcommand\pinjval{\operatorname{\downarrow\!}}
\newcommand\injval[1]{\pinjval #1}

\newcommand\equivc[1]{\simeq^c_{#1}}
\newcommand\equivs[1]{\simeq^s_{#1}}

\newcommand\intty{\mathtt{int}}

\newcommand\fragenv[1]{FCE(#1)}
\newcommand\co[1]{\widehat{#1}}

\newenvironment{subproof}[1][\proofname]{%
  \begin{proof}[$\bullet$ #1]%
}{%
  \end{proof}%
}

\mprset{flushleft}

\ifsubmit
\renewcommand{\TODO}[1]{}
\fi

\usepackage{tikz}
\usetikzlibrary{backgrounds,positioning,shapes,shadings,shadows,arrows,decorations.pathmorphing,calc,fit,fadings}

\tikzstyle{class}=[rectangle, draw=black,scale=0.8, fill=yellow!20, drop shadow, text width=3cm, rectangle split, rectangle split parts=2]
\tikzstyle{uml}=[class, scale=0.68, rectangle split parts=3]
\tikzstyle{interface}=[uml]

\tikzstyle{contenant}=[liaison, open diamond-]

\tikzstyle{impl}=[liaison, -open triangle 45, dashed]
\tikzstyle{manip}=[liaison,-angle 45,dashed]
\tikzstyle{extends}=[liaison, -open triangle 45]
\tikzstyle{navig}=[liaison,-angle 45]

\tikzstyle{flow}=[draw=black,rectangle,drop shadow,text width = 3cm,text centered,scale=0.8,left color=yellow!60, right color = white]
\tikzstyle{flow_algo}=[flow,left color=blue!80]
\tikzstyle{flow_choix}=[flow,diamond,text width=2.2cm]

\tikzstyle{flow_arrow}=[->,>=latex,thick]
\tikzstyle{flow_not_arrow}=[flow_arrow,*->]

\tikzstyle{bloc}=[rectangle, draw=black, fill=yellow!20, drop shadow]
\tikzstyle{etiquette}=[black!60,scale=0.8]

\tikzstyle{link}=[->,>=latex, rounded corners=6pt]
\tikzstyle{Tlink}=[link, thick]
\tikzstyle{vTlink}=[link, very thick]

\tikzstyle{inter}=[<->,>=latex, double, very thick, rounded corners=6pt]

\tikzstyle{boite}=[draw, fill=white, thick, rounded corners=2pt]

\theoremstyle{definition}
\newtheorem{proposition}{Proposition}

\usepackage{newfloat}
\DeclareFloatingEnvironment[
listname={List of Examples},
name={Example},
placement={!ht},
]{ex}
\addto\extrasenglish{}

\newcommand{\aref}[1]{\hyperref[#1]{Appendix~\ref{#1}}}
\usepackage[sort,noabbrev,nameinlink,capitalize]{cleveref}
\crefname{ex}{Example}{Examples}
\crefname{subex}{Example}{Examples}
\crefname{subfigure}{Figure}{Figures}

\begin{document}

\title{\eliom: A Language for Modular Tierless Web Programming}

\author{Gabriel Radanne}
\affiliation{
  \institution{University of Freiburg}
  \country{Germany}
}
\email{radanne@informatik.uni-freiburg.de}
\author{Jérôme Vouillon}
\affiliation{
  \department{IRIF UMR 8243 CNRS}
  \institution{Univ Paris Diderot, Sorbonne Paris Cité}
  \city{Paris}
  \country{France}
}
\affiliation{
  \institution{BeSport}
  \city{Paris}
  \country{France}
}
\affiliation{
  \institution{CNRS}
  \country{France}
}
\email{jerome.vouillon@irif.fr}

\author{Vincent Balat}
\affiliation{
  \department{IRIF UMR 8243}
  \institution{Univ Paris Diderot, Sorbonne Paris Cité}
  \city{Paris}
  \country{France}
}
\affiliation{
  \institution{BeSport}
  \city{Paris}
  \country{France}
}
\email{vincent.balat@irif.fr}

\thanks{This work was partially performed at IRILL, center for
    Free Software Research and Innovation in Paris, France,
    \url{http://www.irill.org}}

\begin{abstract}
  Tierless Web programming languages allow programmers to combine client-side and server-side programming
  in a single program.
  Programmers can then define components with both client and server parts
  and get flexible, efficient and typesafe client-server communications.
  However, the expressive client-server features found in most
  tierless languages are not necessarily compatible with
  functionalities found in many mainstream languages.
  In particular, we would like to benefit from type safety,
  an efficient execution, static compilation, modularity and separate compilation.

  In this paper, we propose
  \eliom, an industrial-strength tierless functional Web programming language
  which extends \ocaml with support for rich client/server interactions.
  It
  allows to build whole applications as a single distributed program,
  in which it is possible to define modular tierless libraries
  with both server and client behaviors and combine them effortlessly.
  \eliom is the only language that
  combines type-safe and efficient client/server communications
  with a static compilation model that supports separate compilation and
  modularity. It also supports excellent integration with
  \ocaml, allowing to transparently leverage its ecosystem.

  To achieve all these features, \eliom borrows ideas not only from distributed
  programming languages, but also from meta-programming and modern module systems.
  We present the design of \eliom, how it can be used in practice
  and its formalization; including its type system, semantics and compilation
  scheme. We show that this compilation scheme preserves typing
  and semantics, and that it supports separate compilation.
\end{abstract}

\maketitle

\section{Introduction}

Writing websites nowadays requires the use of many different technologies:
\mysc{HTML} and \mysc{CSS} for the content and visuals of the website,
\js for client interactions,
any of the numerous server languages (\mysc{PHP}, Ruby, C\#, \dots), a database
language such as SQL, etc.
Not only do programmers need to use all these technologies, but they also
need them to cooperate in an harmonious manner by ensuring
that communications between these various components are correct.
Such verification is often done manually by the programmer, which is both
error-prone and time-consuming.

This juggling of many different technologies does not only make programming more
complicated, it also imposes strong constraints on code organization and
prevents modularity. Let us consider the case where we want to add a comment
widget to a website.
This widget
requires \js code for client interactions, for example a convenient editor.
It also
requires server code that will store and serve the comments and potentially
some associated database queries. This code will thus be split in two codebases and
up to three languages. Furthermore, in order to be properly integrated in the
larger website, the internal communications between the client
and server parts of the widget will be exposed to the rest of the program.
All these
programming constraints, which stem directly from the way Web programming
is done currently, make it impossible to preserve abstraction and
encapsulation for widgets who have both client and server aspects.
This hinders safety and reusability of widgets by
preventing programmers to create self-contained independent libraries that
have both client and server aspects.

\subsection{\ocsigen}

\ocsigen provides a comprehensive set of tools and libraries
for developing Web applications in \ocaml.
\ocaml is an industrial strenght statically typed functional programming language
with a rich ecosystem.
The \ocsigen project includes the compiler
\jsocaml~\citep{SPE:SPE2187}, a Web server, libraries for
concurrency~\citep{DBLP:conf/ml/Vouillon08} and HTML
manipulation~\citep{tyxml}.
It also contains a complete framework to develop complex Web application
with both client and server components.
It includes modules for, among other things, RPCs,
Web GUIs~\cite{ocsigentoolkit},
functional reactive Web programming and
an advanced {\em service identification mechanism}~\cite{Balat:webinteraction:intech2014}.
\ocsigen is already used in production for a wide variety of websites
\citep{besport,gencore,pumgrana}.

Based on our experience building such a framework, we realized we needed
additional tools to bridge the gap between client and server components.
Indeed, while \ocaml provides excellent support for abstraction and modularity
through its module language, it doesn't allow to talk directly about the
relationship between client and server code.
In particular, we needed language constructs that allows us to blur the boundaries
between client and server.
Such language constructs can notably be provided by tierless languages.

\subsection{Tierless Web Programming Languages}

Tierless languages~\citep{ur/web,HopProposal,LinksProposal} allow programmers to
write in a composable way programs that have both client and server aspects.
They provide language constructs to define on the server functions that create
fragments of a Web page together with their associated client-side behavior.
This is done by allowing to freely intersperse client and
server expressions with seamless communication in a unique programming language.
By grouping together client and server parts, they allow to encapsulate
hybrid libraries and create more modular websites.
To be executed, tierless programs are sliced in two:
a part which runs on the server and a part
which is compiled to \js and runs on the client.
This slicing can be done either dynamically, by generating \js code
at runtime, or statically, by cutting the program in two during compilation.
Tierless languages can also leverage static typing to statically ensure that
communications between client and server are always correct.
Such tierless languages can also leverage both functional programming and
static typing.
Functional programming provides increased expressiveness and flexibility while
static typing gives strong guarantees about client-server separation,
in particular ensuring that communications are consistent
across tiers.

To complete the design of \ocsigen, we needed a language that not only
provides such expressive tierless features, but
also provides support for most features already supported by
\ocaml: type safety, efficiency, static compilation, modularity and separate
compilation.
Furthermore, Web programming often relies on numerous
external libraries (for encryption, HTML, logging, \dots). We wanted
to leverage all the existing tools and libraries developed either in the
\ocsigen project or the larger \ocaml ecosystem.

\subsection{The design of a Modular Tierless language}

To build such a language with support for both modular and tierless features,
we outline several goals that will guide the design of our language.

% In order to support so many features, \eliom relies on two main ingredients.
% First, an \emph{expression language} which provides fine-grained modularity by allowing programmers to manipulate
% \emph{on the server}, as first class values, fragments of code which will be executed \emph{on the client}.
% Second, an \ml-style \emph{module language} which provides large scale
% modularity, encapsulation, good \ocaml integration
% and incremental compilation.
% %
% Given the many choices involved in designing a tierless languages,
% we consider 

\paragraph{Explicit communications.}
Manual annotations should determine whether a piece of
code is to be executed server- or
client-side. This design
decision stems from our belief that the programmer must be well aware
of where the code is to be executed, to avoid unnecessary remote
interaction. Explicit annotations also prevent ambiguities in the
semantics, allow for more flexibility, and enable the programmer to
reason about where the program is executed and the resulting
trade-offs. Programmers can thus ensure that some data stays on the
client or on the server, and choose how much communication takes
place.

\paragraph{A simple and efficient execution model.}
The language should not incur additional back-and-forth communications
between client and server. Furthermore, the execution model should
be simple and predictable.
Having a predictable execution model is essential in the context of a language which is not purely functional, like \ocaml.

\paragraph{Leveraging the type system.}
The language should leverages the type system to allows composition and
modularity of client-server programs while preserving type-safety and abstraction.
This ensures, via the type-system, that client functions are not called
by mistake inside server code (and conversely) and ensures the correctness of client-server communications.

\paragraph{Integration with the host language.}
Programmers must be able to leverage both
the language and the ecosystem of \ocaml. \ocaml libraries can be useful
on the server, on the client or on both.
As such, any \ocaml file, even when compiled with the regular \ocaml
compiler, should be a valid module in our language.
Furthermore, we should be able to specify
if we want to use a given library on the client, on the server, or
everywhere.

\paragraph{Modularity and encapsulation.}
Module and type abstractions are very powerful programming tools. By only exposing
part of a library, the programmer can safely hide implementation details
and enforce specific properties.
The language should leverages module abstraction to provide encapsulation
and separation of concern for widgets and libraries.
By combining module abstraction and tierless features, library authors can
provide good APIs that do not expose the fine-grained details of client-server
communication to the users.

\subsection{Contributions and Plan}

We propose \eliom,
a tierless web programming language that supports static typing,
an efficient compilation scheme that avoids extra communications
and a very powerful form of modularity inspired by \ml-style languages.
In particular, \eliom is the only tierless programming language
featuring efficient static separate compilation.
\eliom extends \ocaml
and can transparently leverage its complete ecosystem.
This article presents the design of \eliom, how it can
be used in practice, and formalizes the underlying core language.
Our contributions are the following:

\begin{itemize}[leftmargin=*]
\item
  We designed \eliom, a tierless web programming language that follows the
  set of goals presented above.
  We demonstrate the practicality of our language through
  examples (\cref{sec:howto}).
\item
  We formalized the type system and the semantics of \eliom,
  including both the expression and the module layers (\cref{sec:theory}).
  This formalization provides an \emph{interpreted} semantics that is easy to
  reason with and can be explained to programmers, but still maintains
  \eliom's good properties regarding communications and typesafety.
  Using this semantics, we show that \eliom supports
  separate typechecking (\cref{eliom:separation})
  and complete integration with \ocaml (\cref{sec:base,theorem:base}).
\item
  As highlighted before, \eliom is not an interpreted language.
  We formalized the compilation model of \eliom (\cref{sec:compilation}).
  This compilation scheme turns one \eliom program into two simpler
  \ocaml programs; one for the client and one for the server.
  We show that such compilation scheme
  supports separate compilation and
  preserves both the typing and the semantics of \eliom
  (\cref{thm:simulation:type,thm:simulation}).
\item
  We implemented \eliom as a patch on the \ocaml compiler
  and a runtime library\footnote{\url{https://github.com/ocsigen/ocaml-eliom} and \url{https://github.com/ocsigen/eliomlang}}.
\end{itemize}

%%% Local Variables:
%%% mode: latex
%%% TeX-master: "main"
%%% End:

\section{Programming with \eliom}
\label{sec:howto}

An \eliom application is composed of a single program that has both client and
server behaviors.
During compilation, the \eliom compiler decomposes the program into two parts.
The first part runs on a Web server, and is able to manage several connections and sessions at the same time.
The second part, compiled
statically to \js, is sent to each client by the Web server together with
the \html page, in response to the initial HTTP request.

In this section, we give a tour of the \eliom language through various examples.
As a guiding example for our exploration of the \eliom language, we consider
the case of a commenting system similar to websites such as Reddit or Hackernews,
A comment is a piece of HTML written by a user and identified by
a unique identifier.
Comments are initially stored on the server. Users can then manipulate
them on the client by voting, hiding or searching them.
Such a library features both server
aspects (storing and rendering the comments) and client
interactions (browsing and searching comments).
We start by giving a reminder of some important concepts on \ocaml modules
before presenting the various \eliom concepts.
Even
though \eliom is based on \ocaml, little knowledge of \ocaml is required.
We explicitly write some type annotations for
illustration purposes but they are not mandatory.

\subsection[Of comments and camels]{Of comments and camels -- 
  A short introduction to \ocaml modules}
\label{tuto:modules}

The \ocaml module system forms a second language separate from the expression language.
While the language of expressions allows programming ``in the small'',
the module language allows programming ``in the large''.
The \ml flavor of module systems, which \ocaml is part of, significantly extend
usual module languages by providing module types (called signatures) and
functions from modules to modules (called functors).
The module system
is implicitly used for any kind of \ocaml or \eliom programming:
Each \ocamlc{.ml} and \ocamlc{.eliom} file form a structure containing the list of declarations
included in the file. It is also possible to specify a signature for such
module by adding a \ocamlc{.mli} or \ocamlc{.eliomi} file.
We can do a lot more with \ocaml modules. For example, let us say we
are writing an \ocamlc{HTML} library. We want to gather the event related
attributes in a single module. We can easily do so with the following construction.

\begin{lstlisting}[language=eliom,name=html.ml]
module On = struct
  let click = ...
  let keypress = ...
end
\end{lstlisting}

These functions can then be used through qualified accesses:

\begin{lstlisting}[language=eliom]
open Html
let mywidget = div ~a:[On.click myclickhandler] [ ... ]
\end{lstlisting}

Some users of our HTML library may want to experiment with new, custom-made
HTML elements. They can easily do so by extending the \ocamlc{Html} module:

\begin{lstlisting}[language=eliom]
module HtmlPlus = struct
  include Html
  let blink elems = ...
end
\end{lstlisting}

Here, we declare a new module, \ocamlc{HtmlPlus}, in which we
\emph{include} \ocamlc{Html} and define the new \ocamlc{blink} function.
The include operation simply takes all the fields of a module and adds them
to the enclosing module. This way, we obtain a new module \ocamlc{HtmlPlus}
which can be used anywhere \ocamlc{Html} can, but also includes the new function.

\subsubsection{Abstraction and encapsulation}

\begin{figure}[!b]
  \centering
  \begin{subfigure}{0.32\linewidth}
\begin{lstlisting}[language=eliom]
module type ID = sig
  type t (* type of ids *)

  val compare : t -> t -> int
  val create : unit -> t
  val to_string : t -> string
end
\end{lstlisting}
  \end{subfigure}
  \hfill
  \begin{subfigure}{0.35\linewidth}
\begin{lstlisting}[language=eliom]
module SequentialID : ID = struct
  type t = int

  (* ... *)
end
\end{lstlisting}
  \end{subfigure}
  \hfill
  \begin{subfigure}{0.3\linewidth}
\begin{lstlisting}[language=eliom]
module DateID : ID = struct
  type t = date

  (* ... *)
end
\end{lstlisting}
  \end{subfigure}
  \caption{The \ocamlc{ID} signature and two implementations.}
  \label{ex:id:mod}
\end{figure}

We now want to build a simple library to handle internet comments.
In our library,
comments are pieces of HTML (constructed with the \ocamlc{Html} module)
identified by a unique number. We are not sure yet if we should use
simple sequential IDs, date-based IDs or something else
like UUIDs and \citet{hashids}.
Fortunately, we do not have to make this decision immediately! All we need
in order to
write the rest of our library is an interface for creating and using identifiers.
We can declare such an interface in \ocaml using a \emph{signature}.
In \cref{ex:id:mod}, we declare the \ocamlc{ID} signature
describing what a module implementing unique identifiers should look like.
We then define two modules implementing this specification,
\ocamlc{SequentialID} and \ocamlc{DateID},
which we can switch easily.

To the outer world, these two modules
have exactly the same type and can not be distinguished. The type that implements
the identifiers in the \ocamlc{ID} signature is abstract: its implementation is
only visible inside the module and can not be used outside. It is also useful to
note that such abstraction can be provided after the fact. Declaring a module
and abstracting its interface are completely distinct operations.

Hiding the internal details of our \ocamlc{ID} modules is not only useful
for modularity: it also allows to enforce abstraction boundaries. For example in
the case of \ocamlc{SequentialID}, it is impossible to inadvertently use
the ID as an integer, since the fact that it is an integer is not revealed!
We can use this fact to enforce numerous complex properties, as we see in
the rest of this section.

\subsubsection{Functors}
\label{tuto:functor}

To implement our comment system, we sometimes need to find comments by their ID.
The idiomatic \ocaml solution is to use maps, also called dictionaries. Such
maps are implemented with Binary Search Trees which require a comparison function
on the keys of the map.
\eliomc{Map.Make} is a pre-defined functor in the \ocaml standard library that takes
a module implementing the \eliomc{COMPARABLE} signature as argument
and returns a module that implements dictionaries
whose keys are of the type \ocamlc{t} in the provided module.
In \cref{ex:map}, we use this functor to create the \ocamlc{IDMap} module
which defines dictionaries with IDs as keys.
This is very easy, since the \ocamlc{ID} signature is already a super-set
of the \ocamlc{COMPARABLE} signature.
We then define \ocamlc{register},
a function which associates a fresh id to a comment \ocamlc{c}.

The \ocamlc{Map.Make} functor uses abstraction in two important ways.
First, since the type of the map is abstract, it is impossible to modify it through
means not provided by the module. In particular, this enforces that the binary tree
is always balanced. Second, since the comparison function is provided
in advance by the argument of the functor,
it is impossible to mix different comparison functions by mistake. Indeed,
application of the functor to different modules would yield different types of
\label{ex:id:mod}maps.

\begin{figure}[!h]
  \centering
  \hfill
  \begin{minipage}[c]{0.4\linewidth}
\begin{lstlisting}[language={eliom}]
module type COMPARABLE = sig
  type t
  val compare : t -> t -> int
end

module Make (Key : COMPARABLE) : sig
  type 'a t
  val empty : 'a t
  val add : Key.t -> 'a -> 'a t -> 'a t
  (* ... *)
end
\end{lstlisting}
  \caption{the \ocamlc{Map} module}
  \label{api:comparable}
  \end{minipage}\hfill
  \begin{minipage}[c]{0.5\linewidth}
\begin{lstlisting}[language={eliom}]
module TheID = DateID (* The ID of our choice *)
module IDMap = Map.Make(TheID)

let register c map : Html.t IDMap.t =
  let commentid = TheID.create () in
  IDMap.add commentid c map
\end{lstlisting}
  \caption{Dictionaries from IDs to comments}
  \label{ex:map}
\end{minipage}
\hfill
\end{figure}

% \subsubsection{Going further}
% This was just a taste of modules.
% For a longer introduction to modules, please
% consult the \ocaml manual~\citep{Ocamlmanual}
% or the Real World OCaml book~\citep{RWO}.

%%% Local Variables:
%%% mode: latex
%%% TeX-master: "../main"
%%% End:

\subsection{Tierless widgets}
\label{tuto:small}

Until now, we presented how to write various elements of libraries useful for
our comment system by leveraging the power of the \ocaml module
system in various ways.
We now want to write the widget that presents a comment. We thus
need to define both client and server code, along with some client-server
communication,
which is precisely where tierless languages shine.

\subsubsection{Declarations}
\label{tuto:section}

Locations in \eliom are explicit. Each declaration must be marked with
an annotation that specifies whether a declaration is to be performed
on the server or on the client as follows:

\inputeliom{code/section.eliom}

Every declaration, such as types, values and modules, can be annotated.
These annotations allows to group related code in the
same file, regardless of where it is executed.
In the rest of this article, we use the following color convention:
client is in \textbf{\color{colorclient} yellow} and
server is in \textbf{\color{colorserver} blue}.
Colors are however not mandatory to understand the rest of this article.

\subsubsection{Fragments}
\label{tuto:clientvalue}

While location annotations allow programmers to gather code across locations,
they don't allow convenient communication.
For this purpose, \eliom allows to include
client-side expressions inside server declarations:
an expression placed inside \eliomc{[\%client ... ]} will be computed on
the client when the page is received;
but the eventual client-side value of the expression can
be passed around immediately as a black box on the server.
These expressions are called client \emph{fragments}.

In the example below, the expression \eliomc{1 + 3} will be evaluated on
the client, but it's possible to refer server-side to the future value of this
expression (for example, put it in a list).
The variable \ocamlc{x} is only usable server-side, and has type
\eliomc{int fragment} which should be read ``a fragment containing some integer''.
The value inside the client fragment cannot be accessed on the server.

\inputeliom{code/cvalue.eliom}

\subsubsection{Injections}
\label{tuto:injection}

Fragments allow programmers to manipulate client values on the server.
We also need the opposite direction.
Values that have been computed on the server can be used on the client
by prefixing them with the symbol \eliomc{\~\%}. We
call this an \emph{injection}.
\inputeliom{code/inj.eliom}
Here, the expression \eliomc{1 + 2} is evaluated and bound to variable
\eliomc{s} on the server.
The resulting value \eliomc{3} is transferred to the client together with
the Web page.
The expression \eliomc{\~\%s + 1} is computed client-side.

An injection makes it possible to access client-side a client fragment
which has been defined on the server. The value inside the client fragment is
extracted by \eliomc{\~\%x}, whose value is \eliomc{4} here.
\inputeliom{code/injclientvalue.eliom}

\subsubsection{Comment widget}
\label{tuto:widget}

These three constructions are sufficient to create complex client-server
interactions such as the comment widget.
The comment widget shows a single comment and can be used several times
to show a list of comments.
It not only shows the author and the content of the comment, but
will also hide the content when
the user clicks on it.
Finally, we want the HTML to be generated server-side and
sent to the client as a regular HTML page, which allows the comments to be
accessible even when \js cannot run.
The implementation, the interface and the produced HTML
fragment are shown in \cref{ex:comment}.

In order to implement our comment widget,
we use an HTML DSL~\citep{tyxml} that provides combinators such as
\ocamlc{div} and \ocamlc{a_onclick} (which respectively create an HTML tag
and an HTML attribute).
In \ocaml, \xspace\ocamlc{\~a} denotes a named argument that is used here 
to provide the list of \html attributes.
We first create a \ocamlc{p} element
which contains the text of the comment and a unique id. The text is
included in a \ocamlc{div} which represents the comment.
We then use a handler listening to the \ocamlc{onclick} event: since
clicks are performed client-side, this handler needs to be a client
function inside a fragment.
Inside the fragment, an injection is used to access the
argument \ocamlc{id} which contains the identifier of the comment.
We then use this identifier to fetch the correct element and toggle the
``hidden'' CSS property, which hides it.

The signature of our widget function, shown \cref{ex:comment:intf},
does not expose the internal
details of the widget's behavior. In particular, the
communication between server and client does not leak
in the API: This provides proper encapsulation for
client-server behaviors.
Furthermore, this widget is easily composable: the embedded
client state cannot
affect nor be affected by any other widget and can be used to
build larger widgets.

\begin{figure}[t]
  \centering
  \begin{subfigure}[t]{0.55\linewidth}
    \inputeliom{code/comment.eliom}
    \caption{Implementation}
  \end{subfigure}\hfill
  \begin{subfigure}[t]{0.42\linewidth}
    \inputeliom{code/comment.eliomi}
    \caption{Interface}
    \label{ex:comment:intf}
    \vspace{3mm}

\begin{lstlisting}[language=html]
<div>
  Author: The Ocsigen Team
  <p id=42>I'm a composable widget!</p>
</div>
\end{lstlisting}
    \caption{Resulting HTML}
  \end{subfigure}
  \caption{The comment widget}
  \label{ex:comment}
\end{figure}

\subsubsection{Notes on semantics}
\label{tuto:sem}

In the examples above, we showed that we can interleave client and server
expressions and communications in fairly arbitrary manners.
This would be costly if the communication between client and server were done naively.

Instead, the server only sends data once when the Web page is sent.
In particular, in the comment widget presented above, the id of the
comment is not sent for each click.
This is made possible by the fact that client fragments are not
executed immediately when encountered inside server code. Intuitively,
the semantics is the
following. When the server code is executed, the encountered client code
is not executed right away; instead it is just registered for later
execution. The client code is executed only once the Web page has been sent to the client. We also guarantee that client code, be it either
client declarations or fragments, is executed in the order
that it was encountered on the server.

This presentation might makes it seem as if we dynamically create the client
code during execution of the server code. This is not the case.
Like \ocaml, \eliom is statically compiled and separates client
and server code at compile time. During compilation, we statically
extract the code included inside fragments and compile it as part of the client
code to \js. This allows us to provide both an efficient execution scheme
that minimizes communication and preserves side-effect orders
while still presenting an easy-to-understand semantics.
We also benefits from optimizations done by the \jsocaml compiler,
thus producing efficient and compact \js code.

\subsection{Hybrid data-structures}
\label{sec:eliom:mixeddata}

We want to add some buttons to our comment widget for various possible actions
on comments such as ``reply'', ``permalink'', ``report'', etc.
We want our buttons to be created in a uniform manner
as a list of symbolic actions. Some of our actions are normal links
and some are client-side actions.
Thanks to fragments, we can create hybrid data-structures that have
both client and server parts.
\cref{ex:buttonlist} implements such an hybrid data-structure to specify
button actions. The \ocamlc{action} type
is a server type that is either a normal link or a client-side action (here, a function from unit to bool) in a fragment.
The \ocamlc{attr_of_action} function turns actions into HTML attributes, either
a link or an on-click handler. Finally, \ocamlc{button_list} walks through a list
of pairs of names and actions and returns an unordered list of
\texttt{a} elements. 

Creating such hybrid data-structures allows a great flexibility in how
\eliom code is organized by allowing to meld client and server code
at any point in the application.
Having explicit annotations for client and server code of is essential here.
This would be quite difficult to achieve if the delimitation between
client and server values were implicitly inferred.

\begin{figure}[!ht]
  \centering
  \begin{minipage}{0.6\linewidth}
    \inputeliom{code/buttons.eliom}
  \end{minipage}
  \caption{Function generating a list of buttons}
  \label{ex:buttonlist}
\end{figure}

\subsubsection{Dynamic loading of comments}
\label{sec:rpc}

\begin{figure}[!bt]
  \centering
  \begin{minipage}{0.45\linewidth}
    \inputeliom{code/rpc.eliomi}
    \caption{\ocamlc{Rpc} signature}
    \label{api:rpc}
  \end{minipage}\hfill
  \begin{minipage}{0.5\linewidth}
    \inputeliom{code/dynamic-comment.eliom}
    \caption{Dynamic loading of comments with \ocamlc{Rpc}}
    \label{example:rpc}
  \end{minipage}
\end{figure}

In order to handle pages with many comments, we want to only
present a subset of the comment upfront and dynamically load additional comments
when the user asks for it (or, alternatively, when the page scrolls).
This requires dynamic communications which are not directly provided by
fragments and injections, as indicated in the previous section.
For such purposes, \eliom provides the \ocamlc{Rpc} module whose API is
presented in \cref{api:rpc}.

We implement a button which dynamically loads new comments in \cref{example:rpc}.
We first create server-side an RPC endpoint
\code{fetch_comments} with the function \ocamlc{Rpc.create}.
This endpoints is of type \ocamlc{(page_id, Html.t list) Rpc.t}:
it takes as argument the id of the current page
and return the list of associated comments using the pre-defined function
\ocamlc{Comments.from_page}.
The type \ocamlc{Rpc.t} is abstract
on the server, but is a synonym for a function type on the client. Of
course, this function does not contain the actual implementation of
the RPC handler, which only exists server-side.
To use this API, we leverage injections.  By using an injection on
\eliomc{\~\%fetch_comments} on \cref{example:rpc:inj}, we obtain \emph{on the client} a value of type
\eliomc{Rpc.t}. We describe the underlying machinery that we leverage
for converting RPC endpoints into client-side functions in
\cref{sec:example:conv}.  What matters here is that we end up with a
function that we can call like any other; calling it performs the
remote procedure call.
Once we have a way to fetch new comments, we can
create a button \ocamlc{button_load} which, when clicked, loads
the comments. This is done by creating a client-side handler and using
the \ocamlc{On.click} function.

The RPC API proposed in \cref{api:rpc} is ``blocking'': the execution
waits for the remote call to finish before pursuing,
thus blocking the rest of the client program.
Remote procedure calls should be made asynchronously:
the client program keeps running while the call is made and the
result is used when the communication is done.
The actual implementation uses
\lwt~\citep{DBLP:conf/ml/Vouillon08} to express asynchronous
calls in a programmer-friendly manner through promises.
The use of \lwt is pervasive in
the \eliom ecosystem both on the server and on the client.
In this article, we omit mentions of the \lwt types and
operators for pedagogic purposes.

\subsubsection{Converters}
\label{sec:example:conv}

In the RPC API, we associate two types with different
implementation on the server and on the client. We rely on
injections to transform the datastructure when moving
from one side to the other.
This ability to transform data before sending it to the client via an
injection is made possible by the use of
\emph{converters}.
\cref{api:conv} broadly presents the
converter API.
Given a serialization format \eliomc{serial}, a
converter is a pair of a \emph{server} serialization function and a
\emph{client} de-serialization function.  Note that the client and server
types are not necessarily the same. Furthermore,
we can arbitrarily manipulate the value before returning it.
Several predefined converters are available for fragments,
basic \ocaml datatypes, and tuples in the module \ocamlc{Conv}.
Implementation details about converters can be found in
\cref{sec:converter}.

\begin{figure}[!bt]
  \centering
  \begin{minipage}{0.7\linewidth}
    \inputeliom{code/conv.eliomi}
    \caption{Schematized API for converters}
    \label{api:conv}
  % \end{minipage}
  % \hfill
  % \begin{minipage}{0.5\linewidth}
    \inputeliom{code/rpc.eliom}
    \caption{Simplified RPC implementation corresponding to \cref{api:rpc}.}
    \label{api:rpc:impl}
  \end{minipage}
\end{figure}

We can use converters to implement the RPC API (\cref{api:rpc:impl}).
The server implementation of \eliomc{Rpc.t} is
composed of a handler, which is a server function, and a \mysc{URL}
to which the endpoint answers.
Our serialization function only sends the \mysc{URL} of the endpoint.
The client de-serialization function uses this \mysc{URL} to create a function performing an \http request to the endpoint. This way, an RPC endpoint can be accessed simply with an injection.
Thus, for the \eliomc{create} function,
we assume that we have a function \ocamlc{serve}
of type \ocamlc{string -> (request -> answer) -> unit}
that creates an \http handler at a specified \mysc{URL}.
When \ocamlc{Rpc.create} is called, a unique identifier \ocamlc{id} is created, along with a new \http endpoint \ocamlc{"/rpc/id"} that invokes the specified function.

This implementation has the advantage that code using the \eliomc{Rpc}
module is completely independent of the actual \mysc{URL} used. The
URL is abstracted away. Converters preserve abstraction by only
exposing the needed information.

%%% Local Variables:
%%% mode: latex
%%% TeX-master: "../main"
%%% End:

\subsection{Modular Tierless Programming}
\label{tuto:large}

We are now equipped with two tools.
On one hand,
we have a rich and expressive non-tierless module system,
as presented in \cref{tuto:modules}, which provides abstraction and modularity
at the library level.
On the other hand,
we have a powerful tierless
programming language, as presented in \cref{tuto:small}, which allows us to
describe sophisticated client-server behaviors.
In this section, we present how we can bring those two tools together and
reap the numerous benefits of the \ocaml module system in a tierless setting.

\subsubsection{Interaction with \ocaml}
\label{tuto:base}

In the previous examples, we freely used \ocaml modules in server
and client contexts. In \cref{api:conv}, we even defined
an API that has both normal \ocaml declarations and server declarations!
Web programming is never only about the Web. Web programmers needs external
libraries and a rich ecosystem that can not be provided by a fresh new
language. The ability to transparently leverage the \ocaml ecosystem in
\eliom is essential to write web applications productively.

This close integration is provided through the use of a third
location called \textbf{base}. Code located on
base can be used both on the client and on the server.

\begin{lstlisting}[language=eliom]
let%base f x = "Hello "^x^"!"
let%client a = f "client"
let%server b = f "server"
\end{lstlisting}

\eliom-specific features such as fragments and injections are not allowed inside base
code. In fact, base code corresponds exactly to \ocaml code.
This equivalence holds in theory but also in practice, meaning that any \ocaml library compiled by the vanilla \ocaml compiler can be directly reused by
\eliom as being on the base location.
This allows a very smooth integration with the \ocaml
ecosystem.
Furthermore, a given
\ocaml library can be loaded either on base, on the client or on the server,
depending on what the user wants. For example, an \ocaml library manipulating
file descriptors might be better kept only on the server in order to avoid misuse.
The type-checker then raises an error if the library is mistakenly used
on the client.

\subsubsection{Modules and locations}
\label{tuto:module:locs}

Since \ocaml modules, such as the \ocamlc{Html} module defined earlier,
are immediately available as
\eliom modules located on base, we can also use such modules on the client
or on the server.

\begin{lstlisting}[language=eliom]
module%base TextHtml = Html
module%server ServerHtml = TextHtml
let%client l = Html.p [Html.text "Hello client!"]
\end{lstlisting}

Locations are checked by the compiler.
For example, using a server module on the client is forbidden.

\begin{lstlisting}[language=eliom]
let%client x = ServerHtml.text "hello client!" (* (*@\color{red}\ding{56} Error!@*) *)
\end{lstlisting}

It is also possible to reuse \ocaml module types freely. For example, we might
want to define a client module \ocamlc{DomHtml} which shares the exact same
API as the \ocamlc{Html} module, but is implemented using the
Document Object Model that is available on the client. The type
declaration for such a module would then be very simple, as shown below.

\begin{lstlisting}[language=eliom]
module%client DomHtml : Html.Signature = struct
  (* ... *)
end
\end{lstlisting}

We can easily declare a new structure completely on one location. The constraint
is that all the fields on such modules, including submodules, should be on the
same location. For example, a client structure can only contain fields that are
declared on the client.
The following piece of code declares a \ocamlc{JsMap} client module containing
various fields and implementing a dictionary data-structure with \js strings.

\begin{lstlisting}[language=eliom]
module%client JsMap : sig
  type 'a t

  val empty : 'a t
  val add : Js.string -> 'a -> 'a t -> 'a t
  (* ... *)
end
\end{lstlisting}

We can also use functors in client and server code as we would in regular
\ocaml code. Consider the \ocamlc{JsMap} module above. The simplest way to obtain
such a module would be to use the \ocamlc{Map.Make} functor presented
in \cref{tuto:functor}.
We could for example write a \ocamlc{JsDate} module which uses
\js native support for dates. We can then obtain the \ocamlc{JsDateMap} module
simply by applying \eliomc{Map.Make} to the module
\ocamlc{JsDate} defined in \cref{fig:jstr}.
As expected, the module we obtain is directly on the client. We can thus mix and
match client and server modules using the tierless features
and vanilla \ocaml modules. This also works with all the other
module features such as abstraction, high order functors and module
inclusion. In all these cases, the \eliom typechecker ensures that
modules always end up on the appropriate location.

\begin{figure}[ht]\centering
  \begin{minipage}{0.5\linewidth}
\begin{lstlisting}[language={eliom}]
module%client JsDate = struct
  type t = Js.date

  (** Compare by timestamp *)
  let compare x y = compare x##valueOf y##valueOf
end

module%client JsDateMap = Map.Make(JsDate)
\end{lstlisting}
  \caption{Definition of \ocamlc{JsDate} and \ocamlc{JsDateMap}}
  \label{fig:jstr}
\end{minipage}
\end{figure}

\subsubsection{Abstraction and encapsulation across locations}

A common idiom of web programming is to generate some HTML element
on the server, add an \ocamlc{id} to it, and recover the element on the client
through the \ocamlc{get_element_by_id} function.
Indeed, this is exactly what we did in our comment widget in \cref{tuto:widget}.
This is so common, in fact, that it could be considered the ``id design pattern''.
RPCs, channels and other communication APIs also follow the same mechanisms
through the use of uniquely defined URLs.
In all these cases, the means of identification for a given object is
generally passed around directly, as a string, instead of being abstracted.
Since client and server code are usually
written separately, the programmer \emph{must} expose the internal details
to the outer world, including how to identify objects.

One solution, which we used in \cref{sec:example:conv}, is to use
converters and only use the ID as an implementation detail of the client-server
communication. However, it is sometimes beneficial to keep
an explicit ID. 
By combining tierless annotations and the abstraction capabilities provided
by modules, we can have explicit ID that are abstract and type-safe.
\cref{fig:abstractid} presents
an API that encapsulates unique ids for HTML elements.
This new \ocamlc{HtmlID} is \textbf{\color{colorshared} mixed} and
can contain client, server and base declarations. 
It is composed of an \ocaml abstract type, \ocamlc{id}, and two operations.
The server function \ocamlc{with_id} takes an HTML element, generates a fresh
id and returns a pair composed of the HTML element with that id and the id itself.
The client function \ocamlc{find} takes an id and retrieves the associated element
as a DOM node on the client.
The \ocamlc{id} type is abstract. Both the client and the server functions
can use the real definition of \ocamlc{id} since they are both inside the
module. The outer world, however, can not. Mixed modules allow
abstraction to extends over the client-server boundary.
This can provide further benefits in the case of more complex data-structures,
as we will see in the next section.

\begin{figure}[ht]
  \centering
  \hfill
  \begin{subfigure}{0.45\linewidth}
\begin{lstlisting}[language=eliom]
module%mixed HtmlID : sig
  type id

  val%server with_id : Html.t -> Html.t * id

  val%client get : id -> DomHtml.t
end
\end{lstlisting}
    \caption{Interface}
    \vspace{1mm}
    \label{fig:abstractid:intf}
  \end{subfigure}\hfill
  \begin{subfigure}{0.45\linewidth}
\begin{lstlisting}[language=eliom]
module%mixed HtmlID = struct
  type id = string

  let%server with_id elem =
    let myid = random_string () in
    let elem_with_id = Html.add_id elem myid in
    (elem_with_id, myid)

  let%client find myid = get_element_by_id myid
end
\end{lstlisting}
    \caption{Implementation}
    \label{fig:abstractid:impl}
  \end{subfigure}\hfill
  \caption{Abstract HTML ids for client/server communications}
  \label{fig:abstractid}
\end{figure}

% \subsubsection{Flexible locations}

% \TODO{}

\subsubsection{Multi-tiers comments}
\label{tuto:mixeddata}

We now want to implement a system of client-side search and filtering of comments.
The user should be able to search and filter comments directly on the client,
without the need to reload the page.
For this purpose, we need to maintain the sets of comments both
on the server and on the client. One simple way
to do that is to create a replicated cache of comments which ensures
that all the comments available on the server are also available on the client.

We use the \eliomc{Map} module as inspiration and create a functor that
takes as argument a module describing the keys.
The idea is that adding an entry to a
server-side table also adds the element to the client-side
table. Consequently, the server-side representation of a table needs
to include a client-side one.

The result API is shown in \cref{fig:makeshared:intf}.
The resulting module contains both a client and a server side types, both
named \ocamlc{'a table}, which represent the local table. The module also
exposes traditional \ocamlc{Map} functions.
The implementation, shown in \cref{fig:makeshared:impl}, is more
interesting.
We exploit the fact that client and server namespaces are distinct, and
name both client and server map modules \ocamlc{M}.
On the server, the cache is implemented as a pair of a server-side
and a client-side dictionary.
The server-side \eliomc{add} implementation stores a
new value locally in the expected way, but additionally builds a
fragment that has the side-effect of performing a client-side
addition.
The retrieval operation (\eliomc{find}) returns
a shared value that contains both the server side version and the client side.
On the client, however, we can directly use the local values.
Since the client-side type exactly corresponds to a regular
map, we can directly use the usual definitions for the various
map operations. This is done by including the client \ocamlc{M} module on the
client.

Note that this functor cannot be implemented
in a decomposed way without sacrifying either abstraction or modularity.
Indeed, the server implementation relies on
the client-side version of the functor argument (\eliomc{Comparable}) to
implement proper usage of the keys. Furthermore, the signature of the functor
ensures that the server-side and client-side parts of the cache are in sync without
leaking any implementation details. Separating this mixed
functors in two would require exposing the guts of the data-structure.
Abstraction also makes it easy to extend such modules with new features.
For example, it would possible to add full-blown replication through ``push'' or
``pull'' communications between the client and the server.
Thanks to the abstraction provided by the
signature of the module, this can even be done while keeping the
API of the functor unchanged.

\begin{figure}[!tb]
  \centering
  \hfill
  \begin{minipage}{0.48\linewidth}
    \inputeliom{code/mixedfunctor.eliomi}
    \caption{Interface of \ocamlc{MakeCache}}
    \label{fig:makeshared:intf}
    % \end{figure}
    \vspace{1mm}
    % \begin{figure}[ht]
    \centering
    \inputeliom{code/mixedfunctor.eliom}
    \caption{Implementation of \ocamlc{MakeCache}}
    \label{fig:makeshared:impl}
  \end{minipage}
  \hfill
  \begin{minipage}{0.44\linewidth}
\begin{lstlisting}[language=eliom]
module%mixed DateKey = DateID
module%mixed CommentCache = MakeCache(DateKey)

let%server add_comment id cache =
  let html = make_comment id in
  CommentCache.add id html cache

let%server generate_page cache =
  Html.div 
    ~a:[a_id "comments"] 
    [CommentCache.elements cache]

let%client filter_comments predicate cache =
  let filtered_cache = 
    CommentCache.filter predicate cache
  in
  let comment_container =
    get_element_by_id "comments"
  in
  Dom.replace_children
    comment_container
    (CommentCache.elements filtered_cache)
\end{lstlisting}
    \caption{Using \ocamlc{MakeCache}}
    \label{fig:makeshared:usage}
  \end{minipage}
\end{figure}

We can now use this cache for our comment system by using, for example, the
\ocamlc{DateID} module for the keys.
This is done in \cref{fig:makeshared:usage}.
Adding a new comment to the page is done through the \ocamlc{add_comment} server
function. This function creates the associated HTML using the widget defined in
\cref{tuto:widget} and adds it to the cache.
We can then create the webpage containing all the comments simply
by collecting all the comments and putting them inside a \ocamlc{div}.
This is done by the \ocamlc{generate_page} server function.
Finally, the client function \ocamlc{filter_comments} filters the shown
comments on the client. It takes as argument a predicate function and
the current client cache. It uses this predicate function to filter the
cache, using the function \ocamlc{CommentCache.filter}, which directly uses
the equivalent function from the \ocamlc{Map} module.
We then find the HTML element containing all the elements and replace them
by the updated list.
Since the \ocamlc{CommentCache} module also contains a \emph{client-side}
add function, this approach works very well the dynamic loading of comment
presented in \cref{sec:rpc}: Once the RPC returns the new comments, we can add
them to the client-side cache directly.

\paragraph{A note on mixed functors}
Mixed functors unfortunately have some limitations.
Arguments must themselves be mixed modules and injections inside client-side
bindings can only reference elements outside of the functor. Additionally,
there are some limitations regarding nesting of mixed structures and functors.
These limitations are formalized and discussed more precisely in
\cref{sec:functor}.
Mixed functors are nevertheless sufficient for a large class of complex
client-server APIs that are useful in practice,
as demonstrated with the \ocamlc{MakeCache} functor.

%%% Local Variables:
%%% mode: latex
%%% TeX-master: "../main"
%%% End:

%  LocalWords:  colorclient colorserver colorshared applicative fmap
%  LocalWords:  functor monad composable msg DSL html onclick elt API
%  LocalWords:  parametrized datastructures unordered appendToBody
%  LocalWords:  apis implem

\subsection{Going further}

Through these various examples, we demonstrated how we can combine traditional
tierless features
with an advanced module system to create powerful and
expressive APIs. One one hand, tierless languages traditionally allows for complex
interplay of client and server code.
Module systems, on the other hand, allows to manipulate large pieces of code
while preserving abstraction, encapsulation and modularity.
\eliom allows
to preserve these abstraction capabilities while enjoying the free-form
tierless programming style.
The \ocsigen ecosystem uses these concepts to provide numerous additional tools
such as advanced communication patterns, shared HTML across tiers
or even distributed Functional Reactive Programming \citep{OcsiTutorial,eliomIFL,eliomWWW}.
All these advanced
mechanisms are used in production by \eliom users.

%%% Local Variables:
%%% mode: latex
%%% TeX-master: "main"
%%% End:

\section{The \etiny calculus}
\label{sec:theory}

We now formalize \eliom as an extension of \ml
with both an expression and a module language.
To emphasize the new elements introduced by \eliom, these additional elements will be
colored in \addedx{blue}. This is only for ease of reading and
is not essential for understanding the formalization.

While \eliom is an extension of \ocaml, \etiny is an extension of a
simpler \ml calculus, which we present quickly in
\cref{sec:ml}.
We then present the location system in \cref{sec:eliom:loc},
followed by the expression and the module languages
in \cref{sec:expression,sec:module}.
Finally, we describe its semantics in \cref{sec:eliom:execution}.

\paragraph{Syntactic considerations}

Let us first define some notations and meta-syntactic variables. As a general rule,
the expression language is in lowercase ($e$) and the module language is in
uppercase ($M$). Module types are in calligraphic letters ($\Mm$).
More precisely: $e$ are expressions, $x$ are variables, $p$ are module paths, $X$ are module variables, $\tau$ are type expressions and $\type{t}$ are type constructors.
$x_i$, $X_i$ and $\type{t_i}$ are identifiers (for values, modules and types).
Identifiers (such as $x_i$) have a name part ($x$) and a stamp part ($i$) that
distinguish identifiers with the same name (following \citet{Leroy95}):
$\alpha$-conversion should keep the name intact and change only the stamp.
Sequences are noted with a star; for example $\tau^*$ is a sequence of type expressions.
Indexed sequences are noted $(\tau_i)$, with an implicit range.
Substitution of $a$ by $b$ in $e$ is noted $\subst[]{a}{b}{e}$.
Repeated substitution of each $a_i$ by the corresponding $b_i$ is noted
$\substi[]{a_i}{b_i}{e}$.

\subsection{A crash course in the \ml calculus}
\label{sec:ml}

There are many variants of the \ml calculus. For the purpose of \etiny, we consider
a core calculus with polymorphism, let bindings and parametrized datatypes
in the style of \citet{wright1994syntactic},
accompanied by a fully featured module system with separate compilation and
applicative functors in the style of \citet{Leroy94,Leroy95}.
The goal of this calculus is to closely module the most relevant features of
\ocaml, while being sufficiently
simple to be extended and reasoned on formally.

In this section, we will simple give a quick overview of the syntax and
an informal reminder of the most distinctive or unusual points.
The complete treatment of the language
is given in \cref{appendix:ml}.
Each set of typing or reduction rules can also be obtained by considering
the \etiny rules in the next section and ignoring the parts written in \addedx{blue}.

\subsubsection{Syntax}

The complete syntax is presented in \cref{ml:grammar} and follows the syntax
of \ocaml closely.
The expression language is a fairly simple extension of
the lambda calculus with a fixpoint combinator ($\Y$) and let bindings ($\letin{x}{e_1}{e_2}$).
The language is parametrized by a set of constants $\Const$.
Variables can be qualified by a module path $p$.
Paths can be either module identifiers such as $X_i$,
a submodule access such as $X_i.Y$,
or a path application such as $X_i(Y_j.Z)$.
Note that, as said earlier, that fields of modules are only called by their
name, without stamp.

The module language is composed of functors, module application
($\appm{\mm_1}{\mm_2}$),
type constraints ($\constraint{\mm}{\Mm}$).
and structures containing a list of value, types
or module definitions ($\struct{\letm[]{x_i}{2}}$).
Programs are lists of definitions.
Module types can be either functor types ($\functor{X_i}{\Mm_1}{\Mm_2}$)
or a signature ($\signature{\valm[]{x_i}{\intty}}$), which contains a list
of value, types and module descriptions. Type descriptions can expose their definition
or can be left abstract.
Typing environments are simply module signatures. We note them $\Env$ for convenience.

\subsubsection{Type system}

As a general rule, judgements are denoted with the symbol $\iswt$ for
expressions and $\iswtm$ for modules.
More precisely:

\begin{tabular}{ll}
  $\wt[]{\Env}{e}{\tau}$& Expression $e$ has type $\tau$ in $\Env$\\
  $\equivty[]{\Env}{\tau_1}{\tau_2}$& $\tau_1$ and $\tau_2$ are equivalent in $\Env$\\
  $\wtm[]{\Env}{\mm}{\Mm}$& module $\mm$ has type $\Mm$ in $\Env$\\
  $\submod[]{\Env}{\Mm}{\Mm'}$& Module type $\Mm$ is a subtype of $\Mm'$ in $\Env$\\
  $\wf[]{\Env}{\tau}$ and $\wfm[]{\Env}{\Mm}$& The type $\tau$ (resp. module type $\Mm$) is well formed.
\end{tabular}

The type system for expression is the standard Hindley-Milner type system
with polymorphism, generalization and parametrized datatype.
The module system, which follows \citet{Leroy94,Leroy95}, is less usual. We will focus on three specific features:
qualified accesses, applicative functors and strengthening.

\paragraph{Qualified accesses}

Let us consider the module $X$ with the module type
$\renewcommand\added\addedhide(\signature{\typeabsm{t};\ \valm{a}{t}})$. $X$ contains
a type $t$ and a value $a$ of that type.
We wish to typecheck $X.a$. One expected type for this expression is $X.t$.
However, the binding of $v$ in $X$ gives the type $t$, with no mention of
$X$. In order to make the type of $X.a$ expressible in the current scope,
we prefix the type variable $t$ by the access path $X$.
This is done in the rule \mysc{QualModVar} by the substitution
$\substs[]{n_i}{p.n}{\bv[]{\Sm_1}}{\Mm}$ which prefixes all the bound variables of $\Sm_1$, noted $\bv[]{\Sm_1}$, by the path $p$.
We thus obtain the following type derivation
\begin{mathpar}
  \renewcommand\added\addedhide
  \newcommand\Mx{\bindingm[]{X}{\dots}}
  \inferrule*[Left=QualVar, Right={with $X.t = \subst[]{t}{p.t}{t}$}]
  { \inferrule*[Left=ModVar]
    { \Mx \in \Mx }
    { \wtm{\Mx}{X}{\signature{\typeabsm{t};\ \valm{a}{t}}} }
  }
  { \wt{\Mx}{X.a}{X.t} }
\end{mathpar}

\paragraph{Applicative functors}

In \cref{tuto:modules}, we used the \ocamlc{Map.Make} functor
to create a new dictionary data-structure using the provided type
as keys. One might wonder what happens if \ocamlc{Map.Make} is applied
to the same module twice. Are the dictionaries thus produced compatible?
More formally, Does $\mm_1 = \mm_2$ implies $F(\mm_1) = F(\mm_2)$?
When this holds, we say that functors are ``applicatives''. Otherwise, we
say they are ``generatives''. \ocaml functors are applicatives by default.
In our simple \ml calculus, all functors are applicatives.
This is implemented by enriching type constructors with types
of the form $F(M).t$. If we consider the modules $N_i = \text{Map.Make}(\mm_i)$,
then we have
$N_1.t = \text{Map.Make}(\mm_1).t = \text{Map.Make}(\mm_2).t = N_2.t$.

\paragraph{Strengthening}

Let us consider a module $X$ answering the signature shown
\cref{ex:funcmanifest:env}. We want to pass this module to the functor $F$
shown in \cref{ex:funcmanifest:prog}. This functor application
is expected to succeed, since $X.t = t$, where $t$ comes from $X$.
However, the definition of $t$ in $X$ is abstract, and does not have
a definition allowing us to check that $X$ is indeed compatible with the proposed
signature.
For this purpose, we \emph{strengthen} the type of the module $X$ with
additional type equalities. If $\mathcal X$ is the type of $X$, we note
$\strengthen{\mathcal X}{X}$ the strengthened signature.
The result is the signature
$\renewcommand\added\addedhide(\signature{\typem[]{t}{X.t};\ \valm[]{v}{t}})$.
This signature is now trivially included in the argument type of $F$.

\begin{figure}[!h]
  \renewcommand\added\addedhide
  \mprset{sep=1.7em}
  \setlength{\jot}{0pt}% Remove interline space in align*
  \begin{subfigure}{0.3\linewidth}
  \begin{align*}
    &\moduletym[]{X}{\mathtt{sig}}\\
    &\quad\typeabsm[]{t}\\
    &\quad\valm[]{v}{t}\\
    &\dend
  \end{align*}
  \caption{Typing environment}
  \label{ex:funcmanifest:env}
  \end{subfigure}\hfill
  \begin{subfigure}{0.65\linewidth}
  \begin{align*}
    &\mathtt{module}\ F\ (E:\signature{\typem[]{t}{X.t}\ \valm[]{v}{t}}) = \mathtt{struct}\\
    &\quad\dots\\
    &\dend\\
    &\modulem[]{X}{F(X)}
  \end{align*}
  \caption{Application of multi-argument functor using manifests}
  \label{ex:funcmanifest:prog}
  \end{subfigure}
  \caption{Program using functors and manifest types}
  \label{ex:funcmanifest}
\end{figure}

\subsubsection{Semantics}

We use a rule-based big-step semantics with traces. Big-step semantics
are more amenable to reasoning with modules, as they allow to easily introduced
local bindings. We record traces in order to reason about
execution order.
We note $v$ for values in the expression language and $\vm$ for values
in the module language.
Values in the expression language can be either constants or lambdas.
Module values are either structures, which are list of bindings of
values, or functors.
Execution environments, noted $\envr$,
are a list of value bindings. We
note the concatenation of environment $+$. Environment access is noted
$\bdrin{x}{v}{\envr}$ where $x$ has value $v$ in $\envr$.
The same notation is also used for structures.
Traces are lists of messages.
The empty trace is noted $\niltr$. Concatenation of traces is noted $\ctr$.

Given an expression $e$ (resp. a module $\mm$), an execution
environment $\envr$, a value $v$ (resp. $\vm$) and a trace $\trace$,
\[\renewcommand\added\addedhide
  \redto{\envr}{e}{v}{\trace}
\]
means that $e$ reduces to $v$ in $\envr$ and prints $\trace$.
The rules are composed of traditional call-by-value lambda-calculus
with a fixed execution order that is represented by the traces.

\subsubsection{Formal results}

The \ml calculus considered here has been shown to support
for separate compilation (\cref{ml:separation}) and
representation independence \cite{Leroy94}.
It doesn't have a proof of soundness, which discussed in \cref{ml:soundness}.

\subsection{Locations}
\label{sec:eliom:loc}

Before introducing the \etiny language constructs, we consider
the notation for \emph{locations} annotations
which indicate ``where the code runs''.
The grammar of locations is given in \autoref{loc:grammar}.
There are three core locations:
\textbf{s}erver, \textbf{c}lient or \textbf{b}ase.
The base side represents expressions that are ``location-less'', that is,
which can be used everywhere. We use the meta-variable $\loc$
for an unspecified core location.
There is a fourth location that is only available for modules: \textbf{m}ixed.
A mixed module can have client, server and base components.
We use the meta-variable $\mloc$ for locations that are either \textbf{m} or
one of the core locations. In most contexts, locations are annotated
with subscripts.
\begin{figure}[bt]
  \centering
  \begin{align*}
    \loc ::=&\ s\ |\ c\ |\ \base&
    \mloc ::=&\ \sideX\ |\ \loc
  \end{align*}\vspace{-8mm}
  \caption{Grammar of locations -- $\loc$ and $\mloc$}
  \label{loc:grammar}
  \input{theory/subloc}\vspace{-8mm}
  \caption{``can be used in'' relations on locations -- $\canuseloc{\mloc}{\mloc'}$ }
  \label{loc:canuse}
  \input{theory/subdefineloc}\vspace{-8mm}
  \caption{``can contain'' relation on locations -- $\canbedefloc{\mloc}{\mloc'}$ }
  \label{loc:canbedef}
\end{figure}

We also introduce two relations:
\begin{compactitem}
\item[$\canuseloc{\mloc}{\mloc'}$] defined in \autoref{loc:canuse}, means that
a variable (either values, types or modules) defined on location $\mloc$ can
be used on a location $\mloc'$. For example, a base type can be used in a client context.
Base declarations are usable everywhere. Mixed declarations are not usable in base code.
\item[$\canbedefloc{\mloc}{\mloc'}$] defined in \autoref{loc:canbedef}, means that
a module defined on location $\mloc$ can contain component on location $\mloc'$.
In particular, the mixed location $\sideX$ can contain any component, while other
location can contain only component declared on the same location.
\end{compactitem}
Both relations are reflexive: For instance,
it is always possible to use client declarations when you are on the client.

\subsection{Expressions}
\label{sec:expression}

\etiny's expression language is based on a simple \ml language,
as presented in \cref{sec:ml},
extended with \eliom specific
constructs: fragments and injections.
This expression language can be seen as an alternative formulation of the core language proposed by
\citet{eliomAPLAS}, but that is compatible with modules.

\begin{figure}[b]
  \vspace{-15pt}
  \input{theory/grammarexpr}
  \vspace{-8pt}
  \caption{Grammar for \etiny expressions}
  \label{grammar:eliom}
  \vspace{-8pt}
\end{figure}

A {\em client fragment} $\cv{e}$ can be used on the server to
represent an expression that will be computed on the client,
but whose future value can be referred to on the server.

An {\em injection} $\injf{v}{f}$ can be used on the client to
access values defined on the server. An injection must make explicit
use of a converter $f$ that specifies how to send the value. In
particular, this should involve a serialization step, executed on the
server, followed by a deserialization step executed on the client.
For ease of presentation, injections are only done on variables and
constants.
In the implementation, this restriction is removed by adding
a lifting transformation.
For clarity, we sometime distinguish injections \emph{per se}, which occur outside
of fragments, and escaped values, which occur inside fragments.
The syntax of types is also extended with two constructs. A {\em fragment type} $\cvtype{\tau}$ is the type of a fragment.
A {\em converter type} $\conv{\tau_s}{\tau_c}$ is the type of a converter taking
a server value of type $\tau_s$ and returning a client value of type $\tau_c$.
All type variables $\alpha_\loc$ are annotated with a core
location $\loc$. There are now three sets of constants: client, server and base.

\subsubsection{Typing rules}

Typing judgements for \etiny are annotated with a location that specifies where
the given code should be typechecked. For the expression language, we only consider
core locations $\loc$ which are either base, client or server.
The typing judgment, defined in \cref{core:typing},
is noted $\wt[\loc]{\Env}{e}{\tau}$ where $e$ is of type $\tau$
in the environment $\Env$ on the location $\loc$.
We also define type validity and equivalence judgements in
\cref{core:validity,core:eqtype}.
We note $\Typeof_\loc(c)$ the type of a given constant $c$ on the location $\loc$.
Binding in typing environments, just like in signatures, are annotated with
a location.
The first three kind of bindings, corresponding to the core language, can only
appear on core locations: $s$, $c$ or $b$.
Modules can also be of mixed location $\sideX$.
Names are namespaced by locations, which means it is valid to have different client
and server values with the same name.

Most of the rules are straightforward adaptions of traditional \ml rules
and are described in \cref{sec:ml}.
We focus on rules that demonstrate some particular features of \etiny.

Rule \mysc{Var} contains a significant difference compared
to the traditional \ml rules.
As described earlier, bindings in \etiny are located. Since access across
sides are explicit,
we want to prevent erroneous cross-location accesses.
For example, client variables can not be used on the server.
To use a variable $x$ bound on location $\loc'$ in a context $\loc$,
we check that location $\loc'$ can be used in location $\loc$, noted $\canuseloc{\loc'}{\loc}$ and defined in \autoref{loc:canuse}.
Base elements $\base$ are usable everywhere. Mixed elements $\sideX$ are
usable in both client and server.

Type variables and types constructors are also annotated with a location
and follow the same rules.
Using type variables from the client on the server, for example,
is disallowed. Such constraints are enforced by
the type validity and equivalence judgements though rules such as
\mysc{FragVal} and \mysc{ConvVal} which
tracks locations across type expressions.

Rule \mysc{Fragment} is for the construction of client fragments and can only be applied on the server.
If $e$ is of type $\tau$ {\em on the client}, then $\cv{e}$ is of type $\cvtype{\tau}$ {\em on the server}.
Since no other typing rule involves client fragments, it is
impossible to deconstruct them on the server.

Rule \mysc{Injection} is for the communication from the server to the client
and can only be applied on the client.
If $e$ is of type $\tau_s$ on the server and $f$ is a converter of type $\conv{\tau_s}{\tau_c}$ on the server,
then $\injf{e}{f}$ is of type $\tau_c$ on the client.
Note that injections always require a converter, which we describe in greater details now.

\begin{figure}[!htb]
  \renewcommand\added\addedx
  \renewcommand\addedrule\addedno
  \input{theory/core/typing}\vspace{-3mm}
  \caption{\etiny expression typing rules -- $\wt{\Env}{e}{\tau}$}
  \label{core:typing}
\end{figure}

\begin{figure}[!htb]
  \renewcommand\added\addedx
  \renewcommand\addedrule\addedno
  \input{theory/core/validity}\vspace{-3mm}
  \caption{Type validity rules -- $\wf{\Env}{\tau}$}
  \label{core:validity}
  \input{theory/core/equivty}\vspace{-3mm}
  \caption{Type equivalence rules -- $\equivty{\Env}{\tau}{\tau'}$}
  \label{core:eqtype}
\end{figure}

\subsubsection{Converters}\label{sec:converter}

To transmit values from the server to the client, we need a
serialization format. We assume the existence of a type $\serial$ in
$\Const_b$ which represents the
serialization format. The actual format is irrelevant. For instance,
one could use JSON or XML.

Converters are special values that describe how to move a value from the server to the client.
A converter can be understood as a pair of functions.
A converter $f$ of type $\conv{\tau_s}{\tau_c}$ is composed of a
server-side encoding function of type $\lamtype{\tau_s}{\serial}$, and
a client-side decoding function of type $\lamtype{\serial}{\tau_c}$.
We assume the existence of two built-in converters:
\begin{itemize}[noitemsep]
\item The $\serial$ converter of type $\conv{\serial}{\serial}$. Both sides are the identity.
\item The $\fragment$ converter of type  $\forall\alpha_c.(\conv{\cvtype{\alpha_c}}{\alpha_c})$.
\end{itemize}

\subsubsection{Type universes}\label{eliom:typeuniv}

It is important to note that there is no identity converter (that would be of type $\forall\alpha.(\conv{\alpha}{\alpha})$). Indeed the client and server type universes
are distinct and we cannot translate arbitrary types from one to the other.
Some types are only available on one side: database handles, system types, \js API types.
Some types, while available on both sides
(because they are in $\base$ase for example), are simply not transferable.
For example, functions cannot be serialized in general.
Another example is file handles: they are available both on the server and
on the client, but moving a file handle from server to client seems adventurous.
Finally, some types may share a semantic meaning, but not their actual representation. This is the case where converters are used, as demonstrated in \cref{sec:example:conv}.

\subsubsection{Mixed datatypes}
\label{sec:datatypes}

The version of \ml
we consider supports an interesting
combination of three features: abstract datatypes, parametrized datatypes and
separate compilation at the module level.
\etiny, as an extension of \ml, also supports these features.
These three features have non-trivial interactions that need to be
accounted for, in particular when introducing properties on types, such as locations.

\begin{ex}[!bt]
  \begin{subfigure}[b]{0.5\textwidth}
\begin{lstlisting}[language=eliom]
module M : sig
  type%server ('a, 'b) t
end = struct
  type%server ('a, 'b) t = 'a fragment * 'b
end
\end{lstlisting}\vspace{-4mm}
    \caption{Incorrect abstract datatype}
    \label[figure]{code:datatype:incorrect}
  \end{subfigure}~
  \begin{subfigure}[b]{0.55\textwidth}
\begin{lstlisting}[language=eliom]
module M : sig
  type%server ('a[@client], 'b) t
end = struct
  type%server ('a[@client], 'b) t = 'a fragment * 'b
end
\end{lstlisting}\vspace{-4mm}
    \caption{Correct abstract datatype}
    \label[figure]{code:datatype:correct}
  \end{subfigure}
  \caption{A module with an abstract datatype.}
  \cref{code:datatype:incorrect} does not exposes information about acceptable sides for \ocamlc{'a} and \ocamlc{'b}.
  In \cref{code:datatype:correct}, annotations specifying the side of type variables are exposed in the interface.
  \label{code:datatype}
  \vspace{-5mm}
\end{ex}

Let us consider the module shown in \cref{code:datatype}.
We declare a server datatype \ocamlc{t} with two parameters and we hide the definition in the signature.
We now want to check that \ocamlc{(t1, t2) t} is a correct type expressions.
However, without the type definition, we don't know if \ocamlc{t1} and \ocamlc{t2} are
base, client or server types. In order to type check the type sub-expressions, we
need more information about the definition of \ocamlc{t}.
The solution, much like variance, is to annotate type variables in datatypes with extra information.
This is done in the syntax for type declarations given in \cref{grammar:eliom}.
Each type parameters is annotated with a location. Type variables can only be used on the right location. This ensures proper separation of client and server type variables and their proper usage.

These annotations can be though as a simplistic kind system. One could
also considered \ocamlc{'a} as a constrained type variable, in the style of
\ml{}$^{\text{F}}$~\citep{MLF}.

%%% Local Variables:
%%% mode: latex
%%% TeX-master: "../main"
%%% End:

\subsection{Modules}\label{sec:module}

We now present the module language part of \etiny
as an extension of \citet{Leroy95} which model the bulk
of the \ocaml module language.
The syntax, presented in \cref{grammar:modules}, is composed
of the usual module constructs: functors,
module constraints, functor application and structures. A structure is composed of a list
of components. Similarly, module types are composed of functors and signatures which are
a list of signature components. Components can be declaration of values, types or
modules. A type in a signature can be declared abstract or not.

The main difference between \etiny and \ml is that structure and signature components are annotated
with locations. Value and type declarations can be annotated with a core location $\loc$
which is either $\base$, $s$ or $c$.
Module declarations can also have one additional possible location: the mixed location $\sideX$.
We use $\mloc$ for locations that can be $\sideX$, $\base$, $s$ or $c$.
Only modules on location $\sideX$ can have subfields on different locations.
We also introduce mixed functors, noted $\functor[\sideX]{X}{\Mm}{\Mm}$, which body can
contain both client and server declarations.
A program is a list of declarations including a client value declaration $\ret$ which is the result of the program.

\begin{figure}[bt]
  \renewcommand\added\addedx
  \input{theory/grammarmod}\vspace{-2mm}
  \caption{\emodule's grammar for modules}
  \label{grammar:modules}
\end{figure}

We first introduce the various feature of our module system along with some
motivating examples. We then detail how those features are enforced by the
typing rules.

\subsubsection{Base location and specialization}\label{sec:specialization}

In \cref{tuto:modules}, we presented an example where a base functor
\ocamlc{Map.Make}, is applied to a client module to obtain a new client module.
As \eliomc{Map.Make} is a module provided by the standard library of \ocaml,
it is defined on location $\base$.
In particular, its input signature has components on location $\base$, thus it would seem a module
whose components are on the client or the server should not be accepted.
We would nevertheless like to create maps of elements that are only available on the client.
To do so, we introduce a specialization operation, defined in \cref{module:specialize},
that allows to use a base module in a client or server scope by replacing
instances of the base location with the current location.

The situation is quite similar to the application of a function of type
$\forall\alpha.\alpha\to\alpha$ to an argument of type $int$: we need to
instantiate the function before being able to use it.
Mixed modules only offer a limited version of polymorphism for locations:
there is only one ``location variable'' at a time, and it's always
called $\base$.
The specialization operation simply rewrites a module signature by substituting all
instances of the location $\base$ or $\sideX$ by the specified $c$ or $s$ location.
Note that before being specialized, a module should be accessible according to
the ``can be used'' relation defined \cref{loc:canuse}. This means that we never have to specialize a server module on the client (or conversely).
Specialization towards location $\base$ has no effect since only
base modules are accessible on location base.
Specialization towards the location $\sideX$ has no effect either: since all
locations are allowed inside the mixed location, no specialization is needed.
Mixed functors are handled in a specific way, as we see in the next section.

\subsubsection{Mixed Functors}
\label{sec:functor}

Mixed functors are functors that take as input a mixed module and return
a mixed module.
We note $\functor[\sideX]{X_i}{\Mm}{\mm}$ the mixed functor that takes an
argument $X_i$ of type $\Mm$ and return a module $\mm$.
They can contain both client and server declarations (or mixed submodules).
Mixed functors and regular functors have different types that are not compatible.
We saw in \cref{tuto:mixeddata} an example of usage for mixed functors.
Mixed functors have several restrictions compared to regular functors
which we now detail using various examples.

\paragraph{Specialization}
A naive implementation of specialization of mixed functors
would be to specialize on both side of the arrow and apply the
resulting functor.
Let us see on an example why this solution does not work.
In \cref{code:mixedfunctor}, the functor \eliomc{F} takes as argument
a module containing a base declaration and uses it on both sides. If the type of the functor parameter were specialized, the functor application
in \cref{code:mixedfunctor:wrong} would be well-typed. However, this
makes no sense: \ocamlc{M.y} is supposed to represent a fragment
whose content is the client value of $b$, but this value doesn't exist
since $b$ was declared on the server.
There is no value available to inject in the declaration of \ocamlc{y'}.

The solution here is that specialization on mixed functors
should only specialize the return type, not the argument.

\begin{ex}[!h]
  \begin{subfigure}{0.50\linewidth}
\begin{lstlisting}[language={eliom}]
module%mixed F (A : sig val b : int end) 
= struct
  let%server x = A.b
  let%server y = [%client A.b]
end
\end{lstlisting}
    \caption{A mixed functor using a base declaration}
  \end{subfigure}\hfill
  \begin{subfigure}{0.47\linewidth}
\begin{lstlisting}[language={eliom}]
module%server M =
  F(struct let%server b = 2 end)
let%client y' = ~%M.y
\end{lstlisting}
    \caption{An ill-typed application of \eliomc{F}}
    \label{code:mixedfunctor:wrong}
  \end{subfigure}
  \vspace{-2mm}
  \caption{A mixed functor using base declaration polymorphically}
  \label[ex]{code:mixedfunctor}
  \vspace{-3mm}
\end{ex}

\paragraph{Injections}

Injections inside client sections (as opposed to escaped values inside client fragments) are fairly static: the value might be dynamic, but
the position of the injection and its use sites are statistically known and
does not depend on the execution of the program.
In particular, injections are independent of the control flow.
We can just give a unique identifier to each injection, and use that unique name
for lookup on the client.
This property comes from the fact that injected server identifiers cannot be bound
in a client section.

Unfortunately, this property does not hold in the presence of
mixed functor when we assume the language can apply functor at arbitrary positions,
which is the case in \ocaml.
Let us consider \cref{code:compile:mixedinj}.
The functor $F$ takes a structure
containing a server declaration $x$ holding an integer and returns a structure containing
the same integer, injected in the client.
In \cref{code:compile:inj:use}, the functor is used on $A$ or $B$ conditionally. The issue is that the client integer
depends both on the server integer and on the local \emph{client} control flow.
Lifting the functor application at toplevel would not preserve the semantics of the language, due to side effects.
Thus, we avoid this kind of situation by forbidding
injections that access dynamic names inside mixed functors.

\begin{ex}[!h]
  \begin{subfigure}{0.4\linewidth}
    \centering
\begin{lstlisting}[language={eliom}]
module%mixed F 
  (A : sig val%server x : int end)
= struct
  let%client x' = ~%A.x
end
\end{lstlisting}
    \caption{An problematic mixed functor with an injection}
    \label{code:compile:inj}
  \end{subfigure}\hfill
  \begin{subfigure}{0.53\linewidth}
\begin{lstlisting}[language={eliom}]
module%mixed A = struct let%server x = 2 end
module%mixed B = struct let%server x = 4 end
let%client a =
  if Random.bool ()
  then let module M = F(A) in M.x'
  else let module M = F(B) in M.x'
\end{lstlisting}
    \caption{A pathological functor application}
    \label{code:compile:inj:use}
  \end{subfigure}
  \vspace{-2mm}
  \caption{Problematic example of injection inside a mixed functor}
  \label{code:compile:mixedinj}
  \vspace{-3mm}
\end{ex}

In order to avoid this situation, we add the constraints that injections
inside the body of a mixed functors can only refer to outside of the
functor. Escaped values, which are injections
inside client fragments, are still allowed. The functor
presented in \cref{code:mixedfuninj:wrong} is not allowed while
the one in \cref{code:mixedfuninj:right} is allowed.
Formally, this is guaranteed by the \mysc{MixedFunctor} rule,
where each injection is typechecked in the outer typing environment.

\begin{ex}[!h]
  \begin{subfigure}{0.4\linewidth}
\begin{lstlisting}[language={eliom}]

module%mixed F
  (A:sig val%server x : int end)
= struct
  let%client y = ~%A.x + 2
end
(*@@*)
\end{lstlisting}
  \caption{An ill-typed mixed functor using an injection}
    \label{code:mixedfuninj:wrong}
  \end{subfigure}\hfill
  \begin{subfigure}{0.5\linewidth}
\begin{lstlisting}[language={eliom}]
let%server x = 3
module%mixed F
  (A:sig val%server y : int end)
= struct
  let%client z = ~%x
  let%server z' = [%client ~%A.y + 1]
end
\end{lstlisting}
  \caption{A well-typed mixed functor using an injection}
    \label{code:mixedfuninj:right}
  \end{subfigure}
  \vspace{-2mm}
  \caption{Mixed functor and injections}
  \label{code:mixedfuninj}
  \vspace{-3mm}
\end{ex}

\paragraph{Functor application}
Mixed functors can only be applied to mixed structures.
%, as can be seen in rule \mysc{MixedApplication}.
This means that in a functor application \ocamlc{F(M)},
\ocamlc{M} must be a structure defined by a $\mathtt{module}_\sideX$ declaration.
Note that this breaks the property that the current location of an expression
or a module can be determined syntactically:
The location inside \ocamlc{F(struct ... end)} can be either mixed or not,
depending on \ocamlc{F}. This could be mitigated by using a different syntax
for the application of mixed functor.
The justification for this restriction is detailed in
\cref{sec:eliom:execution}.

\subsubsection{Type rules}\label{sec:module:types}

We now review how these various language constructs are reflected in the
rules of our type system. As before, the \eliom module system is built on
the \ml module system. We extend the typing, validity and subtyping judgments
by a location annotation that specifies the location of the current scope.
The program typing judgments don't have a location, since
a program is always considered mixed.
Most rules are fairly straightforward adaptations of the \ml rules, annotated with
locations.

The typing rules \mysc{ModVar} and \mysc{QualModVar} follow the usual rules of \ml modules
with two modifications: We first check that the module
we are looking up can indeed be used on the current location. This is done
by the side condition $\canuseloc{\mloc'}{\mloc}$ where $\mloc$ is the current location
and $\mloc'$ is the location where the identifier is defined.
This allows, for instance, to use {\em base} identifiers in a {\em client} scope.
We also specialize the module type of the identifier towards the current location $\mloc$.
The specialization operation, which was described in \cref{sec:specialization},
is noted $\specialize{\mloc}{\mloc}{M}$ and
is defined in \cref{module:specialize}.

There are two new typing rules compared to \ml: the rules
\mysc{MixedFunctor} and \mysc{MixedApplication} define mixed functor definition
and application. We use $\injs{\cdot}$ which returns the set of all injections in
client declarations.

\subsubsection{Subtyping and equivalence of modules}

Subtyping rules are given in \cref{module:subtyping}.
For brevity, we note $\withinlocs{\mloc}{\mloc_1}{\mloc_2}$ as
a shorthand for
$\canbedefloc{\mloc}{\mloc_1} \wedge
\canbedefloc{\mloc}{\mloc_2} \wedge
\canuseloc{\mloc_1}{\mloc_2}$, that is, both
$\mloc_1$ and $\mloc_2$ are valid locations for components of a module on location $\mloc$
and location $\mloc_1$ encompasses location $\mloc_2$.
Note that the following holds:
\[\submod{\Env}{\struct{\valm[b]{t_i}{int}}}{\struct{\valm[c]{t_i}{int}}}\]
This is perfectly safe,
since for any identifier $x_i$ on base, $\letm[c]{x'_j}{x_i}$ is always valid.
This allows programmers to declare some code
on base (and get the guarantee that the code is only using usual \ocaml constructs)
but to expose it as client or server in the module type.

\begin{figure}[!htbp]
  \renewcommand\added\addedx
  \input{theory/module/typing}
  \caption{Module typing rules -- $\wtm{\Env}{m}{M}$}
  \label{module:typing}
\end{figure}
\begin{figure}[!htbp]
  \renewcommand\added\addedx
  \input{theory/module/subtyping}
  \caption{Module subtyping rules -- $\submod{\Env}{M}{M'}$}
  \label{module:subtyping}
\end{figure}
\begin{figure}[!htbp]
  \centering
  \input{theory/module/specialize}
  \caption{Module specialization operation -- $\specialize{\mloc'}{\mloc}{M}$}
  \label{module:specialize}
  % \vspace{5mm}
\end{figure}
\begin{figure}[!htbp]
  \centering
  \renewcommand\added\addedx
  \input{theory/module/strengthen}\vspace{-3mm}
  \caption{Module strengthening operation -- $\strengthen{M}{p}$}
  \label{module:strengthen}
  % \vspace{3mm}
\end{figure}

%%% Local Variables:
%%% mode: latex
%%% TeX-master: "../main"
%%% End:

\subsection{Separate typechecking}

Our \etiny calculus inherit all the properties of the \ml calculus regarding
separate compilation. In particular, we show here that it trivially supports
separate typechecking, where each module is typechecked independently and only
requires knowledge of the module \emph{types} (but not the implementations!) of its
dependencies.

\begin{theorem}[Separate Typechecking]\label{eliom:separation}
  Given a list of module declarations that form a typed program, there exists
  an order such that each module can be typechecked with only knowledge
  of the type of the previous modules.

  More formally,
  given a location $\mloc$
  and a list of $n$ declarations $\dm_i$ and a signature $\Sm$ such that
  \[
    \wtm[\mloc]{}{(\dm_1;\dots;\dm_n)}{\Sm}
  \]
  then there exists $n$ definitions $\Dm_i$ and a permutation $\pi$ such that
  \begin{align*}
    \forall\mloc\forall i < n,\ &
    \wtm[\mloc]{\Dm_{1};\dots;\Dm_{i}}{\dm_{i+1}}{\Dm_{i+1}}&
    \submod[\mloc]{}{\Dm_{\pi(1)};\dots;\Dm_{\pi(n)}}{\Sm}
  \end{align*}
\end{theorem}
\begin{proof}
  Proceed exactly as the \ml proof for \cref{ml:separation} in \cref{sec:ml:module:typing}.
\end{proof}

%%% Local Variables:
%%% mode: latex
%%% TeX-master: "../main"
%%% End:

\subsection{Interpreted semantics}\label{sec:eliom:execution}

While \eliom, just like \ocaml, is a compiled language, it is desirable
to present a semantics that does not involve complex program transformation.
The reason is two-fold: First, this simple semantics should be reasonably
easy to explain to users. Indeed, this semantics is the one used
to present \eliom in \cref{sec:howto}. However, we must also show
that this semantics is correct, in that it does actually
corresponds to our compilation scheme. This is done in \cref{sec:simulation}.
As presented in \cref{sec:howto}, \eliom execution proceeds
in two steps: The server part of the program is executed first. This
creates a client program, which is then executed.

Let us first introduce a few notations.
Generated client programs are noted $\clientprog$. Server
expressions (resp. declarations) that do not contain injections
are noted $\noinj{e}$ (resp. $\noinj{\dm}$).
Values are the same as for \ml: constants, closures, structures
and functor closures.
We consider a new class of identifiers called ``references'' and noted in bold, such
as $\rf{r}$ or $\rf{R}$.
We assume the existence of a name generator
that can create arbitrary new fresh $\rf{r}$ identifiers at any point of the
execution. References are used as global identifiers that ignore
scoping rules.
References can also be qualified as ``reference paths'', noted $\rf{X.r}$.
This is used for mixed functors, in particular.
We use $\envg$ to note the global environment where such
references are stored.

We now introduce a new reduction relation,
$\pred{}_{\mloc}$, which is the reduction over \eliom constructs
on side $\mloc$.
The notation $\pred{}_{\mloc}$ actually represents several reduction relations
which are presented in \cref{eliom:semclient,eliom:semantics,eliom:semmixed}.
Four of these relations reduce the server part of the code and emit
a client program.
We note
$\redel[\iota]{\envr}{e}{v}{\clientprog}{\trace}$
the reduction of a server expression $e$ inside a context $\iota$ in the
environment $\envr$.
It returns the value $v$, the client program $\clientprog$ and
emits the trace $\trace$.
The context $\iota$ can be either base
($\base$), server ($s$), server code inside client contexts ($\cins$) or
server code inside mixed contexts ($\sideX$).
We also have a client reduction, noted
$\redc{\envr}{e}{v}{\trace}{\envg}{\envg'}$
which reduces a client expression $e$ inside an environment $\envr$, returns
a value $v$ and emits a trace $\trace$. It also updates a global environment
from $\envg$ to $\envg'$.

Note that the first family of relation executes \emph{only} the
server part of a program and returns a client program, which is then executed
by $\pred{}_c$.
This is represented formally by the \mysc{Program} rule. In order to
reduce an \eliom program $\Pm$, we first reduce the server part using
$\pred{}_\sideX$. This returns no value and a client program $\clientprog$
which we execute.
We now look into each specific feature in greater detail

\subsubsection{Generated client programs}
\label{sec:eliom:execution:client}

Let us first describe evaluation rules for generated client programs.
Generated client programs are \ml programs with some additional constructions
which are described in \cref{eliom:clientgrammar}.
The new evaluation rules are presented in \cref{eliom:semclient}.
The construction $\bindenv{f}$ binds the current accessible environment
to $\rf{f}$ in the global environment $\envg$. This is implemented
by the \mysc{BindEnv} rule.
$\bindw{r}{f}{e}$ computes $e$ in the environment previously associated to
$\rf{f}$. The result is then stored as $\rf{r}$ in $\envg$.
This construction is also usable for module expressions
and is implemented by the \mysc{Bind} and \mysc{Bind$_m$} rules.
All these constructions also accept paths of references such as $\rf{R.f}$.

The new $\pbind$ constructs are similar to the ones
used in languages with continuations in the catch/throw style.
Instead of storing both an environment and the future computation, we
store only the environment. This will allow us to implement closures
across locations, in particular the case where fragments
are used inside a server closure.

The client reduction relation also inherits the \ml rules
(rule \mysc{ClientCode}). In such a case, the global environment
is passed around following an order compatible with traces.
For example, the \mysc{LetIn} rule for let expression would
be modified like so:
\vspace{-2mm}
\begin{mathpar}
  \renewcommand\added\addedx
  \inferrule
  { \redc{\envr}{e'}{v'}{\trace}{\envg}{\envg'} \\
    \redc{\envr+\bdr{x}{v'}}{e}{v}{\trace'}{\envg'}{\envg''} }
  { \redc{\envr}{(\letin{x}{e'}{e})}{v}{\trace\ctr\trace'}{\envg}{\envg''} }
\end{mathpar}
Here, $e'$ is evaluated first (since $\trace$ is present first
in the resulting traces), hence it uses the initial environment $\envg$ and
returns the environment $\envg'$, which is then passed along.

From now on, we use $\rf{f}$ to denote
the reference associated to fragments closures and $\rf{r}$ to denote
the reference associated to a specific value of a fragment.

\begin{figure}[htb]
  \vspace{-4mm}
  \renewcommand\added\addedx
  \input{theory/clientgrammar}\vspace{-7mm}
  \caption{Grammar of client programs}
  \label{eliom:clientgrammar}
  \vspace{-2mm}
  \mprset{sep=1.5em}
  \input{theory/semclient}\vspace{-5mm}
  \caption{Semantics for client generated programs --
    $\redc{\envr}{e}{v}{\trace}{\envg}{\envg'}$ }
  \label{eliom:semclient}
  \vspace{-2mm}
\end{figure}

\subsubsection{Base, Client and Server declarations}

We now consider the case of base, client and server declarations.
The rules are presented in \cref{eliom:semantics}.
Let us first describe the execution of complete \etiny programs
(rule \mysc{Program}). A program $P$ reduces to a client value
$v$ if and only if we can first create a \emph{server}
reduction of $P$ that produces no value, emits a client program $\clientprog$
and a trace $\trace_s$. We can then create a reduction of $\clientprog$
that reduces in $v$ with a trace $\trace_c$. The trace of the program
is the concatenation of the traces.
We see that the execution of \etiny program is split in two as described
earlier. Let us now look in more details at various construction
of the \etiny language.

\paragraph{Base} The base reduction relation corresponds exactly to the \ml
reduction relation, and always returns empty programs (rule \mysc{BaseCode}).
When reducing a base declaration in a mixed context, we both reduce the
declaration using $\pred{}_\base$, but also add the declaration to the emitted
client program (rule \mysc{BaseDecl}).
As we can see, base declarations are executed twice: once on the server
and once on the client.

\paragraph{Client contexts and injections}
The goal of the client reduction relation $\pred{}_{\cins}$
is \emph{not} to reduce client programs. It only reduces server
code contained by injections inside client code.
It returns a client expression without injections, a client program
and a trace.
Since we don't want to execute client code,
it does not inherit the reduction rules for \ml.
Given an injection $\injf{e}{f}$, the rule \mysc{Injection}
reduces the server side expression $\app{f^s}{e}$ to a value $v$.
We then transform the server value $v$ into a client value using the
$\pinjval{}$ operator presented \cref{eliom:injval}.
We then returns the client expression $\app{f^c}{\pinjval{v}}$ without executing it.
This expression will be executed on the client side, to deserialize the value.
The value injection operator, noted $\pinjval{}$ represents the serialization of values
from the server to the client
and is the identity over constants in $\Const_b$ and references, and fail on any other
values.
According to the definition of converters,
if $f$ is a converter $\conv{\tau_s}{\tau_c}$,
then $f^s$ is the server side function of type $\lamtype{\tau_s}{\serial}$
and $v$ should be of type $\serial$. Since $\serial$ is defined on $\base$, the injection
of values should be the identity.

The rule \mysc{ClientContext} defines the evaluation of server expression up
to client contexts. Client contexts are noted $\C{e_1,\dots,e_n}$ and are defined in
\cref{eliom:clientcontext}. A client context can have any number of holes which must
all contain injections. The rest of the context can contain arbitrary \ml
syntax that are not injections. Evaluation under a multi-holed context
proceed from left to right. The resulting programs and traces are constructed by
concatenation of each program and trace.

In order to evaluate client declarations, the rule \mysc{ClientDecl} uses $\pred{}_{\cins}$
to evaluate the server code present in the declaration $\dm_c$ which returns
a declaration without injections $\noinj{\dm_c}$ and a client program $\clientprog$.
We then return the client program composed by the concatenation of the two.
We demonstrate this in \cref{ex:eliom:clientdecl}.
The \etiny program is presented on the left side.
It first declares the integer $a$ on the server then inject it
on the client and returns the result.
The emitted code, shown in the middle, contains
an explicit call to the $\intty^c$ deserializer while the rest of the client
code is unchanged.
The returned value is shown on the right.

\begin{ex}[!htb]
  \setlength{\jot}{0pt}
  \[\begin{aligned}
      &\letm[s]{a}{3}\\
      &\letm[c]{\ret}{\injf{x}{\intty} + 1}
    \end{aligned}
    \quad\pred{\quad}_{\sideX}\quad
    \begin{aligned}
      &\letm[]{\ret}{\app{\intty^c}{3} + 1}
    \end{aligned}
    \quad\pred{\quad}_c\quad
    4,\niltr
  \]
  \vspace{-7mm}
  \caption{Execution of a client declaration}
  \label{ex:eliom:clientdecl}
  \vspace{-2mm}
\end{ex}
\begin{figure}[!htb]
  \vspace{-4mm}
  \begin{minipage}[b]{0.58\linewidth}
    \input{theory/context}\vspace{-5mm}
    \caption{Execution contexts for injections -- $\C{\cdot}$}
    \label{eliom:clientcontext}
  \end{minipage}\hfill
  \begin{minipage}[b]{0.41\linewidth}
    \input{theory/injval}\vspace{-5mm}
    \caption{Injections\\ of values -- $\injval{v}$}
    \label{eliom:injval}
  \end{minipage}
\end{figure}

\paragraph{Server code and fragments}
The server reduction relation reduces server code and emits
the appropriate client program associated to client fragments.
Since client program are mostly \ml programs, it inherits the
\ml reduction rules (rule \mysc{ServerCode})
where client programs are concatenated in the same order as traces.
Client fragments are handled by the rule \mysc{Fragment}.
Let us consider a fragment $\cv{e}$,
this evaluation proceeds in two steps: first, we evaluate all the injections
inside the client expression $e$ using the relation $\pred{}_{\cins}$
described in the previous section. We thus
obtain an expression without injection $\noinj{e}$ and a client program
$\clientprog$.

The second step is to register $\noinj{e}$ to be evaluated in the client program.
One could propose to simply consider client fragments as values. This is however
quite problematic, as it could lead to duplicated side effects. Consider
the program presented on the left side of \cref{ex:eliom:fragmentprint}.
If we were simply to
simply pass fragments along, the $\print$ statement would be evaluated twice.
Instead, we create a fresh identifier $\rf{r}$ that will be
globally bound to $\noinj{e}$
in the client program, as shown in rule \mysc{Fragment}. This way, the client expression
contained inside the fragment will be executed once, in a timely manner.
The execution rule for fragment is demonstrated in \cref{ex:eliom:fragmentprint}.
As before, the \etiny program is presented on the left, the emitted
client program in shown in the middle and the returned value is on the right.
Note that both $\fragment^s$ and $\fragment^c$ are the identity function.
\begin{ex}[h]
  \setlength{\jot}{0pt}
  \[\begin{aligned}
      &\letm[s]{x}{\cv{\app{\print}{3}}}\\
      &\letm[c]{\ret}{}\\
      &\quad\injf{x}{\fragment} + \injf{x}{\fragment}
    \end{aligned}
    \quad\pred{\quad}_{\sideX}\quad
    \begin{aligned}
      &\bindenv{\rf{f}}\\
      &\bindw{r}{f}{\app{\print}{3}}\\
      &\letm[]{\ret}{}\\
      &\quad\app{\fragment^c}{\rf{r}} + \app{\fragment^c}{\rf{r}}
    \end{aligned}
    \quad\pred{\quad}_c\quad
    6,\singltr{3}
  \]
  \vspace{-2mm}
  \caption{Execution of a fragment containing side-effects}
  \label{ex:eliom:fragmentprint}
\end{ex}

\paragraph{Closures and fragments}
In the client program above, we also use a reference $\rf{f}$ and the
$\bindenv{}$ construct. To see why this is necessary, we now consider
a case where fragments are used inside closures. This is presented
in \cref{ex:eliom:fragmentclos}. The \etiny program, presented
on the left, computes $1+3+2$ on the client (although in a fairly
obfuscated way). We first define the client variable $a$ as $1$.
We then define a server closure $f$ containing a client fragment capturing $a$.
We then define a new variable also named $a$ and call $\app{f}{3}$, inject the results and returns.
When evaluating the definitions of $f$, since it contains syntactically
a client fragment, we will emit the client instruction $\bindenv{f}$,
where $\rf{f}$ is a fresh identifier. This will capture the local
environment, which is $\bdr{a}{1}$ at this point of the client program.
When we execute $\app{f}{3}$, we will finally reduce the client
fragment and emit the $(\bindw{r}{f}{\app{\intty^c}{3} + a})$ instruction.
On the client, this will be executed in the $\rf{f}$ environment,
hence $a$ is $1$ and the result is $4$.
Once this is executed, we move back to the regular environment, where $a$
is $2$, and proceed with the execution.

Thanks to this construction, the capturing behavior of
closures is preserved across
location boundaries. The $\bindenv{}$ construct is
generated by the \mysc{ServerDecl} rule. $\frags{\dm_s}$ returns
the fragments syntactically present in $\dm_s$.
For each fragment, the local environment is bound to the associated reference.
\begin{ex}[h]
  \setlength{\jot}{0pt}
  \[\begin{aligned}
      &\letm[c]{a}{1}\\
      &\letm[s]{f\ x}{\cv{\injf{x}{\intty} + a}}\\
      &\letm[c]{a}{2}\\
      &\letm[s]{y}{\app{f}{3}}\\
      &\letm[c]{\ret}{\injf{y}{\fragment} + a}
    \end{aligned}
    \quad\pred{\quad}_{\sideX}\quad
    \begin{aligned}
      &\letm[]{a}{1}\\
      &\bindenv{\rf{f}}\\
      &\letm[]{a}{2}\\
      &\bindw{r}{f}{\app{\intty^c}{3} + a}\\
      &\letm[]{\ret}{\app{\fragment^c}{\rf{r}} + a}
    \end{aligned}
    \quad\pred{\quad}_c\quad
    6,\niltr
  \]
  \vspace{-2mm}
  \caption{Execution of a fragment inside a closure}
  \label{ex:eliom:fragmentclos}
\end{ex}

\begin{figure}[!b]
  \mprset{sep=1.5em}
  \setlength{\jot}{-3pt}% Remove interline space in align*
  \renewcommand\added\addedx
  \input{theory/semantics}\vspace{-5mm}
  \caption{Semantics for base, client and server sections --
    $\redel{\envr}{e}{v}{\clientprog}{\trace}$ }
  \label{eliom:semantics}
\end{figure}

\paragraph{Fragment annotations}
\label{sec:eliom:execution:annotate}

In the previous examples, we presented the server reduction rules
where, for each syntactic fragment, a fresh reference $\rf{f}$ is
generated and bound to the environment. In the rest of this thesis,
we will simply assume that all fragments syntactically present in the
program are annotated with a unique reference. Such annotation
is purely syntactic and can be done by walking the syntax tree of the program.
Annotated fragments are noted $\cv{\dots}_{\rf{f}}$.

Mixed structures syntactically present in the program are also annotated
in a similar manner with a unique module reference. Annoted
mixed structures are noted $\struct{\dots}_{\rf{F}}$.

\subsubsection{Mixed modules}
\label{sec:eliom:execution:modules}

Let us now describe the reduction relation for mixed modules.
The mixed reduction relation is presented in \cref{eliom:semmixed} and,
just like the server relation, has for goal
to evaluate all the server code and emit a client program to be later
evaluated by the client relation.
Mixed modules can be composed of either mixed functors, functor applications
or structures. The mixed relation contains various rules that are similar
to the \ml reduction rules for modules. The notable novel aspect
of mixed functor is that they both have a client part and a server part.
This is different from client fragments, which only have a client part that
can be manipulated on the server via an identifier.
The server part of mixed modules also need to indicate its client part. In order
to do this, each mixed structure will contains an additional field called
$\dyn$ which contains a module identifier. The identifier points to a
globally bound module on the client which is the result of the client-side
evaluation.

Let us first demonstrate these features in \cref{ex:eliom:mixed}.
In this example, we declare a mixed module $X$ containing
a fragment $x$ and an integer $y$. We then declare another mixed module $Y$
containing a submodule. The structure of the emitted client code
mimics closely the structure of the server code. In particular, the $\pbind$
operation is nested inside the mixed module $X$ that is emitted on the client.
The exact same names are reused on the client. We also register
each structure in the global environment using the annotated identifier of
the structure. Here, we use the $\pbind$ construct
as a shorthand for $\pbind\ \mathtt{with}$ that doesn't change the
environment.
The shape of the program is kept intact thanks to the
\mysc{MixedModVar}, \mysc{MixedQualModVar} and \mysc{MixedStruct} rules.
The first two are similar to the non mixed version, but the last one
deserves some explanation.
First, it prefixes all the fragment references inside the body of the
structure. This is for consistency with functors, as we will see later.
It then adds the $\dyn$ field to the returned structure, as discussed before.
Finally, it emits a $\pbind$ on the client
and returns the module reference. Each structure is thus bound
appropriately, even when nested.

Module identifiers are not used in the present program, but they are
used in the case of mixed functors, as we will see now.

\begin{ex}[!hbt]
  \setlength{\jot}{0pt}% Remove interline space in align*
  \[\begin{aligned}
      &\modulem[\sideX]{X}{\mathtt{struct}}\\
      &\quad\letm[s]{x}{\cv{1}}\\
      &\quad\letm[c]{y}{2 + \injf{x}{\fragment}}\\
      &\dend_{\rf{X}}\\
      &\modulem[\sideX]{Y}{\mathtt{struct}}\\
      &\quad\modulem[\sideX]{A}{X}\\
      &\dend_{\rf{Y}}\\
      &\letm[c]{\ret}{Y.A.y}
    \end{aligned}
    \quad\pred{\quad}_{\sideX}\quad
    \begin{aligned}
      &\bind{X}{\mathtt{struct}}\\
      &\quad\bindenv{X.f}\\
      &\quad\bindw{r}{X.f}{1}\\
      &\quad\letm[]{y}{2 + \app{\fragment^c}{\rf{r}}}\\
      &\dend\\
      &\modulem[]{X}{\rf{X}}\\
      &\bind{Y}{\mathtt{struct}}\\
      &\quad\modulem[]{A}{X}\\
      &\dend\\
      &\modulem[]{Y}{\rf{Y}}\\
      &\letm[]{\ret}{Y.A.y}
    \end{aligned}
    \quad\pred{\quad}_c\quad
    3,\niltr
  \]
  \vspace{-4mm}
  \caption{Execution of mixed modules}
  \label{ex:eliom:mixed}
  \vspace{-2mm}
\end{ex}

\paragraph{Mixed functors, injections and client side application}
\label{sec:eliom:execution:modules:client}

\begin{ex}[!htbt]
  \setlength{\jot}{0pt}% Remove interline space in align*
  \[\begin{aligned}
      &\letm[s]{x}{1}\\
      &\modulem[\sideX]{F(X:\Mm)}{\dstruct}\\
      &\quad\letm[c]{b}{X.a + \injf{x}{\intty}}\\
      &\dend_{\rf{Y}}\\
      &\modulem[\sideX]{Y}{\mathtt{struct}}\\
      &\quad\letm[c]{a}{2}\\
      &\dend_{\rf{Y}}\\
      &\modulem[c]{Z}{\appm{F}{Y}}\\
      &\letm[c]{\ret}{Z.b}
    \end{aligned}
    \quad\pred{\quad}_{\sideX}\quad
    \begin{aligned}
      &\modulem[]{F(X:\Mm)}{\dstruct}\\
      &\quad\letm[]{b}{X.a + \app{\intty^c}{1}}\\
      &\dend_{\rf{Y}}\\
      &\bind{Y}{\mathtt{struct}}\\
      &\quad\letm[]{a}{2}\\
      &\dend\\
      &\modulem[]{Y}{\rf{Y}}\\
      &\modulem[]{Z}{\appm{F}{Y}}\\
      &\letm[]{\ret}{Z.b}
    \end{aligned}
    \quad\pred{\quad}_c\quad
    3,\niltr
  \]
  \vspace{-4mm}
  \caption{Execution of mixed functors with injections}
  \label{ex:eliom:functclient}
\end{ex}

Before exposing the complex interaction of mixed functors and fragment, let us
illustrate various details about mixed functors in \cref{ex:eliom:functclient}.
The server code proceed in the following way: we first define
a server variable $x$ followed by a mixed functor $F$ containing an
injection. We then define a mixed module $Y$ and executes \emph{on the client}
the functor application $\appm{F}{Y}$.

First, let us recall that injections inside mixed functors
can only refer to elements outside the functor. This means
that injections inside functors can be reduced as soon
as we consider a functor. In particular, we do not wait
for functor application. This can be seen in
the \mysc{ModClosure} rule which returns
a functor closure on the server side and emit the client
part of the functor on the client side.
We then take the client part of
the body of the functor (noted $\restrict{c}{\mm}$)
and applies the $\pred{}_{c/s}$ reduction relation,
which executes injections inside client code.
In this example, it results in the injection $\injf{x}{\intty}$ being
resolved immediately in the client-side version of the functor.

Mixed functor application can be done in client and server contexts.
When it is done in a
client context, we simply call the client-side definition
and omits the server-side execution completely. Hence we can simply
emit the client-code $\appm{F}{Y}$. Execution is done through the
usual rules for client sections. This is always valid since
each mixed declaration emits a client declaration with the same name
and the same shape.

\paragraph{Mixed functors and fragments}\label{sec:eliom:execution:functor}

\begin{ex}[!bt]
  \setlength{\jot}{0pt}% Remove interline space in align*
  \[\begin{aligned}
      &\modulem[\sideX]{F(X:\Mm)}{\mathtt{struct}}\\
      &\quad\letm[s]{x}{\cv{X.a + \injf{X.b}{\intty}}_{\rf{f_x}}}\\
      &\dend_{\rf{F}}\\
      &\modulem[\sideX]{Y}{\mathtt{struct}}\\
      &\quad\letm[c]{a}{4}\\
      &\quad\letm[s]{b}{2}\\
      &\dend_{\rf{Y}}\\\\
      &\modulem[\sideX]{Z}{\appm{F}{Y}}\\\\
      &\letm[c]{\ret}{\injf{Z.x}{\fragment}}
    \end{aligned}
    \ \pred{\ }_{\sideX}\ %
    \begin{aligned}
      &\\
      &\bindenv{F}\\
      &\modulem[]{F(X:\Mm)}{\struct{}}\\
      &\bind{Y}{\mathtt{struct}}\\
      &\quad\letm[]{a}{4}\\
      &\dend\\
      &\modulem[]{Y}{\rf{Y}}\\\\
      &\bind{R_Z}{\dstruct}\\
      &\quad\modulem[]{X}{\rf{Y}}\\
      &\quad\bindenv{R_Z.f_x}\\
      &\quad\bind{r_x}{Y.a + \app{\intty^c}{2}}\\
      &\quad\ \ \mathtt{with}\ \rf{R_Z.f_x}\\
      &\dend\ \mathtt{with}\ \rf{F}\\
      &\modulem[]{Z}{\appm{F}{Y}}\\\\
      &\letm[]{\ret}{\app{\fragment^c}{\rf{r_x}}}
    \end{aligned}
    \ \pred{\ }_c\ 6,\niltr
  \]
  \vspace{-3mm}
  \caption{Execution of mixed functors with fragments}
  \label{ex:eliom:functor}
\end{ex}

The difficulty of the reduction of mixed functor containing
fragments is that the
server-side application of a mixed functor should result in both
server and client effects. This makes the reduction
rules for mixed functor application quite delicate.
We illustrate this with \cref{ex:eliom:functor}.
In this example, we define a functor $F$ contains only the server
declaration $x$. The argument of the functor simply
contains two integers, one on the server and one on the client.
In the fragment bound to $x$, we add the two integers (using
an escaped value). The interesting aspect here is that the body of
the client fragment depends on both the client and the server side
of the argument, even if there is no actual client side for the functor $F$.
The rest of the program is composed of a simple mixed module $Y$ and the
mixed functor application $F(Y)$.

\begin{figure}[!btp]
  \mprset{sep=1.5em}
  \setlength{\jot}{-3pt}% Remove interline space in align*
  \renewcommand\added\addedx
  \input{theory/semmixed}\vspace{-5mm}
  \caption{Semantics for mixed modules --
    $\redel{\envr}{M}{\vm}{\clientprog}{\trace}$ }
  \label{eliom:semmixed}
\end{figure}

The first step of the execution is to define the client side part of $F$ and
$Y$, as demonstrated in the previous example.
In this case, since $F$ only contains a server side declaration, the client
part of the functor returns an empty structure.
We then have to
execute $\appm{F}{Y}$. This is done with the \mysc{StructBeta} rule.
When reducing a mixed functor application, we first generate a fresh
identifier ($\rf{R_Z}$ here) and prefix all the fragment closure identifiers.
We then evaluate the body of the functor on the server, which gives us
both the server module value and the generated client code.
In this case, we simply obtain the binding of $\rf{r_x}$. Note
that this reference is not prefixed by $\rf{R_Z}$ since it
is freshly generated at runtime. If the functor was applied
again, we would simply generated a new one.
In order for functor arguments to be properly available on the client,
we need to introduce additional bindings. For this purpose, we lookup
the $\dyn$ field for each module argument and insert the
additional binding. In this case, $\modulem[]{X}{\rf{Y}}$.
This gives us a complete client structure which we can bind to $\rf{R_Z}$.

We see here that the body of functors allows to emit client code
in a dynamic but controlled way. Generated module references used on the
client are remembered on the server using the $\dyn$ field while
closure identifiers ensure that the proper environment is used.
One problematic aspect of this method is that it leads to two
executions of the client side. A partial work-around is to keep track of
loaded client module to avoid duplicating side-effects.

%%% Local Variables:
%%% mode: latex
%%% TeX-master: "../main"
%%% End:

\section{Integration with \ocaml}\label{sec:base}

We now formalize more precisely the interaction between \etiny, its location
system, and the vanilla \ml language.
In particular, \cref{theorem:base} shows that \ml modules
can be completely embedded in \etiny programs without changes.

Let us note
$\substloc{\loc}{\loc'}{\Mm}$ the
substitution on locations in an \eliom module type $\Mm$.
Given an \ml module $\mm$, we note $\substloc{\ml}{\loc}{\mm}$ the \eliom
module where all the module components have been annotated with location
$\loc$.
Given en \eliom module $\mm$, we note $\substloc{}{\ml}{\mm}$ the \ml module
where all the location have been erased.
We extend these notations to module types and environments.

\subsection{Properties of specialization}

Let us first clarify the behavior of specialization with regard to mono-located \etiny module types.

\begin{proposition}\label[proposition]{prop:spec:identity}
  Given an \eliom module type $\Mm$ and a location $\loc\in \{\base,c,s\}$,
  if $\wfm[\loc]{\Env}{\Mm}$, then $\specialize{\loc}{\loc}{\Mm}=\Mm$.
\end{proposition}
\begin{proof}
  By definition of $\inclloc$, $\Mm$ can only contain declarations on $\loc$.
  This means that, by reflexivity of $\subloc$, only specialization rules
  that leave the declaration unchanged are involved.
\end{proof}

\begin{proposition}
  Given an \eliom module type $\Mm$ and a location $\loc\in \{c,s\}$,
  if $\wfm[\base]{\Env}{\Mm}$,
  then $\specialize{\base}{\loc}{\Mm}=\substloc{\base}{\loc}{\Mm}$.
\end{proposition}
\begin{proof}
  We remark that for all $\loc\in\{c,s\}$, $\canuseloc{\base}{\loc}$.
  Additionally, mixed functors cannot appear on base
  (since $\cannotuseloc{\sideX}{\base}$).
  We can then proceed by induction over the rules for specialization.
\end{proof}

\subsection{Interaction between \ml and \etiny}

We can now relate \ml code to equivalent \etiny code that has been annotated
with a single location.

\begin{proposition}\label[proposition]{prop:mltoeliom}
  Given \ml type $\tau$, expression $e$, module $\mm$ and module type $\Mm$ and
  locations $\loc$, $\loc'$:
  \begin{align*}
    \wf[\ml]{\Env}{\tau}
    &\implies \wf[\loc]{\substml[\loc']\Env}{\tau}
    &\text{Where }\canuseloc{\loc'}{\loc}\\
    \wt[\ml]{\Env}{e}{\tau}
    &\implies \wt[\loc]{\substml[\loc']\Env}{e}{\tau}
    &\text{Where }\canuseloc{\loc'}{\loc}\\
    \wfm[\ml]{\Env}{\Mm}
    &\implies \wf[\loc]{\substml[\loc']\Env}{\substml{\Mm}}
    &\text{Where }\canuseloc{\loc'}{\loc}\\
    \wtm[\ml]{\Env}{\mm}{\Mm}
    &\implies \wtm[\loc]{\substml[\loc']\Env}{\substml{\mm}}{\substml{\Mm}}
    &\text{Where }\canuseloc{\loc'}{\loc}
  \end{align*}
\end{proposition}
\begin{proof}
  We remark that each syntax, typing rule or well formedness rule for
  \ml has a direct equivalent rule in \eliom.
  We can then simply rewrite the proof tree of the hypothesis to use
  the \eliom type and well-formedness rules.
  We consider only some specific cases:
  \begin{itemize}
  \item By \cref{prop:spec:identity} and since the modules are of uniform
    location, the specialization operation in \mysc{Var} and \mysc{ModVar}
    are the identity.
  \item The side conditions $\canbedefloc{\loc'}{\loc}$ are always respected
    since the modules are of uniform location and by reflexivity of $\inclloc$.
  \item The side conditions $\canuseloc{\loc'}{\loc}$ are respected by hypothesis.\qedhere
  \end{itemize}
\end{proof}

\begin{proposition}\label[proposition]{prop:basetoml}
  Given \ml type $\tau$, expression $e$, module $m$ and module type $\Mm$:
  \begin{align*}
    \wf[\base]{\Env}{\tau}
    &\implies \wf[\ml]{\substloc{}{\ml}\Env}{\tau}&
    \wfm[\base]{\Env}{\Mm}
    &\implies \wf[\ml]{\substloc{}{\ml}\Env}{\substloc{}{\ml}\Mm}\\
    \wt[\base]{\Env}{e}{\tau}
    &\implies \wt[\ml]{\substloc{}{\ml}\Env}{e}{\tau}&
    \wtm[\base]{\Env}{\mm}{\Mm}
    &\implies \wtm[\ml]{\substloc{}{\ml}\Env}
      {\substloc{}{\ml}{\mm}}{\substloc{}{\ml}\Mm}
  \end{align*}
\end{proposition}
\begin{proof}
  We first remark that the following features are forbidden in the base part
  of the language: injections, fragments, mixed functors and any other location
  than base. The rest of the language contains no tierless features and
  coincides with \ml. We can then proceed by induction over the proof trees.
\end{proof}

\begin{proposition}\label[proposition]{prop:execbase}
  Given an \ml module $\mm$ (resp. expression $e$),
  an execution environment $\envr$, a location $\loc$, a value $\vm$ (resp. $v$) and
  a trace $\trace$:
  \begin{align*}
    \renewcommand\added\addedhide
    \redel[]{\envr}{\mm}{\vm}{\emptym}{\trace}
    &\iff
      \redel[\loc]{\substloc{\ml}{\loc}\envr}
      {\substloc{\ml}{\loc}{\mm}}
      {\substloc{\ml}{\loc}{\vm}}
      {\emptym}{\trace}\\
    \renewcommand\added\addedhide
    \redel[]{\envr}{e}{v}{\emptym}{\trace}
    &\iff
      \redel[\loc]{\substloc{\ml}{\loc}\envr}
      {\substloc{\ml}{\loc}{e}}
      {\substloc{\ml}{\loc}{v}}
      {\emptym}{\trace}
  \end{align*}

  Furthermore, given an \ml program $\Pm$,
  an execution environment $\envr$, a value $v$ and a trace $\trace$:
  \begin{align*}
    \renewcommand\added\addedhide
    \redel[]{\envr}{\Pm}{v}{\emptym}{\trace}
    &\iff
      \redp{\substloc{\ml}{\csloc}\envr}
      {\substloc{\ml}{\csloc}{\Pm}}
      {\substloc{\ml}{\csloc}{v}}
      {\trace}
    &\csloc\in\{c,s\}\\
    \renewcommand\added\addedhide
    \redel[]{\envr}{\Pm}{v}{\emptym}{\trace}
    &\iff
      \redp{\substloc{\ml}{\base}\envr}
      {\substloc{\ml}{\base}{\Pm}}
      {\substloc{\ml}{\base}{v}}
      {\trace\ctr\trace}
  \end{align*}
\end{proposition}
\begin{proof}
  Let us first note that the \ml reduction relation is included in the base, the server
  and the client-only relations.
  Additionally, the considered programs, modules or expressions can not contain fragments,
  injections or binds.
  The additional rules in the server and client-only relations are only used
  for these additional syntactic constructs.
  For the first three statements, we can then proceed by induction.
  For the last statement, we remark that base code is completely copied to the client
  during server execution. Using rule \mysc{Program}, we execute the program
  twice, which returns the same value but duplicates the trace.
\end{proof}

\begin{theorem}[Base/\ml correspondance]
  \label[theorem]{theorem:base}
  \eliom modules, expressions and types on base location $\base$ correspond
  exactly to the \ml language.
\end{theorem}
\begin{proof}
  By \cref{prop:mltoeliom,prop:basetoml,prop:execbase}.
\end{proof}

Thanks to \cref{theorem:base}, we can completely identify the language
\ml and the part of \eliom on base location.
This is of course by design: the base location allows us to reason about
the host language, \ocaml, inside the new language \eliom.
It also provides
the guarantee that anything written in the base location does not contain any
communication between client and server.
In the rest of the thesis, we omit location substitutions of the form
$\substml[\base]{}$ and $\substloc{\base}{\ml}{}$.

\Cref{prop:mltoeliom} also has practical consequences:
Given a file previously typechecked by an \ml
typechecker, we can directly use the module types either on base, but also
on the client or on the server, by simply annotating all the signature
components.
This give us the possibility, in the implementation, to completely
reuse compiled objects from the \ocaml typechecker and load them
on an arbitrary location.
In particular, it guarantees that we can reuse pure \ocaml libraries
safely and freely.

\section{Compilation of \eliom programs}
\label{chap:compilation}

In \cref{sec:theory}, we gave a tour of the \etiny language
from a formal perspective, providing a type system and an interpreted semantics.
\eliom, however, is not an interpreted language. The interpreted semantics
is here both as a formal tool and to make the semantics of the language more
approachable to users, as demonstrated in \cref{sec:howto}.
In the implementation, \eliom programs are compiled to two programs: one server program
(which is linked and executed on the server) and a client program (which is
compiled to \js and executed in a browser). The resulting programs are efficient
and avoid unnecessary back-and-forth communications between the client and the server.

Description of the complete compilation toolchain, including emission of
\js code, is out of scope of this article (see \citet{SPE:SPE2187}).
Instead, we describe
the compilation process in term of emission of client and server
programs in an \ml{}-like language equipped with additional primitives.
Hence, we present
the typing and execution of compiled programs, in \cref{sec:target}
and the compilation process, in \cref{sec:compilation}.
We then show that our compilation process preserves both the typing
and the semantics in \cref{sec:simulation}.
% Finally, we discuss the design
% of mixed functors from a compilation perspective in \cref{sec:discussion}.

\subsection{Target languages \ocsis and \ocsic}
\label{sec:target}

We introduce the two target languages \ocsic and \ocsis as extensions of \ml.
The additions in these two new languages are highlighted in \cref{target:grammar}.
Typing is provided in \cref{sec:target:typing}.
The semantics is provided in \cref{sec:target:semantics}.
As before, we use globally bound identifiers, which we call ``references''
and note in bold: $\rf{r}$. References can also be paths, such as $\rf{X.r}$.
In some contexts, we accept a special form of reference path,
noted $\dyn.\rf{x}$ which we explain in \cref{sec:target:modules}.
In practice, these references are implemented
with uniquely generated names and associative tables.
Contrary to the interpreted semantics, references are also used to
transfer values from the server to the client and can appear
in expressions.
A reference used inside an expression is always of type $\serial$.

\begin{figure}[hbt]
  \input{theory/target/grammar.tex}
  \caption{Grammar for \ocsis and \ocsic as extensions of \ocsi}
  \label{target:grammar}
\end{figure}

\subsubsection{Converters}\label{sec:target:conv}
For each converter $f$, we note $f^s$ and $f^c$ the server side encoding
function and the client side decoding function. If $f$ is of type
$\conv{\tau_s}{\tau_c}$, then $f^s$ is of type $\lamtype{\tau_s}{\serial}$ and
$f^c$ is of type $\lamtype{\serial}{\tau_c}$.
We will generally assume that if the converter $f$ is available in
the environment, then $f^c$ and $f^s$ are available in the client
and server environment respectively.

\subsubsection{Injections}
For injections, we associate server-side the injected
value $e$ to a reference $\rf{v}$ using the construction $\pinj{\rf{v}}{e}$,
where $e$ is of type $\serial$.
When the server execution is over, a mapping from references to injected values
is sent to the client. $\rf{v}$ is then used client-side to access the value.

An example is given in \cref{ex:target:inj}. In this example, two integers are
sent from the server to the client and add them on the client.
We suppose the existence of a base abstract type $\intty$, a converter
$\intty$ of type $(\conv{\intty}{\intty})$ and the associated encoding and decoding functions.
The server program, in \cref{ex:target:inj:server}, creates two injections, $\rf{v_1}$
and $\rf{v_2}$ and does not expose any bindings nor return any values. These injections
hold the serialized integers $4$ and $2$.
The client program, in \cref{ex:target:inj:client}, uses these two injections, deserialize
their values, adds them, and returns the result.
Note that $\injd$ \emph{is not a network operation}. It simply stores a mapping
between references (\ie names) and serialized values. The mapping generated at the
end of the server execution is shown in \cref{ex:target:inj:serial}. After the
execution of the server code, this mapping is sent to the client, and used to execute
the client code.

\begin{ex}[hbt]
  \centering
  \begin{subfigure}[t]{.3\linewidth}
    \small\setlength{\jot}{0pt}% Remove interline space in align*
    \begin{align*}
      &\pinj{\rf{v_1}}{\app{\intty^s}{4}};\\
      &\pinj{\rf{v_2}}{\app{\intty^s}{2}};
    \end{align*}\vspace{-5mm}
    \caption{Server program}
    \label{ex:target:inj:server}
  \end{subfigure}\hfill
  \begin{subfigure}[t]{.28\linewidth}
    \small\setlength{\jot}{0pt}% Remove interline space in align*
    \begin{align*}
      &\letm[]{\ret}{}\\
      &\quad\app{\intty^c}{\rf{v_1}} + \app{\intty^c}{\rf{v_2}};
    \end{align*}\vspace{-5mm}
    \caption{Client program}
    \label{ex:target:inj:client}
  \end{subfigure}\hfill
  \begin{subfigure}[t]{.27\linewidth}
    \small\setlength{\jot}{0pt}% Remove interline space in align*
    \begin{align*}
      \rf{v_1}&\mapsto\ 4\\
      \rf{v_2}&\mapsto\ 2
    \end{align*}\vspace{-5mm}
    \caption{Mapping of injections}
    \label{ex:target:inj:serial}
  \end{subfigure}
  \caption{Client-server programs calling and using injections}
  \label{ex:target:inj}
\end{ex}

\subsubsection{Fragments}
\label{sec:target:fragment}

The primitive related to fragments also relies on shared references between
the server program and the client program. However, these references
allow to uniquely identify functions that are defined on the client but
are called on the server.
To implement this, we use the following primitives:
\begin{itemize}
\item In \ocsic structures, $\bind{p}{e}$ declares a new client function bound
  to the reference $\rf{p}$.
  The function $e$ takes an arbitrary amount of argument of type
  $\serial$ and returns any type.
\item In \ocsis expressions, $\pfragment{p}{e_1 \dots e_n}$ is a delayed function application
  which registers that, on the client, the function associated to $\rf{p}$ will
  be applied to the
  arguments $e_i$. All the arguments must be of type $\serial$. It returns a
  value of type $\fragty$, which holds a unique identifier refering to the result of this application.
\end{itemize}

Here again, none of these
primitives are network communication primitives. While the API
is similar to Remote Procedure Calls, the execution is very different:
$\fragmentd$ only accumulates
the function call in a list, to be executed later. When the server execution
is over,
the list of calls is sent to the client, and used during the client
execution.
\ocaml, and consequently \eliom, are impure languages: the order of execution
of statement is important. In order to control the order of execution, we
introduce two additional statements: $\pend{}$, on the server, introduces
an $\dend$ marker in the list of calls. $\pexec{}$, on the client, executes
all the calls until the next $\dend$ token.

\cref{ex:target:frag} presents a pair of programs which
emit the client trace $\singltr{2;3;3}$, but in such a way that, while
the client does the printing, the values and the execution
order are completely determined by the server.
The server code
(\cref{ex:target:frag:server}) calls $\rf{f}$ with $2$ as argument, injects
the result and then calls $\rf{f}$ with $3$ as argument.
The client code, in \cref{ex:target:frag:client}, declares a fragment
closure $\rf{f}$, which simply adds one to its arguments, and exec both fragments.
In-between both executions, it prints the content of the injection $\rf{v}$.
During the execution of the server, the list of calls
(\cref{ex:target:frag:list}) and the mapping of injections
(\cref{ex:target:frag:injs}) are built.
First, when $\pfragment{f}{\app{\intty^s}{2}}$ is executed, a fresh
reference $\rf{r_1}$ is generated, the call to the fragment is added
to the list and $\rf{r_1}$ is returned.
The injection adds the association $\rf{v_1}\mapsto\rf{r_1}$ to the
mapping of injections.
The call to $\pend$ then adds the token $\tokend$ to the list of fragments.
The second fragment proceeds similarly to the first, with a fresh
identifiers $\rf{r_2}$.
Once server execution is over, the newly generated list of fragments
and mapping of injections are sent to the client.
During the client execution, the
execution of the list is controlled by the $\dexec$ calls.
First, $\app{\rf{f}}{2}$ emits $\singltr{2}$ and
is evaluated to $3$, and the mapping $\rf{r_1}\mapsto 3$ is added to a global
environment. Then $\rf{v_1}$ is resolved to $\rf{r_1}$ and printed
(which shows $\singltr{3}$).
Finally $\app{\rf{f}}{3}$ emits $\singltr{3}$ and is evaluated to $4$.

The important thing to note here is that both the injection mapping and the
list of fragments are completely dynamic. We could add complicated control
flow to the server program that would drive the client execution
according to some dynamic values.
The only static elements are
the names $\rf{f}$ and $\rf{v_1}$,
the behavior of $\rf{f}$
and the number of call to $\pexec$.
We cannot, however, make the server-side control flow depend on
client-side values, since we obey a strict phase separation between
server and client execution.

Finally, remark that we do not need the $\bindenv{}$ construct
introduced in \cref{sec:eliom:execution:client}. Instead, we directly
capture the environment using closures that are extracted in advance.
We will see how this extraction works in more details while studying
the compilation scheme, in \cref{sec:compilation}.

\begin{ex}[hbt]
  \centering
  \begin{subfigure}[t]{.32\linewidth}
    \small\setlength{\jot}{0pt}% Remove interline space in align*
    \begin{align*}
      &\letm[]{x_1}{\pfragment{f}{\app{\intty^s}{2}}};\\
      &\pinj{v_1}{\app{\fragment^s}{x_1}};\\
      &\pend{};\\
      &\letm[]{x_2}{\pfragment{f}{\app{\intty^s}{3}}};\\
      &\pend{};
    \end{align*}\vspace{-5mm}
    \caption{Server program}
    \label{ex:target:frag:server}
  \end{subfigure}\hfill
  \begin{subfigure}[t]{.3\linewidth}
    \small\setlength{\jot}{0pt}% Remove interline space in align*
    \begin{align*}
      &\bind{f}{\lam{x}{(\app{\print}{\app{\intty^c}{x}} + 1)}};\\
      &\pexec{};\\
      &\letm[]{a}{\app{\print}{\app{\fragment^c}{\rf{v_1}}}};\\
      &\pexec{};\\
      &\letm[]{\ret}{a}
    \end{align*}\vspace{-5mm}
    \caption{Client program}
    \label{ex:target:frag:client}
  \end{subfigure}\hfill
  \begin{subfigure}[t]{.27\linewidth}
    \small\setlength{\jot}{0pt}% Remove interline space in align*
    \begin{align*}
      \tokfrag{r_1}{\app{\rf{f}}{2}};\tokend;\\
      \tokfrag{r_2}{\app{\rf{f}}{3}};\tokend;
    \end{align*}\vspace{-7mm}
    \caption{List of fragments}
    \label{ex:target:frag:list}
    \vspace{-3mm}
    \begin{align*}
      \rf{v_1}&\mapsto\ \rf{r_1}
    \end{align*}\vspace{-7mm}
    \caption{Mapping of injections}
    \label{ex:target:frag:injs}
  \end{subfigure}
  \caption{Client-server program defining and calling fragments}
  \label{ex:target:frag}
\end{ex}

\subsubsection{Modules}\label{sec:target:modules}

We introduce three new module-related construction that
are quite similar to fragment primitives:
\begin{itemize}
\item $\binddyn{\rf{p}}{\mm}$ is equivalent to $\pbind$ for modules. It is a client
  instruction that associates the module or functor $\mm$ to the reference $\rf{p}$.
\item $\fragm{p}{(\rf{R_1})\dots(\rf{R_n})}$ is analogous to $\pfragment{p}{e}$ for modules.
  It is a delayed functor application that is used on the server to register that the functor associated to $\rf{p}$ will have to be applied to the modules
  associated to $\rf{R_i}$. It returns
  a fresh reference that represents the resulting module. Contrary to $\mathtt{fragment}$,
  it can only be applied to module references.
\item $\getdyn{p}$ returns a reference that represents the client part of a server
  module $p$. This is used for \etiny mixed structure that have both a server
  and a client part.
\end{itemize}

The first argument of $\fragmentd$, $\fragmentd_m$, $\pbind$ and $\pbind_m$
can also be a reference path $\dyn.\rf{f}$, where $\dyn$ is the locally bound
$\dyn$ field inside a module. This allows us to isolate some bound
references inside a fresh module reference. This is useful
for functors, as we will now demonstrate in \cref{ex:target:module}.

In this example, we again add integers\footnote{%
  But better! or at very least, more obfuscated.}
on the client
while controlling the values and the control flow on the server.
We want to define server modules that contain server values but also
trigger some evaluation on the client, in a similar way to fragments.
The first step is to define a module $X$ on the server and to bind
a corresponding module $\rf{X}$ on the client. Similarly to the
interpreted semantics presented in \cref{sec:eliom:execution}, we
add a $\dyn$ field to the server module that points to the client
module. Plain structures such as $X$ are fairly straightforward,
as we only need to declare each part statically and add the needed
reference. $\pbind_m$ allows to declare modules globally.

We then declare the functor $F$ on the server and bind
the functor $\rf{F}$ on the client. The server-side functor contains a call
to a fragment defined in the client-side functor. The difficulty here
is that we should take care of differentiating between fragment
closures produced by different functor applications.
For this purpose, we use a similar technique than the one presented
in \cref{sec:eliom:execution:modules}, which is to prefix
the fragment closure identifier $\rf{f}$ with the reference of the
client-side module. This reference is available on the server
side as the $\dyn$ field and is generated by a call to $\fragmentd_m$.
When $F$ is applied to $X$ on the server, we generate a fresh reference
$\rf{R}$ and add $\tokfrag{R_1}{\rf{F}\ \rf{X_0}}$ to the execution queue.
When $\pexec$ is called, We introduce the additional binding
$\bdr{\dyn}{\rf{R_1}}$ in the environment and apply
$\rf{F}$ to $\rf{X_0}$,
which will register the $\rf{R_1.f}$ fragment closure.
Since it is the result of this specific functor application, the
closure $\rf{R_1.f}$ will always add $4$ to its argument.
The rest of the execution proceed as shown in the previous section:
we call a new fragment, which triggers the client-side addition $2+4$
and use an injection to pass the results around.

\begin{ex}[!hb]
  \centering
  \begin{subfigure}[t]{.33\linewidth}
    \small\setlength{\jot}{0pt}% Remove interline space in align*
    \begin{align*}
      &\modulem[]{X}{\mathtt{struct}}\\
      &\quad\modulem[]{\dyn}{\rf{X_0}};\\
      &\quad\letm[]{a}{2}\\
      &\mathtt{end};\\
      &\modulem[]{F(Y:\Mm_s)}{\mathtt{struct}}\\
      &\quad\modulem[]{\dyn}{\fragm{\rf{F}}{(\getdyn{Y})}};\\
      &\quad\letm[]{b}{\pfragment{\dyn.f}{\app{\intty^s}{Y.a}}};\\
      &\mathtt{end};\\
      &\modulem[]{Z}{\appm{F}{X}};\\
      &\pend{};\\
      &\pinj{v_1}{(\fragment^s\ Z.b)};
    \end{align*}\vspace{-6mm}
    \caption{Server program}
    \label{ex:target:module:server}
  \end{subfigure}\hfill
  \begin{subfigure}[t]{.36\linewidth}
    \small\setlength{\jot}{0pt}% Remove interline space in align*
    \begin{align*}
      &\binddyn{\rf{X_0}}{\mathtt{struct}}\\
      &\quad\letm[]{c}{4}\\
      &\mathtt{end};\\\\
      &\binddyn{\rf{F}(Y:\Mm_c)}{\mathtt{struct}}\\
      &\quad\bind{\dyn.f}{}\\
      &\quad\ \ \lam{a}{(\app{\intty^c}{a} + Y.c)};\\
      &\mathtt{end};\\\\
      &\pexec{}\\
      &\letm[]{\ret}{\app{\fragment^c}{\rf{v_1}}};
    \end{align*}\vspace{-6mm}
    \caption{Client program}
    \label{ex:target:module:client}
  \end{subfigure}\hfill
  \begin{subfigure}[t]{.21\linewidth}
    \small\setlength{\jot}{0pt}% Remove interline space in align*
    \begin{align*}
      &\tokfrag{R_1}{\rf{F}\ \rf{X_0}};\\
      &\tokfrag{r_2}{\rf{R_1.f}\ 2};\\
      &\tokend
    \end{align*}\vspace{-7mm}
    \caption{List of fragments}
    \label{ex:target:module:list}
    \vspace{-3mm}
    \begin{align*}
      \rf{v_1}&\mapsto\ \rf{r_2}
    \end{align*}\vspace{-7mm}
    \caption{Mapping of injections}
    \label{ex:target:module:injs}
  \end{subfigure}
  \caption{Client-server program using module fragments}
  \label{ex:target:module}\vspace{-2mm}
\end{ex}

\subsubsection{Type system rules}\label{sec:target:typing}

\begin{figure}[!tb]
  \vspace{-3mm}
  \input{theory/target/typing.tex}
\end{figure}

The \ocsis and \ocsic typing rules are presented in
\cref{target:typing:server,target:typing:client} as
a small extension over the \ml typing rules presented in \cref{sec:ml:typing}.
Note that the typing rules for the new primitives are weakly typed
and are certainly not sound with respect to serialization and
deserialization. Given arbitrary \ocsis and \ocsic programs, there is no
guarantee that (de)serialization will not fail at runtime. This is on purpose.
Indeed, all these guarantees are provided by \eliom itself. \ocsis and \ocsic
are target languages that are very liberal by design, so that all patterns
permitted by \eliom are expressible with them.
Furthermore, from an implementation perspective. \ocsis and \ocsic are simply
\ocaml libraries and do not rely on further compiler support.
Note that $\dyn$ fields are not reflected in signatures.
The fragment \mysc{Fragment$_m$} rule
does not enforce that the $\dyn$ field is present in all the arguments.
This is enforced by construction during compilation.

\subsubsection{Semantics rules}\label{sec:target:semantics}

We define two reduction relations as extensions of the \ml reduction
rules (see \cref{sec:ml:semantics}).
The $\pred{}_{\ocsis}$ reduction for \ocsis server programs is presented
in \cref{target:semantics:server}.
The $\pred{}_{\ocsic}$ reduction for \ocsic client programs is presented
in \cref{target:semantics:client}.
Let us consider a server structure $\sm_s$ and a client structure $\sm_c$.
A paired
execution of the two structures is presented below:
\begin{align*}
  \redmls{\envr_s}{\sm_s}{\vm_s}{\envfrag}{\envinj}{\trace_s}
  &&\redmlc{\envr_c}{\sm_c}{\vm_c}{\envfrag}{\envfrag'}
     {\envg\cup\envinj}{\envg'}{\trace_c}
\end{align*}
Let us now detail these executions rules.
As with the \ml reduction,
$\envr_s$ and $\envr_c$ are
the local environments of values while $\trace_s$ and $\trace_c$ are
the traces for server and client executions respectively. $\vm_s$ and
$\vm_c$ are the returned values.
Similarly to the interpreted semantics for \etiny, the client reduction uses
global environments noted $\envg$.

As we saw in the previous examples, server
executions emits two sets of information during execution: a queue of fragments
and a map of injections.
Mapping of injection is a traditional (global) environment where bindings are
noted $\bdr{\rf{v}}{\dots}$.
The queue of fragments is noted $\envfrag$ and
contains end tokens $\tokend$ and fragment calls
$\tokfrag{r}{\rf{f}\ v_1\dots v_n}$. Concatenation of fragment queues is
noted $\cfrag$. We now see the various rules in more details.

\paragraph{Injections}

Injection bindings are collected on the server
through the \mysc{Injection} rule.
When creating a new injection binding, we inject the server-side value
using the injection of value operator, noted $\injval{v}$ and presented
in \cref{eliom:injval}. This models the serialization of values
before transmission from server to client by ensuring that only
base values and references are injected. Other kinds of values should be
handled using converters explicitly.

The injection environment $\envinj$ forms a valid client-side
global environment. When executing the client-side program, we simply assume
that $\envinj$ is included in the initial global environment $\envg$.

\paragraph{Fragments and functors}

On the server, fragments and functors calls are added to the queue through
the \mysc{Fragment} and \mysc{Fragment$_\sideX$} rules.
In both rules, the reference of the associated closure or functor
is provided, along with a list of arguments. A fresh reference symbolizing
the fragment is generated and the call is added to the queue $\envfrag$.
Note that in the case of regular fragments, the arguments
are expressions which can themselves contains fragment calls.
The module rule \mysc{Fragment$_\sideX$} is similar,
the main difference being that it only
accept module references as arguments of the call.

Fragment closures and functors are bound on the client through
the \mysc{Bind} and \mysc{Bind$_\sideX$} rules, which simply binds a reference
to a value or a module value in the global environment $\envg$.
Since $\pbind$ accepts references of the form $\dyn.\rf{f}$, it must first
resolves $\dyn$ to the actual reference. This is done through the \mysc{Dyn}
rule.

\paragraph{Segmented execution}

\begin{figure}[!hbtp]
  \mprset{sep=1.5em}
  \input{theory/target/semserver}\vspace{-3mm}
  \caption{Semantics rules for \ocsis
  -- $\redmls{\envr}{\sm}{\vm}{\envfrag}{\envinj}{\trace}$}
  \label{target:semantics:server}
  \input{theory/target/semclient}\vspace{-3mm}
  \caption{Semantics rules for \ocsic
  -- $\redmlc{\envr}{\sm}{\vm}{\envfrag}{\envfrag'}{\envg}{\envg'}{\trace}$}
  \label{target:semantics:client}
\end{figure}

In \ocsis and \ocsic programs, the execution of fragments is segmented
through the use of the $\pexec/\pend$ instructions. On the
server, $\dend$ instructions are handled through the \mysc{End} rule, which
simply adds an $\tokend$ token to the execution queue $\envfrag$.
On the client, we use the \mysc{Exec} rule, associated to the \mysc{Frag}
and \mysc{Frag$_\sideX$} rules.
When $\dexec$ is called, the \mysc{Exec} rule triggers
the execution of the segment of the queue until the next $\tokend$ token.
Each token $\tokfrag{r}{\rf{f}\ v_1\dots v_n}$ is executed
with the \mysc{Frag} rule as
the function call $\app{\rf{f}}{v_1\dots v_n}$. The result of
this function call is bound to $\rf{r}$ in the global environment $\envg$.
Similarly, functor calls are executed using the \mysc{Frag$_m$} rule.
Note that for functors we also introduce the $\dyn$ field in the local
environment, which allows local $\pbind$ definitions.
Once all the tokens have been executed in the considered fragment queue,
we resume the usual execution.

%%% Local Variables:
%%% mode: latex
%%% TeX-master: "../main"
%%% End:

\subsection{Slicing}\label{sec:compilation}

In \cref{sec:target}, we presented the two target languages \ocsis and \ocsic.
We now present the compilation process transforming \emph{one} \etiny program
into two distinct \ocsis and \ocsic programs. Before giving a more formal
description in \cref{sec:compilation:rules},
we present the compilation process through three examples of increasing
complexity.

\paragraph{Injections and fragments}

\cref{compilation:example:expression} presents an \etiny program
containing only simple declarations involving fragments and
injections without modules.
The \etiny program is presented on the left, while the compiled
\ocsis and \ocsic programs are presented on the right.
In this example, a first fragment is created. It only contains an integer and is
bound to $a$. A second fragment that uses $a$ is created and bound on the server to $b$.
Finally, $b$ is used on the client via an injection. The program returns $4$.

For each fragment, we emit a $\pbind$ declaration on the client.
The client expression contained
in the fragment is abstracted and transformed in a closure that is bound to a fresh
reference. The number of arguments of the closure corresponds to the number of
injections inside the fragment. Similarly to the interpreted
semantics, we use the client part of the converter on the client.
In this case, $\cv{1}$ is turned
into $\lam{()}{1}$ and $\cv{\injf{a}{\fragment}+1}$ is turned
into $\lam{v}{((\fragment^c\ v)+1)}$.
On the server, each fragment is replaced by a call to the primitive $\fragmentd$.
The arguments of the call are the identifier of the closure and all the injections
that are contained in the fragment.
The $\fragmentd$ primitive, which was presented in \cref{sec:target:fragment},
registers that the closure declared on the client should be executed later on.
Since all the arguments of $\fragmentd$ should be of type $\serial$, we apply the
client and server parts of the converters at the appropriate places.
The $\dexec$ and $\dend$ primitives synchronize the execution so that the order
of side effects is preserved. When $\dexec$ is encountered, it executes queued
fragment up to an $\dend$ token which was pushed by an $\dend$ primitive.
We place an $\dexec$/$\dend$ pair at each server section. This is enough
to ensure that client code inside server fragment and client code
in regular client declaration is executed in the expected order.

Note that injections, which occur outside of fragments,
and escaped values, which occur inside fragments,
are compiled in a very different way.
Injections have the useful property that the use site and number of injections is
completely static: we can collect all the injections on the server, independently of
the client control flow and send them to the client. This is the property that allows
us to avoid communications from the client to the server.

\begin{ex}[!ht]
  \small
  \setlength{\jot}{0pt}% Remove interline space in align*
  \centering
  \begin{tabular}{@{}c|c|c@{}}
    \etiny & \ocsis & \ocsic \\\hline
    $\begin{aligned}
       &\letm[s]{a}{\cv{{1}}};\\
       &\\
       &\letm[s]{b}{\cv{\injf{a}{\fragment}+1}};\\
       &\\
       &\letm[c]{\ret}{\injf{b}{\fragment}+2};
     \end{aligned}$&
    $\begin{aligned}
      &\letm[]{a}{\pfragment{\idclosure_0}{()}};\\
      &\pend{};\\
      &\letg b = \pfragment{\idclosure_1}{(\fragment^s\ a)};\\
      &\pend{};\\
      &\pinj{x}{(\fragment^s\ b)};
    \end{aligned}$
      &$\begin{aligned}
        &\bind{f_0}{\lam{()}{1}};\\
        &\pexec{};\\
        &\bind{f_1}{\lam{v}{((\fragment^c\ v)+1)}};\\
        &\pexec{};\\
        &\letm[]{\ret}{(\fragment^c\ \rf{x})+2};
      \end{aligned}$
  \end{tabular}
  \caption{Compilation of expressions}
  \label{compilation:example:expression}
\end{ex}

\paragraph{Sections and modules}
\label{sec:compilation:sections}

We now present an example with client and server modules in
\cref{compilation:example:monofunctor}. The lines of the various programs
have been laid out in a way that should highlight their correspondence.

We declare a server module $X$ containing a client fragment,
a client functor $F$ containing an injection,
a client functor application $Z$
and finally the client return value, with another injection.
The compilation of the server module $X$ proceeds in the following way: on the
server, we emit a module $X$ similar to the original declaration but
where fragments have been replaced by a call to the $\fragmentd$ primitive.
On the client, we only emit the call to $\pbind$, without any of the server-side
code structure.
Compilation for the rest of the code proceeds in a similar manner.

This compilation scheme is correct thanks to the following insight:
In client and server modules or functors, the special instructions
for fragments and injection can be freely lifted to the outer scope.
Indeed, the fragment closure bound in $\rf{f_0}$ can only reference client
elements. Since the server $X$ can only introduce server-side variables, the body
of the fragment closure is independent from the server-side code.
A similar remark can be made about the client functor $F$: the functor argument
must be on the client, hence it cannot introduce new server binding. The server
identifier that is injected must have been introduced outside of functor and
the injection can be lifted outside the functor.

Using this remark, the structure of the \ocsis and \ocsic programs is fairly
straightforward:
Code on the appropriate side has the same shape as in the original \etiny program
and code on the other side only contains calls to the appropriate primitives.

\begin{ex}[!ht]
  \small
  \setlength{\jot}{0pt}% Remove interline space in align*
  \centering
  \begin{tabular}{@{}c|c|c@{}}
    \etiny & \ocsis & \ocsic \\\hline
    $\begin{aligned}
      &\modulem[s]{X}{\mathtt{struct}}\\
      &\quad\letm[s]{a}{\cv{2}}\\
      &\quad\letm[s]{b}{4}\\
      &\mathtt{end};\\\\
      &\modulem[c]{F(Y:\Mm)}{\mathtt{struct}}\\
      &\quad\letm[c]{a}{Y.b + \injf{X.b}{\mathtt{int}}}\\
      &\mathtt{end};\\
      &\modulem[c]{Z}{}\\
      &\quad\appm{F}{\struct{\letm[c]{b}{2}}};\\
      &\letm[c]{\ret}{}\\
      &\quad\injf{X.a}{\fragment}+Z.a;
    \end{aligned}$&
    $\begin{aligned}
      &\modulem[]{X}{\mathtt{struct}}\\
      &\quad\letm[]{a}{\pfragment{\idclosure_0}{()}}\\
      &\quad\letm[]{b}{4}\\
      &\mathtt{end};\\
      &\pend{};\\\\
      &\pinj{x_0}{(\mathtt{int}^s\ X.b)};\\ \\\\\\\\
      &\pinj{x_1}{(\fragment^s\ X.a)};
    \end{aligned}$
      &$\begin{aligned}
        &\\
        &\bind{f_0}{\lam{()}{2}};\\\\\\
        &\pexec{};\\
        &\modulem[]{F(Y:\Mm)}{\mathtt{struct}}\\
        &\quad\letm[]{a}{Y.b + (\mathtt{int}^c\ \rf{x_0})}\\
        &\mathtt{end};\\
        &\modulem[]{Z}{}\\
        &\quad\appm{F}{\struct{\letm[]{b}{2}}};\\
        &\letm[]{\ret}{}\\
        &\quad(\fragment^c\ \rf{x_1})+Z.a;
      \end{aligned}$
  \end{tabular}
  \caption{Compilation of client and server modules and functors}
  \label{compilation:example:monofunctor}
\end{ex}

\paragraph{Mixed modules}

Finally, \cref{compilation:example:mixed} presents the compilation of mixed modules.
In this example, we create a mixed structure $X$ containing a server
declaration and a client declaration with an injection. We define a functor $F$
that takes a module containing a client integer and use it both inside a client
fragment, and inside a client declaration. We then apply $F$ to $X$ and
use an injection to compute the final result of the program.

The compilation of the mixed module $X$ is similar to the procedure for
programs: we compile each declaration and use the $\mathtt{injection}$
primitive as needed.
Additionally, we add a $\dyn$ field on the server-side version of the
module. The content of the $\dyn$ field is determined statically
for simple structures (here, it is $\rf{X_0}$).
The client-side version of the module is first bound to $\rf{X_0}$ using
the $\pbind_m$ primitive. We then declare $X$ as a simple alias.
This alias ensures that $X$ is also usable in client sections transparently.

For functors, the process is similar. One additional complexity is
that the $\dyn$ field should be dynamically generated.
For this purpose, we add a call to the $\fragmentd_m$ primitive.
Each call to $\fragmentd_m$ generates a new, fresh identifier.
We also prefix each call to $\fragmentd$ by the $\dyn$ field.
On the client, we emit two different functors.
The first one is called $F$ and contains only the client
declarations to be used inside the rest of the client code.
It is used for client-side usage of mixed functors. An example
with the interpreted semantics was presented in
\cref{sec:eliom:execution:modules:client}.
The other one is bound to a new reference (here $\rf{F_1}$) and contains both
client declaration, along with calls to the $\pbind$
and $\dexec$ primitives.
This function is used to perform client side effects:
when the server version of $F$ is applied, a call to $\rf{F_1}$ is registered
and will be executed when the client reaches the associated $\dexec$ call
(here, the last one).

\begin{ex}[!hb]
  \small
  \setlength{\jot}{0pt}% Remove interline space in align*
  \centering
  \begin{tabular}{@{}c|c|c@{}}
    \etiny & \ocsis & \ocsic \\\hline
    $\begin{aligned}
      &\modulem[\sideX]{X}{\mathtt{struct}}\\
      &\quad\letm[s]{a}{2}\\
      &\quad\letm[c]{b}{4 + \injf{a}{\mathtt{int}}}\\
      &\mathtt{end};\\\\[2mm]
      &\modulem[\sideX]{F(Y:\Mm)}{\mathtt{struct}}\\
      &\quad\letm[s]{c}{\cv{Y.b}}\\
      &\quad\letm[c]{d}{2 * Y.b}\\
      &\mathtt{end};\\\\\\\\\\[2mm]
      &\modulem[\sideX]{Z}{\appm{F}{X}};\\\\[2mm]
      &\letm[c]{\ret}{}\\
      &\quad\injf{Z.c}{\fragment}+Z.d;
    \end{aligned}$&
    $\begin{aligned}
      &\modulem[]{X}{\mathtt{struct}}\\
      &\quad\modulem[]{\dyn}{\rf{X_0}};\\
      &\quad\letm[]{a}{2};\ \pend{};\\
      &\quad\pinj{x_0}{(\mathtt{int}^s\ a)};\\
      &\mathtt{end};\\[2mm]
      &\modulem[]{F(Y:\Mm)}{\mathtt{struct}}\\
      &\quad\modulem[]{\dyn}{}\\
      &\qquad\fragm{\rf{F_1}}{(\getdyn{Y})};\\
      &\quad\letm[]{c}{\pfragment{\dyn.\idclosure_0}{()}};\\
      &\quad\pend{};\\
      &\mathtt{end};\\\\\\[2mm]
      &\modulem[]{Z}{\appm{F}{X}};\\\
      &\pend{};\\\\[2mm]
      &\pinj{x_1}{(\fragment^s\ Z.c)};
    \end{aligned}$
      &$\begin{aligned}
      &\binddyn{\rf{X_0}}{\mathtt{struct}}\\
      &\quad\pexec{};\\
      &\quad\letm[]{b}{4 + (\mathtt{int}^c\ \rf{x_0})}\\
      &\mathtt{end};\\
      &\modulem[]{X}{\rf{X_0}};\\[2mm]
      &\binddyn{\rf{F_1}(Y:\Mm)}{\mathtt{struct}}\\
      &\quad\binddyn{\dyn.\rf{f}_0}{\lam{()}{(Y.b)}};\\
      &\quad\pexec{};\\
      &\quad\letm[]{d}{2 * Y.b}\\
      &\mathtt{end};\\
      &\modulem[]{F(Y:\Mm)}{\mathtt{struct}}\\
      &\quad\letm[]{d}{2 * Y.b}\\
      &\mathtt{end};\\[2mm]
      &\modulem[]{Z}{\appm{F}{X}};\\
      &\pexec{}\\[2mm]
      &\letm[]{\ret}{}\\
      &\quad(\fragment^c\ \rf{x_1})+Z.d;
      \end{aligned}$
  \end{tabular}
  \caption{Compilation of a mixed functor}
  \label{compilation:example:mixed}
\end{ex}

\subsection{Slicing rules}
\label{sec:compilation:rules}

Given an \etiny module $\mm$ (resp. module type $\Mm$, structure $\sm$, \dots)
and a location $\csloc$ that is either client or server, we note $\compile{\csloc}{\mm}$ the
result of the compilation of $\mm$ to the location $\csloc$.
The result of $\compile{s}{\mm}$ is a module of \ocsis and the result of
$\compile{c}{\mm}$ is a module in \ocsic.

Let us defines a few notation.
As before, we use $\subst[]{a}{b}{e}$ to denote the substitution of $a$ by $b$
in $e$. $\substi[]{a_i}{b_i}{e}$ denotes the repeated substitution of $a_i$ by $b_i$ in $e$.
We note $\frags{e}$ (resp. $\injs{e}$) the fragments (resp. injections)
present in the expression $e$. We note $(e_i)_i$ the sequence of elements $e_i$.
For ease of presentation, we use $\dm_{\mloc}$ (resp. $\Dm_{\mloc}$)
for definitions (resp. declarations) located on location $\mloc$.
In order to simplify our presentation of mixed functors, both
in the slicing rules and in the simulation proofs, we
consider a sliceability constraint which dictates which programs
can be sliced.

\begin{definition}[Sliceability]
\label{sec:compilation:sliceability}
  A program is said sliceable if mixed structures are only defined at top level, or
directly inside a toplevel mixed functor.
\end{definition}

We give an example of this constraint in
\cref{ex:compilation:sliceable}. The program presented on the left is not sliceable, since
it contains a structure which is nested inside a structure in a functor. The semantically
equivalent program on the right is sliceable, since structures are not nested.
This restriction can be relaxed by using a transformation similar
to lambda-lifting on mixed functors.
In the rest of this section, we assume programs
are sliceable and well typed.
\begin{ex}[!h]
  \setlength{\jot}{0pt}%
  \centering%
  \vspace{-5mm}
  \begin{subfigure}{0.4\linewidth}
    \begin{align*}
      &\modulem[\sideX]{F(X:\Mm)}{\mathtt{struct}}\\
      &\quad\modulem[\sideX]{Y}{\dstruct}\\
      &\qquad\dots\\
      &\quad\dend\\
      &\mathtt{end}
    \end{align*}\vspace{-4mm}%
    \caption{An unsliceable functor}
  \end{subfigure}\qquad
  \begin{subfigure}{0.4\linewidth}
    \begin{align*}
      &\modulem[\sideX]{Y'(X:\Mm)}{\mathtt{struct}}\\
      &\quad\dots\\
      &\dend\\
      &\modulem[\sideX]{F(X:\Mm)}{\mathtt{struct}}\\
      &\quad\modulem[\sideX]{Y}{\appm{Y'}{X}}\\
      &\mathtt{end}
    \end{align*}\vspace{-4mm}%
    \caption{A sliceable functor}
  \end{subfigure}
  \caption{The sliceability constraint}
  \label{ex:compilation:sliceable}
  \vspace{2mm}
\end{ex}

We now describe how to slice the various constructions of our language.
The slicing rules for modules and expressions are defined in
\cref{module:compilation}. The slicing rules for structures and declarations are presented in
\cref{module:compdecl}.

Base structure and signature components are left untouched. Indeed, according
to \cref{prop:basetoml}, base elements are valid \ocsi elements. We do not
need to modify them in any way.
Signature components that are not valid on the target location are simply
omitted. Signature components that are valid on the target have their
type expressions translated.
The translation of a type expression to the client is the identity: indeed, there are
no new \etiny type constructs that are valid on the client. Server types, on the
other hand, can contains pieces of client types inside fragments $\cvtype{\tau_c}$
and inside converters $\conv{\tau_s}{\tau_c}$. Fragments in \ocsis are represented
by a primitive type, $\mathtt{fragment}$, without parameters.
The type of converters is represented by the type of their server part, which is
$\lamtype{\tau_s}{\serial}$.
Module and module type expressions are traversed recursively. Functors and functor
applications have each part sliced. Mixed functors are turned into normal
functors.

\begin{figure}[bt]
  \setlength{\jot}{-1pt}% Remove interline space in align*
  \input{theory/module/compile}
  \caption{Slicing -- $\compile{\csloc}{\cdot}$}
  \label{module:compilation}
\end{figure}

\begin{figure}[!btp]
  \setlength{\jot}{-1pt}% Remove interline space in align*
  \input{theory/module/compdecl}
  \caption{Slicing of declarations -- $\compile{\csloc}{\dm}$}
  \label{module:compdecl}
\end{figure}

Slicing of structure components inserts additional primitives that were
described in \cref{sec:target}. In client structure components, we need
to handle injections.
We associate each injection to a new fresh reference noted $\rf{x}$.
In \ocsis, we use the $\mathtt{injection}$ primitive to register the fact that the
given server value should be associated to a given reference. In \ocsic,
we replace each injection by its associated reference.
This substitution is applied both inside expressions and structures.
Note that for each injection
$\injf{x}{f}$, we use the encoding part $f^s$ and decoding part $f^c$ for
the server and client code, respectively.
For server structure components, we apply a similar process to handle fragments.
For each fragment, we introduce a reference noted $\rf{f}$.
In \ocsis, we replace each fragment by a call to $\mathtt{fragment}$ with
argument the associated reference and each escaped value inside the fragment
(with the encoding part of the converters).
We also add, after the translated component, a call to $\dend$ which indicates that the
execution of the component is finished.
In \ocsic, we use the $\pbind$ primitives to associate to each reference a closure
where all the escaped values are abstracted. We also introduce the decoding part of
each converter for escaped values.
We then call $\dexec$, which executes all the pending fragments
until the next $\pend$. This allows to synchronize interleaved side effects
between fragments and client components.

Given the constraint of sliceability, a mixed module is either a multi-argument
functor returning a structure, or it does not contain any structure at all.
For each structure, we use the reference annotated on structures,
as described in \cref{sec:eliom:execution:annotate}.
Mixed modules without structures can simply be sliced by leaving the module
expression unchanged. Mixed module types are also straightforward to slice.
Mixed structures (with an arbitrary number of arguments) need special care.
In \ocsis, we add a $\dyn$ field to the structure.
The value of this field is the result of a call to the primitive
$\mathtt{fragment}_m$ with arguments $\rf{F}$ and all the $\dyn$ fields
of the arguments of the functor.
In \ocsic, we create two structures for each mixed structure. One is simply
a client functor where all the server parts have been removed. Note here
that we don't use the slicing operation. The resulting structure does not
contain any call to $\pbind$ and $\dexec$.
We also create another structure that uses the regular slicing operation. This
structure is associated to $\rf{F}$ with the $\pbinddyn$ primitive

\subsection{Typing preservation}

One desirable property is that the introduction of new elements in the language
and the compilation operation does not compromise the guarantees provided by
the host language. To ensure this, we show that slicing a well typed
\etiny program provides two well typed \ocsic and \ocsis programs.

We only consider typing environments $\Env$ containing the primitive
types $\fragty$ and $\serial$ \ie
$\bindingabs[s]{\fragty}\in\Env$ and
$\bindingabs[\base]{\serial}\in\Env$.
We also extend the slicing operation to typing environments. Slicing a typing
environment is equivalent to slicing a signature with additional rules for
converters.
Converters, in \etiny, are not completely first class: they
are only usable in injections and not manipulable in the expression language.
As such, they must be directly provided by the environment. We add the two
following slicing rules that ensures that converters are properly present
in the sliced environment:
\begin{align*}
  \compile{s}{\valm[]{f}{\conv{\tau_s}{\tau_c}}} &=
    \valm[]{f^s}{\conv{\compile{s}{\tau_s}}{\serial}}
  \\
    \compile{c}{\valm[]{f}{\conv{\tau_s}{\tau_c}}} &=
    \valm[]{f^c}{\conv{\serial}{\tau_c}}
\end{align*}

\begin{theorem}[Compilation preserves typing]
  \label{thm:simulation:type}
  Let us consider $\mm$ and $\Mm$ such that $\wtm[\sideX]{\Env}{\mm}{\Mm}$.
  Then
  $\wtm[]{\compile{s}{\Env}}{\compile{s}{\mm}}{\compile{s}{\Mm}}$ and
  $\wtm[]{\compile{c}{\Env}}{\compile{c}{\mm}}{\compile{c}{\Mm}}$
\end{theorem}
\begin{proof}
  We proceed by induction over the proof tree of $\wtm[\sideX]{\Env}{\mm}{\Mm}$.
  The only difficult cases are client and server structure components and
  mixed structures. For brevity, we only detail the case of client
  structure components with one injection.

  Let us consider $\dm_c$ such that $\wtm[\sideX]{\Env}{(\dm_c;\sm)}{\Sm}$ and
  $\injs{\dm_c} = \injf{x}{f}$. We note $\rf{x}$ the fresh reference.
  By definition of the typing relation on \etiny,
  there exists $\Env'$ and $\tau_c$, $\tau_s$ such that
  $\Env\subset\Env'$,
  $\wt[s]{\Env'}{f}{\conv{\tau_s}{\tau_c}}$ and $\wt[s]{\Env'}{x}{\tau_s}$.
  We observe that there cannot be any server bindings in $\dm_c$, Hence
  we can assume $\Env' = \Env$. This is illustrated on the proof tree
  below.
  \begin{mathpar}
    \mprset{sep=1em}
    \inferrule
    { \inferrule*[vdots=1.2em]
      { \wt[s]{\Env}{f}{\conv{\tau_s}{\tau_c}} \\
        \inferrule*
        { }
        { \wt[s]{\Env}{x}{\tau_s} }
      }
      { \wt[c]{\Env}{\injf{x}{f}}{\tau_c} }
    }
    { \wtm[\sideX]{\Env}{(\dm_c;\sm)}{\Sm} }
  \end{mathpar}

  By definition of slicing on typing environments,
  $\binding[]{f^s}{\lamtype{\compile{s}{\tau_s}}{\serial}}\in\compile{s}{\Env}$
  and
  $\binding[]{f^c}{\lamtype{\serial}{\tau_c}}\in\compile{c}{\Env}$.
  By definition of $\ocsic$ and $\ocsis$
  typing rules, we have
  $\wt[\ocsis]{\compile{s}{\Env}}{\app{f^s}{x}}{\serial}$
  and
  $\wt[\ocsic]{\compile{c}{\Env}}{\app{f^c}{\rf{x}}}{\tau_c}$.

  We easily have that
  $\wtm[\ocsis]{\compile{s}{\Env}}{\pinj{\rf{x}}{\app{f^s}{x}}}{\emptym}$,
  as seen on the proof tree below.
  \begin{mathpar}
    \inferrule
    { \inferrule*
      { \binding[]{f^s}{\lamtype{\compile{s}{\tau_s}}{\serial}}
        \in\compile{s}{\Env} }
      { \wt[\ocsis]{\compile{s}{\Env}}{f^c}
        {\conv{\compile{s}{\tau_s}}{\serial}} } \\
      \inferrule*
      { }
      { \wt[\ocsis]{\compile{s}{\Env}}{x}{\compile{s}{\tau_s}} }
    }
    { \wtm[\ocsis]{\compile{s}{\Env}}
      {\pinj{\rf{x}}{\app{f^s}{x}}}{\emptym} }
  \end{mathpar}

  By induction hypothesis on
  $\wt[\sideX]{\Env,\binding[c]{x_j}{\tau_c}}
  {\subst[]{\injf{x}{f}}{x_j}{d_c}}{\emptym}$
  where $v_j$ is fresh, we have\\
  $\wt[\ocsic]{\compile{c}{\Env},\binding[]{x_j}{\tau_c}}
  {\subst[]{\injf{x}{f}}{x_j}{d_c}}{\emptym}$.
  We can then replace the proof tree of $v_j$ by the one of
  $\app{f^c}{\rf{x}}$. We simply need to ensure that the environments
  coincide. This is the case since $f^c$ cannot be introduced by new bindings.
  We can then remove the binding of $v_j$ from the environment, since it is
  unused. We obtain that
  $\wtm[\ocsic]{\compile{c}{\Env}}
  {\subst[]{\injf{x}{f}}{\app{f^c}{\rf{x}}}{d_c}}{\emptym}$
  which allows us to conclude.
  \begin{mathpar}
    \inferrule
    { \inferrule*[vdots=1.2em]
      { \inferrule*
        { \binding[]{f^c}{\lamtype{\serial}{\tau_c}}\in\compile{c}{\Env} }
        { \wt[\ocsic]{\compile{c}{\Env}}{f^c}{\conv{\serial}{\tau_c}} } \\
        \inferrule*
        { }
        { \wt[\ocsic]{\compile{c}{\Env}}{\rf{x}}{\serial} }
      }
      { \wt[\ocsic]{\compile{c}{\Env}}{\app{f^c}{\rf{x}}}{\tau_c} }
    }
    { \wtm[\ocsic]{\compile{c}{\Env}}
      {\subst[]{\injf{x}{f}}{\app{f^c}{\rf{x}}}{\dm_c};\sm}{\Sm} }
  \end{mathpar}

  % \TODO{Write the other cases}
\end{proof}

% \begin{corollary}[Compilation of programs preserves typing]
%   Let us consider $p$ and $\tau$ such that $\wtp{}{p}{\tau}$.\\
%   Then
%   $\wtm[]{}{\compile{s}{p}}{()}$ and
%   $\wtm[]{}{\compile{c}{p}}{\tau}$
% \end{corollary}

% \begin{figure}[ht]
%   \input{module/rcontexts}\vspace{-5mm}
%   \caption{Rewriting Contexts}
%   \label{rewrite:context:modules}
% % \end{figure}
% % \begin{figure}[ht]
%   \input{module/rewriting}
%   \caption{Rewrite rules -- $\rw{\Env}{m}{m'}$}
%   \label{rewrite:modules}
% \end{figure}

% \begin{lemma}[Functor lifting preserve typing]
%   Let us consider $\Env$, $m$, $m'$ and $M$ such that $\wtm{\Env}{m}{M}$ and
%   $\rw{\Env}{m}{m'}$. Then $\wtm{\Env}{m'}{M}$
% \end{lemma}

%%% Local Variables:
%%% mode: latex
%%% TeX-master: "../main"
%%% End:

%  LocalWords:  sliceability

% \input{theory/simulation}\clearpage
\subsection{Semantics preservation}\label{sec:simulation}

We now state that the compilation process preserves
the semantics of \etiny programs. In order to do that, we show
that, given an \etiny program $\Pm$, the trace of its execution is the same
as the concatenation of the traces of $\compile{s}{\Pm}$ and
$\compile{c}{\Pm}$.

First, let us put some constraints on the constants of
the \etiny, \ocsis and \ocsic language:

\begin{hypothesis}[Well-behaved converters]
  Converters are said to be well-behaved if for each
  constant $c$ in $\Const$ such that
  $\Typeof(c) = \conv{\tau_s}{\tau_c}$,
  then $c^s \in \Const_s$ and $c^c\in\Const_c$.
\end{hypothesis}

We now assume that converters in \etiny, \ocsis and
\ocsic are \emph{well-behaved}. We can then state the following
theorem.

\begin{theorem}[Compilation preserves semantics]
  \label{thm:simulation}
  Given sets of constants where converters are well-behaved,
  given an \etiny program $\Pm$ respecting the slicability hypothesis and
  such that
  $ \redp{\emptyr}{\Pm}{v}{\trace} $
  then
  \begin{align*}
    \redmls{\emptyr}{\compile{s}{\Pm}}{()}{\envfrag}{\envinj}{\trace_s}&&
    \redmlc{\emptyr}{\compile{c}{\Pm}}{v}{\envfrag}{\envfrag'}{\envinj}{\envg}{\trace_c}&&
    \trace = \trace_s\ctr\trace_c
  \end{align*}
\end{theorem}
\begin{proof}
  The complete proof is given in \cref{sec:simulation:proof}.
  The proof proceed in the following way: First, we simplify the problem
  by applying a code transformation that hoist injections to the top level.
  We then proceed with a proof by simulation for client
  code (\cref{redequivc}),
  server code (\cref{simulation:serverexpr})
  and finally mixed code (\cref{lemma:simulation:struct}).
\end{proof}

This theorem not only show that the return value is the same, but also that the trace is identical. This means that side effects happen in the same order and in the same location as specified by the interpreted semantics.
While our simplified calculus does not have side effects, \ocaml (and thus \eliom)
do, making such guarantee essential.

%%% Local Variables:
%%% mode: latex
%%% TeX-master: "../main"
%%% End:

% \input{theory/discussion}

%%% Local Variables:
%%% mode: latex
%%% TeX-master: "main"
%%% End:

\section{State of the art and comparison}
\label{sec:stateoftheart}

%% Custom paragraph level that doesn't change font/add dot.
\makeatletter
\newcommand\paralang{\@startsection{paragraph}{4}{\parindent}%
{-.5\baselineskip \@plus -2\p@ \@minus -.2\p@}%
{-3.5\p@}%
{}}
\makeatother

\eliom takes inspiration from many sources.
The two main influences are, naturally,
the extremely diverse ecosystem of web programming
languages and frameworks, which we explore in \cref{sec:soa:web},
and the long lineage of \ml programming languages
% , which we described in \cref{sec:ml:soa}
.
One of the important contributions of \eliom is the use of a programming
model similar to languages for distributed systems (\cref{sec:soa:distributed})
while using an execution model inspired by staged
meta-programming (\cref{sec:soa:staged}).

\subsection{Web programming}
\label{sec:soa:web}

Various directions have been explored to simplify Web
development and to adapt it to current needs.
\eliom places itself in one of these directions,
which is to use the same language on the server and the client.
Several unified client-server languages have been proposed. They
can be split in two categories depending on their usage of \js.
\js can either be used on the server, with \mysc{Node.js},
or as a compilation target, for example with \mysc{Google Web Toolkit} for Java
or \mysc{Emscripten} for C.
The approach of compiling to \js was also used to develop new client
languages aiming to address the shortcomings of \js. Some of them are new languages, such as
\mysc{Haxe}, \mysc{Elm} or \mysc{Dart}. Others are only \js extensions, such as
\mysc{TypeScript} or \mysc{CoffeeScript}.%
\footnote{A fairly exhaustive list of languages compiling to \js can be found
in \url{https://github.com/jashkenas/coffeescript/wiki/List-of-languages-that-compile-to-JS}}

However, these proposals only address the fact that \js is an inadequate language for
Web programming. They do not address the fact that the model of Web programming itself --
server and client aspects of web applications are split in two
distinct programs with untyped communication -- raises usability, correctness
and efficiency issues.
A first attempt at tackling these concerns is to specify the communication between
client and server. Such examples includes SOAP\footnote{\url{https://en.wikipedia.org/wiki/SOAP}}
(mostly used for RPCs) and REST\footnote{\url{https://en.wikipedia.org/wiki/Representational_state_transfer}}
(for Web APIs).
A more recent attempt is the GraphQL~\citep{graphql} query language
which attempts to describe, with a type system, the communications
between the client and server parts of the application.
These proposals are very powerful and convenient ways to check and
document Web-based APIs. However, while making the contract between
the client and the server more explicit, they further separate
web applications into distinct tiers.

Tierless languages attempt to go in the opposite direction: by removing
tiers and allowing web applications to be expressed in one single program,
they make the development process easier and restore the possibility
of encapsulation and abstraction without compromising correctness.
In the remainder of this section, we attempt to give a
fairly exhaustive taxonomy of tierless programming languages. We first
give a high-level overview of the various trade-offs involved, then
we give a detailed description of each language.

\subsubsection{Code and data location}

In \eliom, code and data locations are specified through syntactic annotations.
Other approaches for determining locations have been proposed.
The first approach
is to infer locations based on known elements through a
control flow analysis (Stip.js, \opa, \urweb):
database access is on the server, dynamic DOM interaction is done on the client, etc.
Another approach is to extend the type system with
locations information (\links, \ml{}5).
Locations can then be determined by relying on simple type inference and
checking.

These various approaches present a different set of compromises:
\begin{itemize}[leftmargin=*]
\item
  We believe that the semantics of a language should be easy to
  predict by looking at the code,
  which is why \eliom uses syntactic annotations to specify locations.
  This fits well within the \ocaml language,
  which is specifically designed to have a predictable behavior.
  On the other end of the spectrum, languages with inferred location
  sacrifice predictability for a very light-weight syntax which provides
  very little disruption over the rest of the program.
  Typed based approaches sit somewhere in the middle: locations
  are not visible in the code but are still accessible through types.
  Such approaches benefit greatly from IDEs allowing exploring
  inferred types interactively.
\item
  Naturally, explicit approaches are usually more expressive than implicit
  approaches. Specifying locations manually gives programmers greater
  control over the performance of their applications.
  Furthermore, it allows to express mixed \emph{data structures},
  \ie data structures that contain both server and client parts as
  presented in \cref{sec:eliom:mixeddata}.
  Such idioms are difficult to express when code locations are inferred.
  We demonstrate this with an example in the \urweb description.
\item
  Type-directed approaches to infer code location is an extremely
  elegant approach.
  It can be employed either in an algebraic effect
  setting (\links) or as modal logic annotations (\ml{}5).
  By its type-directed nature, error messages can be expressed
  in term of the source language and it should lend itself naturally
  to separate compilation (although this has not yet been achieved).
  However, such novel type systems significantly extend traditional
  general purpose type systems (the \ml one, in this case) to the point
  where it seems difficult to retrofit them on an existing languages.
  One lead
  would be to provide a tight integration using a form of Foreign
  Function Interface. Such integration has yet to be proposed.
\item
  \eliom, as a language based on \ocaml, is an effectful language.
  Marrying inference of locations and a side-effecting semantics is
  delicate. The Stip.js library~\citep{Philips2014} attempts to solve this by
  automatically providing replication and eventual consistency on shared
  references. This might cause many more communications that necessary
  if not done carefully.
  We believe that such decision is better left in the hands of the programmer.
\end{itemize}

\subsubsection{Slicing}

Once code location has been determined, the tierless program must be sliced
in (at least) two components. In \eliom, slicing is done statically
at compile time in a modular manner: each module is sliced independently.
Another common approach is to use a static whole-program slicing transformation (\urweb, Stip.js). This is most common for languages where code
location is inferred, simply due to the fact that such inference is often
non-modular.
This allows precise analysis of location that can benefit from useful
code transformations such as CPS transformation~\citep{Philips2016},
inlining and defunctionalization.
However, this can make it difficult for the users to know where each piece of code is executed and hinder error messages,.
It also prevents any form of separate compilation.

Finally, slicing can be done at runtime simply by generating client
\js code ``on the fly'' during server execution (\links, \hop, \php).
Such solution has several advantages: it is easier to implement and provides a
very flexible programming
style by allowing programmers to compose the client program in arbitrary ways.
The downside is that it provides less guarantees to the users.
Furthermore, it prevents generating and optimizing a single \js file
in advance, which is beneficial for caching and execution purposes.

\paragraph{Separate and incremental compilation}

Most current mainstream compiled language support some form of incremental compilation. Indeed, incremental compilation avoids recompiling
files of which no dependency has changed. This accelerates the feedback
loop between development and testing greatly and allow very fast
recompilation times. In the case of statically typed languages, it also
allows immediate checking of the modified file thus providing developers
very fast iteration cycles.
The easiest way to implement incremental compilation is through
separate compilation, where each file can be compiled completely independently.
Furthermore, separate compilation is compatible with link-time
optimization and thus does not prevent generation of
heavily optimized code, as demonstrated by nearly every C compiler.
As a consequence, we consider languages that do not support incremental
compilation completely unusable for practical usages.

\subsubsection{Communications}

\eliom uses asymmetric communication between client and server
(see \cref{tuto:sem}):
everything needed to execute the client code is sent during the initial communication
that also sends the Web page.
It also exposes a convenient API for
symmetric communications using RPC (\cref{sec:rpc}) and broadcasts, which must be used manually.

We thus distinguish several kind of communications.
First, manual communications are exposed through normal APIs and are performed explicitly by programmers. Of course, the convenience and safety of such
functions vary a lot depending on the framework.
Then, we consider automatic communications, that are inserted automatically
by the language at appropriate points, as determined by code locations
and slicing.
We can further decompose automatic communications further
in two categories. In static asymmetric communications,
information is sent from the server to the client automatically,
when sending the page.
In dynamic symmetric communications, information is sent back and forth
between the client and the server dynamically through some
form of channel (AJAX, websockets, Comet, \dots).

While symmetric communications are very expressive, they impose
a significant efficiency overhead: a permanent connection must be established
and
each communication imposes a round trip between client and server.
Furthermore, such communication channel must be very reliable.
On the other hand, asymmetric communications are virtually free: data
is sent with the web page (and is usually much smaller). Only a thin
instrumentation is needed.
Of course, the various communication methods can be mixed in arbitrary manner.
\eliom, for example, uses both automatic asymmetric and manual communications.

\paragraph{Offline usage}

Many web applications are also used on Mobile phones, where connection
is intermittent at best. As such, we must consider the case where
the web application produced by a tierless language is used offline.
In this context, asymmetric communication offer a significant advantage:
given the initially transmitted information by the server, the client
program can run perfectly fine without connection.
This guarantee, however,
does not extend to dynamic manual communications done by the use of RPCs
and channels.
\citet{Philips2014} explore this question for symmetric communications
through their R5 requirement.

\subsubsection{Type systems}

Type safety in the context of tierless languages can encompass several notions.
The first notion is the traditional distinction between weakly and strongly
typed languages. In the interest of avoiding a troll war among the jury, we
will not comment further.
A more interesting question is whether communication errors between
client and server are caught by the typechecker. This is, surprisingly,
not the case of \urweb since location inference and slicing is done very late
in the compilation process, far after type checking.
One consequence of this is that slicing errors are
fairly difficult to understand
\citep[page 10]{ur/web}.\footnote{
  ``However, the approach we adopted instead, with ad-hoc static analysis on
  whole programs at compile time, leads to error messages that confuse even
  experienced Ur/Web programmers.''}
While the \eliom formalization is type safe, the \eliom implementation
is not, due to the use of wrapping and Marshall,
which will fail at runtime on functional values.

Another remark is the distinction between client and server universes.%
\footnote{Or, more philosophically:
  Is your favorite language platonist or nominalist ?
}
\eliom has separate type universes for client and server types (see \cref{eliom:typeuniv}). Most tierless languages do not provide such distinction,
notably for the purpose of convenience. Distributed systems such as
\textsc{Acute}, however, do make such distinction to provide a solution
for API versionning and dynamic reloading of code. In this case, there
are numerous distinct type universes.

\paragraph{Module systems}

The notion of module system varies significantly depending
on the language. In \eliom we consider an \ml-style module system
composed of a small typed language with structures and functors.
We believe modules are essential for building medium to large sized
programs: this has been demonstrated for general purpose languages but
also holds for web programming languages, as demonstrated by
the size of large modern websites (the web frontend of facebook alone is
over 9 \emph{millions} lines of code). Even \js recently obtained
a module system in ES6.
In the context of tierless languages, an interesting question is the
interaction between locations and modules. In particular, can modules contain
elements of different locations and, for statically typed languages,
are locations reflected in signatures?

\paragraph{Types and documentation}

Type systems are indisputably very useful for correctness purposes, but they
also serve significant documentation purposes. Indeed, given a function,
its type signature provides many properties. In traditional languages,
this can range from very loose (arguments and return types) to very precise
(with dependent types and parametricity~\citep{Theoremforfree}).
In the context of tierless languages, important questions we might want
to consider are ``Where can I call this function?'' and
``Where should this argument come from?''.
The various languages exposes this information in different ways:
\eliom does not expose
location in the types, but it is present in the signature. \ml{}5 exposes
this information directly in the types. \urweb and \links do not expose that
information at all.

\subsubsection{Details on some specific approaches}
We now provide an in-depth comparison with the most relevant approaches.
A summary in \cref{fig:soa} classifies each approach according
to the main distinctive features described
in the previous paragraphs. Each language or
framework is also described below.

\begin{figure}[ht]
  \centering
  \begin{tabular}{r|c|c|c|c|c}
          &Locations&Slicing&Communications&Type safe&Host language\\\hline
    \eliom&Syntactic&Modular&Asymmetric&\checkmark&\ocaml\\
    \links&Type-based*&Dynamic*&Symmetric&\checkmark&-\\
    \urweb&Inferred&Global&(A)symmetric$\sim$&\checkmark*&-\\
    Haste&Type-based&Modular&Symmetric&\checkmark&\mysc{Haskell}\\
    \hop&Syntactic&Dynamic*&(A)symmetric$\sim$&\texttimes&\js{}*\\
    Meteor.js&Syntactic&Dynamic&Manual&\texttimes&\js\\
    Stip.js&Inferred&Global&Symmetric*&\texttimes&\js\\
    \ml{}5&Type-based&Global*&Symmetric&\checkmark&-\\
    Acute&Syntactic&Modular&Distributed&\checkmark&\ocaml\\
  \end{tabular}
  \caption{Summary of the various tierless languages}
  \smaller
  See previous sections for a description of each headline.
  A star * indicates that details are available
  in the description of the associated language.
  A tilde $\sim$ indicates that we are unsure, either because the information
  was not specified, or because we simply missed it.
  \label{fig:soa}
\end{figure}

\paralang{\urweb}\citep{ur/web,ur/web-icfp} is a new statically typed language especially designed for Web programming.
It features a rich \ml-like type and module system and a fairly original
execution model where programs only execute as part of a web-server
request and do not have any state (the language is completely pure).
While similar in scope to \eliom, it follows a very different approach:
Location inference and slicing are done through a whole-program transformation
operated on a fairly low level representation. Notably, this transformation
relies on inlining and removal of high-order functions (which are not supported
by the runtime).
The advantages of this approach are twofold: It makes \urweb applications
extremely fast (in particular because it doesn't use a GC: memory
is trashed after each request) and it requires very little syntactic
overheads, allowing programs to be written in a very elegant manner.

The downsides, however, are fairly significant.
\urweb's approach is incompatible with any form of separate compilation.
Many constructs are hard-coded into the language, such as RPCs
and reactive signals and it does not seem possible to implement them
as libraries.
The language is clearly
not general and has a limited expressivity, in particular
when trying to use mixed data-structures (see \cref{sec:eliom:mixeddata}).
For example, \cref{fig:soa:urweb}. presents the server function
\ocamlc{button_list}
which takes a list of labels and client functions and generates a list of
buttons. We show the \eliom implementation and a tentative \urweb
implementation. The \urweb version typechecks but slicing fails. We
are unable to write a working version and do not believe it to be possible:
indeed, in the \eliom version we use a client fragment to build the
list \ocamlc{l} as a mixed data-structure. This annotation is essential
to make the desired semantics explicit.
Some examples are expressible only using reactive signals, which present a very
different semantics.

\begin{ex}[ht]
  \begin{subfigure}[t]{0.48\linewidth}
\begin{lstlisting}[language=eliom]
let%client handler _ = alert "clicked!"
let%server l =
  [ ("Click!", [%client handler]) ]

let%server button_list lst =
  ul (List.map (fun (name, action) ->
    li [button
      ~button_type:`Button
      ~a:[a_onclick action]
      [pcdata name]])
    lst)

let main () =
  body (button_list l)
\end{lstlisting}
    \caption{\eliom version}
  \end{subfigure}\hfill
  \begin{subfigure}[t]{0.48\linewidth}
\begin{lstlisting}[language=ml]
fun main () : transaction page =
 let
  fun handler _ = alert "clicked!"
  val l = Cons (("Click!", handler), Nil)

  fun button_list lst =
   case lst of
     Nil => <xml/>
   | Cons ((name, action), r) =>
     <xml>
      <button value={name}
              onclick={action}/>
      {button_list r}
     </xml>
 in
 return <xml>
   <body>{button_list l}</body>
 </xml>
end
\end{lstlisting}
    \caption{Tentative \urweb version. Typechecks but does not compile.}
  \end{subfigure}
  \caption{Programs building a list of buttons from a list of client side actions}
  \label{fig:soa:urweb}
\end{ex}

\paralang\hop\citep{HopProposal} is a dialect of Scheme for programming Web
applications. Its successor, \hop.js~\citep{SerranoP16}, takes the same concepts and brings them to \js.
The implementation of \hop.js is very complete and allow them to run
both the \js and the scheme dialect while leveraging the complete
node.js ecosystem.
\hop uses very similar language constructions to the one provided
by \eliom: $\sim$-expressions are fragments and \textdollar-expressions are
injections. All functions seem to be \eliomc{shared} by default.
Communications are asymmetric when possible and use channels otherwise.
However, contrary to \eliom, slicing is
done dynamically during server execution~\citep{HopClientCompilation}.
In the tradition of Scheme, \hop only uses a minimal type system for
optimizations and does not have a notion of location. In particular
\hop does not provide static type checking and does not statically enforce the
separation of client and server universes (such as preventing the use of database
code inside the client).
The semantics of \hop has been formalized~\citep{HopSemantic,HopSemantic2}
and does present similarities to the interpreted \eliom
semantics (\cref{sec:eliom:execution}). \hop is however significantly
more dynamic than \eliom: it allows dynamic communication patterns
through the use of channels and allows nested fragments in the
style of Lisp quotations which allows to generate client code
inside client code.

For dynamically-typed inclined programmers, \hop currently presents the most
convincing approach to tierless Web programming. In particular given
its solid implementation, great flexibility and support for the \js ecosystem.

\paralang\links\citep{LinksProposal}
is an experimental functional language for client-server Web programming
with a syntax close to \js and an \ml-like type system. Its
type system is extended with a notion of {\em effects}, allowing
a clean integration of database queries in the language~\citep{LinksDatabase}.
In \cref{ex:links},
we highlight two notable points of \links: the function \ocamlc{adults}
takes as argument a list \ocamlc{l} and returns the name of all the
person over 18. This function has no effect and can thus run on the client,
the server, but can also be transformed into SQL to run in a database query.
On the other hand, the \ocamlc{print} function has an effect called ``wild''
which indicates it can't be run inside a query. Effects are also used
to provide type-safe channel-based concurrency.

\links also allows
to annotate functions by indicating on which location they should run.
Those annotations, however, are not reflected in the type
system. Communications are symmetric and completely dynamic through
the use of AJAX. Client-server slicing
is dynamic (although some progress has been made towards static
\emph{query} slicing~\citep{LinksQuotations}) and can introduce
``code motion'', which can moves closures from the server
to the client. This can be extremely problematic in practice, both from
an efficiency and a security point of view.
The current implementation of \links is interpreted but a compilation
scheme leveraging the Multicore-\ocaml efforts has been recently added.

Although \links is very seducing, the current
implementation presents many shortcomings given its statically typed
nature: slicing is dynamic and produces fairly large
\js code and the type system does not really track client-server locations.

\begin{ex}[hb]
  \centering
\begin{lstlisting}[language=ml]
links> fun adults(l) { for (x <- l) where (x.age >= 18) [(name = x.name)] } ;;
adults = fun : ([(age:Int,name:a|_)]) -> [(name:a)]

links> print ;;
print : (String) {wild}-> ()
\end{lstlisting}
  \caption{Small pieces of \links code}
  \label{ex:links}
\end{ex}

\paralang{\textsc{Meteor.js}}\citep{meteorjs} is a framework where both the
client and the server sides of an application are written in \js.
It has no built-in mechanism for sections and fragments but relies on conditional $\mathtt{if}$ statements on the \verb+Meteor.isClient+ and
\verb+Meteor.isServer+ constants. It does not perform any slicing. This means that there are no static guarantees over the respective execution of server and client code. Besides, it provides no facilities for client-server communication such as fragments and injections.
Compared to \eliom, this solution only provides coarse-grained composition.

\paralang{\textsc{Stip.js}}\citep{Philips2014} allows to slice tierless \js
programs with a minimal amount of annotations. It emits \textsc{Meteor.js}
programs with explicit communications. Annotations are optionally provided
through the use of comments, which means that \textsc{Stip.js} are actually
perfectly valid \js programs. Location inference and slicing are
whole-program static transformations. Communications are symmetric, through
the use of fairly elaborate consistency and replication mechanisms for
shared references.
This approach allows the programmer to write code with very little annotations.
As opposed to \urweb, manual annotations are possible, which might allow
to express delicate patterns such as mixed data-structures and prevents
security issues.

% \TODO{WebSharper,\dots}

\subsection{Distributed programming}
\label{sec:soa:distributed}

Tierless languages in general are very inspired by distributed programming languages. The
main difference being that distributed programs contain an arbitrary number of locations
while tierless web programs only have two: client and server. Communications
are generally symmetric and dynamic, due to the multi-headed aspect of distributed
systems. There are of course numerous programming languages dedicated to distributed
programming. We present here two relevant approaches that put
greater emphasis on the typing and tierless aspects.

\paralang{Haste}\citep{Ekblad17} is an \textsc{Haskell} EDSL 
for distributed programming.
This DSL allows to express complex orchestrations of multiple nodes and external components
(for example databases and IoT components), with handling of distinct type universes when necessary.
Instead of using syntactic annotations, locations are determined through typing.
This approach works particularly well in the context of \textsc{Haskell},
thanks to the advanced type system and the syntactic support for monads and functors.
Multiple binaries are produced from one program. Slicing relies on type information
and dead code elimination, as provided by the GHC compiler.
Explicit slicing markers similar to \eliom's section annotations are the subject
of future work.
Communications are dynamic and symmetric through the use of websockets.
One notable feature of this DSL is that it offers a client-centric view:
The control flow is decided by the client which pilots the other nodes. This
is the opposite of \eliom where the server can assemble pieces of
client code through fragments.
This work also inherits
the \textsc{Haskell} and GHC features in term of modules, data abstraction and
separate-compilation.
A module language has been developed for \textsc{Haskell} by \citet{backpack}.

An earlier version, \textsc{Haste.App}~\citep{haste}, was limited to only one
client and one server and used a monadic approach
to structure tierless programs.

\paralang{\ml{}5}\citep{ML5} is an \ml language that introduces new constructs for type-safe communication between distributed actors through the use of location annotations
inside the types called ``modal types''.
It is geared towards a situation where all actors have similar capabilities.
It uses dynamic communication, which makes the execution model very different from \eliom.
\ml{}5 provides a very rich type system that allows to precisely export
the capabilities of the various locations. For example, it is possible
to talk about addresses on distant locations and pass them around arbitrary.
\eliom only supports such feature through the use of fragments, for client
code.

Unfortunately, \ml{}5 does not have a module system.
However, we believe that
\ml{}5's modal types can play a role similar to \eliom's location
annotations on declarations, including location polymorphism.
\ml{}5 uses a global transformation for slicing.
Given the rich typing information present
in \ml{}5's types, it should lend itself fairly well to a modular
slicing approach, but this has not been done.

\paralang{\textsc{Acute}}\citep{acute} is an extension of \ocaml for distributed programming.
It provides typesafe serialization and deserialization and also allows arbitrary loading
of modules at runtime.
Like \eliom, it provides a full-blown module system.
However, it takes an opposite stance on the execution model:
each actor runs independent programs and communications
are completely dynamic.

Handling of multiple type universes is done by providing
a description of the type with each message and by versioning APIs.
In particular, great care is taken to provide type safe serialization by
also transmitting the type of messages alongside each message.
This gives \textsc{Acute} very interesting capabilities, such as reloading
only part of the distributed system in a type-safe way.

\subsection{Staged meta-programming}
\label{sec:soa:staged}

An important insight regarding \eliom is that, while it is a tierless programming language
and tries to disguise itself as a distributed programming language, \eliom
corresponds exactly to a staged meta-programming language.
\eliom simply provides only two stages: stage 0 is the server, stage 1 is the
client.
\eliom's client fragments are the equivalent of stage quotations.

Most approaches to partial evaluation are done implicitly (not unlike tierless
languages with implicit locations).
We take inspiration from several approaches that combine staged meta-programming
with explicit stages annotations that are reflected in the type system.
We only look at the most relevant approaches but a longer description of
the history of staged meta-programming approaches
can be found in \citet[Chapter 7]{TahaThesis}.

\paralang{\textsc{MetaOCaml}}\citep{DBLP:conf/flops/Kiselyov14} is an extension of \ocaml for meta programming. It introduces a quotation annotation for staged expressions, whose execution is delayed.
Quotations and antiquotations corresponds exactly to fragments and injections. The
main difference is that \textsc{MetaOCaml} is much more dynamic: quoted code
does not have to be completely closed when produced and well-scopedeness
is checked dynamically, just before running the quoted code. This allows
very dynamic behaviors such as automatic insertion of let-bindings \citep{KiselyovGenLet}
and dynamically determining staged stream pipelines \citep{KiselyovStream}.
One difference is the choice of universes: \eliom has two universes, client and server, which are distinct. \textsc{MetaOCaml} has a single type universe but a series
of scopes, for each stage, included in one another.

\textsc{MetaOCaml} itself provides no support for modules and
only leverages the \ocaml module system. Staging annotations
are only on expressions, not on declarations.

\paralang{Modular macros}\citep{modularmacros,modularmacrosth} are another extension of \ocaml.
It uses staging to implement macros. It provides both a quotation-based expression
language along with staging annotations on declarations. It also aims to support modules
and functors.
The
slicing can be seen as dynamic (since code is executed at compile time to produce
pieces of programs). In particular, this allows to lift most of the restriction imposed on multi-stage functors.
They also use a notion similar to converters, except that the serialization format
here is simply the \ocaml AST.

The main difference compared to \eliom is how the asymmetry between stage 0 and stage 1 is
treated. Only one type universe is used and there is no notion of slicing that would
allow a distant execution.

\paralang{\citet{FeltmanAAF16}} presents a slicing technique for a two-staged simply typed lambda calculus. Their technique is similar to the one used in \eliom.
They distinguish their language it three parts: $1\mathcal{G}$, which corresponds
to base code; $1\mathcal{M}$, which corresponds to server code; and $2\mathcal{M}$,
which corresponds to client code.
They also provide a proof of equivalence between the dynamic semantics and the slicing
techniques. This proof has been mechanized in Twelf.
While their work is done in a more general settings, they do not specify how to
transfer rich data types across stages (which is solved in \eliom using converters).
They also do not propose a module system.

%%% Local Variables:
%%% mode: latex
%%% TeX-master: "main"
%%% End:

%  LocalWords:  js Emscripten Haxe TypeScript CoffeeScript Tierless
%  LocalWords:  Opa DOM RPC effectful buttonlist
%  LocalWords:  namespace Haskell monads monadic isClient isServer
%  LocalWords:  MetaOCaml

\section{Conclusion}

We presented \eliom, a Modular Tierless Web programming
language that supports static typing, an efficient execution model,
abstraction and modularity.
We showed its design through numerous examples and formalized it.
In our formalization, we first gave a simple interpreted
semantics, that is suitable to explain to programmers. We then gave
a more efficient compilation scheme and showed that it
preserves both typing and semantics. The compilation scheme follows precisely
the current implementation of \eliom.

Many design choices of \eliom were inspired by practical concerns.
Indeed, in order to be useful, a language must have an ecosystem. The simplest way to have an ecosystem
is to reuse the one of an existing language, in our case the \ocaml one.
Since separate compilation and modularity are indispensable for any non-trivial programming projects,
we needed to make modularity, abstraction and the tierless annotations interact.
\eliom is thus inspired by many different elements
such as \ml languages, tierless
web programming and staged meta-programming.
In particular, \eliom aims to combine a semantics inspired by staged meta-programming
and an \ml-style module system.

We believe that such combination can be used in a wider context
than simply Web programming. Indeed, \ocsigen has also been used to create
Mobile applications~\citep{besport}. We would like to extend our language
to be usable to any kind of client-server applications
that requires dynamicity, type safety and modularity.
We also consider extending our model to explicitely support
multiple distinct clients. 
Finally, we aim to further explore the programming patterns made
possible by \eliom, for instance multi-tiers functional reactive programming.

% While the notion of locations and mixed modules achieve the desired modularity,
% several questions remain open regarding mixed functors.
% Functors are very expressive, combining this expressivity with a static
% slicing operation and a one-way communication scheme is challenging.
% %
% We proposed to impose careful restrictions over the shape of mixed functors
% that allow to slice them statically, associated with some code transformations
% that help the slicing operation.
% We believe these restrictions can be further relaxed.
% User feedback should give us insight on how mixed functors are used and
% which kind of programming idioms and library organization they allow.

%%% Local Variables:
%%% mode: latex
%%% TeX-master: "main"
%%% End:

%% Acknowledgments
% \begin{acks}
% \end{acks}

\bibliography{biblio}

\clearpage
\appendix
\normalsize
\section{The \ml calculus}
\label{appendix:ml}

This section describes our version of the \ml calculus, which contains the
minimal amount of ingredients that allows us to describe the \eliom extensions:
a core calculus with polymorphism, let bindings and parametrized datatypes
in the style of \citet{wright1994syntactic},
accompanied by a fully featured module system with separate compilation
and applicative functors in the style of \citet{Leroy95}.
We first present the syntax of the language in \cref{sec:ml:syntax} and
its type system in \cref{sec:ml:typing}.
We then present the semantics of this language in
\cref{sec:ml:semantics}.
The description in this appendix is self-contained, and thus restate the
basic description found in \cref{sec:ml}.

\subsection{Syntax}
\label{sec:ml:syntax}

\begin{figure}[!hb]
  \vspace{-2mm}
  \renewcommand\added\addedhide
  \renewcommand\addedrule\addedhide
  \vspace{-3.5mm}
  \input{theory/ml/grammar}
  \caption{\ml grammar}
  \label{ml:grammar}
  \vspace{-3mm}
\end{figure}

Let us first define some notations and meta-syntactic variables. As a general rule,
the expression language is in lowercase ($e$) and the module language is in
uppercase ($M$). Module types are in calligraphic letters ($\Mm$).
More precisely: $x$ are variables, $p$ are module paths, $X$ are module variables, $\tau$ are type expressions and $\type{t}$ are type constructors.
$x_i$, $X_i$ and $\type{t_i}$ are identifiers (for values, modules and types).
Identifiers (such as $x_i$) have a name part ($x$) and a stamp part ($i$) that
distinguish identifiers with the same name. Only the name part of
identifiers is exposed in module signature.
$\alpha$-conversion should keep the name intact and change only the stamp,
which allow to preserve module signatures.
Sequences are noted with a star; for example $\tau^*$ is a sequence of type expressions.
Indexed sequences are noted $(\tau_i)$, with an implicit range.
Substitution of $a$ by $b$ in $e$ is noted $\subst[]{a}{b}{e}$.
Repeated substitution of each $a_i$ by the corresponding $b_i$ is noted
$\substi[]{a_i}{b_i}{e}$.
The syntax is presented in \cref{ml:grammar}.

\paragraph{Expressions}
The expression language is a fairly simple extension of
the lambda calculus with a fixpoint combinator $\Y$ and let bindings $\letin{x}{e_1}{e_2}$.
The language is parametrized by a set of constants $\Const$.
Variables can be qualified by a module path $p$.
Paths can be either module identifiers such as $X_i$,
a submodule access such as $X_i.Y$,
or a path application such as $X_i(Y_j.Z)$.
Note that, as said earlier, that fields of modules are only called by their
name, without stamp.

\paragraph{Types}
Types are composed of type variables $\alpha$, function types $\lamtype{\tau_1}{\tau_2}$ and
parametrized type constructors $\appty{\tau_1,\tau_2,\dots,\tau_k}{t_i}$. Type constructors
can have an arbitrary number of parameters, including zero.
Type constructors can be qualified by a module path $p$.
Type schemes, noted $\sigma$, are type expressions that are universally
quantified by a list of type variables. Type schemes can also have
free variables.
For example: $\forall\alpha. \appty{\alpha, \beta}{t_i}$.

\paragraph{Modules}
The module language is quite similar to a simple lambda calculus: Functors are functions
over module (except that arguments are annotated with their types). Module application
is noted $\appm{\mm_1}{\mm_2}$.
Modules can also be constrained by a module type: $\constraint{\mm}{\Mm}$.
Finally, a module can be a structure which contains a list of value, types
or module definitions: $\struct{\letm[]{x_i}{2}}$.
Programs are lists of definitions.

\paragraph{Module types}
Module types can be either the type of a functor or a signature, which contains a list
of value, types and module descriptions. Type descriptions can expose their definition
or can be left abstract.
Typing environments are simply module signatures. We note them $\Env$ for convenience.

\subsection{Type system}
\label{sec:ml:typing}

We now present the \ml type system. For ease of presentation, we proceed
in two steps: we will first forget that the module language exists, and
present a self-contained type system for the expression language.
We then extend the typing relation to handle modules.

\subsubsection{The expression language}
\label{sec:ml:expr}

We introduce the following judgments:

\begin{tabular}{ll}
  $\wt[]{\Env}{e}{\tau}$& The expression $e$ has type $\tau$ in the environment $\Env$. See \cref{ml:expr:typing}.\\
  $\equivty[]{\Env}{\tau_1}{\tau_2}$&Types $\tau_1$ and $\tau_2$ are equivalent in environment $\Env$. See \cref{ml:expr:equiv}.\\
  $\wf[]{\Env}{\tau}$& The type $\tau$ is well formed in the environment $\Env$. See \cref{ml:expr:validity}.
\end{tabular}

We note $\Typeof(c)$ the type scheme of a given constant $c$.
The instanciation relation is noted $\instanceof{\tau}{\sigma}$
for a type scheme $\sigma$ and a type $\tau$.
The converse operation which closes a type according to an environment is
noted $\Close{\tau}{\Env}$.
We use $(D)\in\Env$ to test if a given type or value is declared in the environment $\Env$. Note that for types,
$(\typeabsm[]{\appty{\alpha_i}{t}})\in\Env$ holds also if $t$ is not abstract
in $\Env$.

\paragraph{Polymorphism}

One of the main benefit of programming language of
the \ml family is the ability to easily define and use functions
that operate on values of various types. For example, the \ocamlc{map} function
can applies to all lists, regardless of the type of their content.
Indeed, the type of \ocamlc{map} is polymorphic:
\[\text{\ocamlc{map}}: \forall\alpha\beta.\ %
  \lamtype{(\lamtype{\alpha}{\beta})}{
    \lamtype{\appty{\alpha}{list}}{
      \appty{\beta}{list}}}
\]

From a type checking point of view, this is possible thanks to two operations:
instanciation and abstraction.
Instanciation takes a type scheme, which is a type where some variables
have been universally quantified, and replace all the quantified type variables
by some type. It is used when looking up a variable (rule \mysc{Var}) or
typechecking a constant (rule \mysc{Const}).
For example, the type of map can be instantiated to the following type.
\[\lamtype{(\lamtype{int}{bool})}{
    \lamtype{\appty{int}{list}}{
      \appty{bool}{list}}}
\]
Once instantiated, the \ocamlc{map} function can be applied on
a list with concretes types.
Naturally, we also need the converse operation: constructing a type scheme given a type
containing some type variables. Closing a type depends on the current
typing environments, we only abstract type variables that have not been
introduced by previous binders. $\Close{\tau}{\Env}$ returns the type scheme
$\forall\alpha_1\dots\alpha_n.\tau$ where the $\alpha_i$ are free variables of $\tau$
that are not present in $\Env$.
While it is possible to apply the closing operation at any step of
a typing derivation, it is only useful at the introduction point of type
variables, in let bindings (rule \mysc{LetIn}).
In the following example, we derive a polymorphic type for a function that
constructs a pair with an element from the environment.
We first use the close operation to obtain a type scheme for $f$. Note
that since $\alpha$ is present in the environment, it is not universally
quantified. We then use the instance operation to apply $f$ to an integer
constant.

\begin{mathpar}
  \renewcommand\added\addedhide
  \newcommand\Ea{\valm{a}{\alpha}}
  \newcommand\Eb{\valm{b}{\beta}}
  \mprset{sep=1em}
  \inferrule*
  { \inferrule*
    { \inferrule*{\vdots}{\wt{(\Ea;\Eb)}{(a,b)}{\alpha * \beta}} }
    { \wt{(\Ea)}{\lam{b}{(a,b)}}{\lamtype{\beta}{\alpha * \beta}} } \\
    \inferrule*
    { \inferrule*
      { \instanceof
        {\lamtype{int}{\alpha * int}}
        {\forall\beta.\ \lamtype{\beta}{\alpha * \beta}}
      }
      { \wt{\Env}{f}{\lamtype{int}{\alpha * int}} }
      \\
      \inferrule*{\instanceof{int}{\Const(3)}}{\wt{\Env}{3}{int}}
    }
    { \wt
      {(\Ea;\valm{f}{\forall\beta.\ \lamtype{\beta}{\alpha * \beta}})}
      {f\ 3}{(\alpha * int)}
    }
  }
  { \wt{(\Ea)}{\letin{f}{\lam{b}{(a,b)}}{f\ 3}}{(\alpha * int)} }
\end{mathpar}

\paragraph{Parametric datatypes}
\label{sec:ml:datatypes}

Parametric polymorphism introduces type variables in type expressions.
In the presence of type definitions, it is natural to expect
the ability to write type definitions which can contain type
variables. This leads us to parametric datatypes: datatypes
which are parametrized by a set of variables.
$\appty{\alpha}{list}$ is of course an example of such datatype.
Note that care must be taken when deciding the equivalence of types.
It the type is not abstract, \ie its definition is available, we can
always unfold the definition, as shown in rule \mysc{DefTypeEq}.
However, when considering an abstract type, we cannot unfold the type
definition. Instead, we check that head symbols are compatible and
that parameters are equivalent pairwise%
\footnote{This is similar to the handling of free symbols in the unification literature.}
.
This is done in rule \mysc{AbsTypeEq}.

\begin{figure}[htb]
  \renewcommand\added\addedhide
  \renewcommand\addedrule\addedhide
  \input{theory/core/validity}\vspace{-3mm}
  \caption{Type validity rules -- $\wf[]{\Env}{\tau}$}
  \label{ml:expr:validity}
\end{figure}
\begin{figure}[!hbt]
  \renewcommand\added\addedhide
  \renewcommand\addedrule\addedhide
  \input{theory/ml/typingexpr}%\vspace{-3mm}
  \caption{\ml expression typing rules -- $\wt[]{\Env}{e}{\tau}$}
  \label{ml:expr:typing}
% \end{figure}
% \begin{figure}[htb]
  \mprset{sep=1em}
  \renewcommand\added\addedhide
  \renewcommand\addedrule\addedhide
  \input{theory/core/equivty}\vspace{-3mm}
  \caption{Type equivalence rules -- $\equivty[]{\Env}{\tau}{\tau'}$}
  \label{ml:expr:equiv}
\end{figure}

\subsubsection{The module language}
\label{sec:ml:module}

We introduce the following judgments:

\begin{tabular}{ll}
  $\wtm[]{\Env}{\mm}{\Mm}$
  &The module $\mm$ is of type $\Mm$ in $\Env$.
  See \cref{ml:module:typing}.\\

  % $\wtp{\Env}{P}{\tau}$&
  % The program $P$ returns a value of type $\tau$.
  % Defined in \cref{module:typing}.\\

  $\submod[]{\Env}{\Mm}{\Mm'}$&
  The module type $\Mm$ is a subtype of $\Mm'$ in $\Env$.
  See \cref{ml:module:subtyping}.\\

  $\wfm[]{\Env}{\Mm}$&
  The module type $\Mm$ is well-formed in $\Env$.
  See \cref{ml:module:validity}.
\end{tabular}

The typing rules for \ocaml-style modules are quite complex. In particular, the
inner details of the rules are not well known,
even by \ocaml programmers. Before presenting the typing rules
in details, we will attempt to give
insight on why some features are present in the languages and what are their
advantages. For this purpose, we present two examples illustrating
the need for applicative functors and strengthening, respectively.
We assume that readers are familiar with simpler usages
of \ml modules.

\paragraph{Applicative Functors}
Let us consider the following scenario: we are given a module $G$
implementing a graph data-structure and would like to implement a simple
graph algorithm
which takes a vertex and returns all the accessible vertices.
We would like the returned module to contain
a function of type
$\lamtype{G.graph}{\lamtype{G.vertex}{set\_of\_vertices}}$.
How to implement $\type{set\_of\_vertices}$? An easy but inefficient way
would be to use lists.
A better way is to use proper sets (implemented with balanced binary tree,
for example).
In \ocaml, this is provided in the standard library
by the functor $Set.Make$, presented in \cref{tuto:modules},
which takes a module
implementing comparison functions for the given type.
We would obtain a signature similar to the one below.

\begin{align*}
  &\moduletym[]{Access(G:\mathcal{G}raph)}{\mathtt{sig}}\\
  &\quad\moduletym[]{VerticesSet}{\signature{\dots}}\\
  &\quad\valm[]{run}
    {\lamtype{G.graph}{\lamtype{G.vertex}{VerticesSet.set}}}\\
  &\dend
\end{align*}
However, this means we need to expose a complete module implementing
set of vertices that is independent from any other set module. This prevents
modularity, since any usage of our new function must use this specific
set implementation. Furthermore, this make the signature bigger than strictly
necessary. What we really want to expose is that the return type comes
from an application of $Set.Make$. Fortunately, we can do so by
using the following signature.
\begin{align*}
  &\moduletym[]{Access(G:\mathcal{G}raph)}{\mathtt{sig}}\\
  &\quad\valm[]{run}
    {\lamtype{G.graph}{\lamtype{G.vertex}{Set.Make(G.Vertex).set}}}\\
  &\dend
\end{align*}
Here, we export the fact that the set type must be the result of a functor
application on a module that is compatible with $\type{G.Vertex}$.
The type system guarantees that any such functor application
will produces types that are equivalent. In particular, if multiple libraries
uses the $Access$ functor, their sets will be of the same types, which
make composition of libraries easier. This behavior of functors is usually
called \emph{applicative}.

\paragraph{Strengthening}
Let us now consider the program presented in \cref{ex:funcmanifest}.
We assume the existence of two modules, presented in
\cref{ex:funcmanifest:env}.
The module $Showable$ exposes the abstract type $\type{t}$, along with
a $show$ function that turns it into a string.
The module $Elt$ exposes a type $\type{t}$ equal to
$\type{Showable.t}$ and a value that inhabits this type.
The program is presented in \cref{ex:funcmanifest:prog}.
We define a functor $F$ taking two arguments $E$ and $S$ whose signature
are similar to $Elt$ and $Showable$, respectively.
The main difference is that $E$ comes first and $\type{S.t}$ is defined as an
alias of $\type{E.t}$. The functor uses the $show$ function on the element
in $E$ to create a string.
It is natural to expect the functor application $F(Elt)(Showable)$ to
type check, since $\type{Elt.t} = \type{Showable.t}$. We must, however,
check for module inclusion. While $Elt$ is clearly included in the
signature of the argument $E$, the same is not clear for $Showable$.
We first need to enrich its type signature with additional type equalities.
We give $Showable$
the type $\signature{\typem[]{t}{Showable.t}\ \dots}$. It makes sense to enrich
the signature in such a manner since $Showable$ is already in the environment.
Given this enriched signature, we can now deduce that
$\submod[]{}{(\typem[]{t}{Showable.t})}{(\typem[]{t}{E.t})}$ since
$\type{E.t} = \type{Elt.t} = \type{Showable.t}$.

\begin{ex}[!htb]
  \vspace{-2mm}
  \renewcommand\added\addedhide
  \mprset{sep=1.7em}
  \setlength{\jot}{0pt}% Remove interline space in align*
  \begin{subfigure}{0.3\linewidth}
  \begin{align*}
    &\moduletym[]{Showable}{\mathtt{sig}}\\
    &\quad\typeabsm[]{t}\\
    &\quad\valm[]{show}{\lamtype{t}{string}}\\
    &\dend\\
    &\moduletym[]{Elt}{\mathtt{sig}}\\
    &\quad\typem[]{t}{Showable.t}\\
    &\quad\valm[]{v}{elt}\\
    &\dend
  \end{align*}
  \caption{Typing environment}
  \label{ex:funcmanifest:env}
  \end{subfigure}\hfill
  \begin{subfigure}{0.6\linewidth}
  \begin{align*}
    &\mathtt{module}\ F\\
    &\quad(E:\signature{\typeabsm[]{t}\ \valm[]{v}{t}})\\
    &\quad(S:\signature{\typem[]{t}{E.t}
      \ \valm[]{show}{\lamtype{t}{string}}})\\
    & = \mathtt{struct}\\
    &\quad\letm[]{s}{\app{S.show}{E.v}}\\
    &\dend\\
    &\modulem[]{X}{F(Elt)(Showable)}
  \end{align*}
  \caption{Application of multi-argument functor using manifests}
  \label{ex:funcmanifest:prog}
  \end{subfigure}
  \caption{Program using functors and manifest types}
  \label{ex:funcmanifest}
  \vspace{-2mm}
\end{ex}

The operation that consists in enriching type signatures of module identifiers
with new equalities by using elements in the environment is called
\emph{strengthening} \citep{Leroy94}.

\subsubsection{Typing rules}
\label{sec:ml:module:typing}

In the previous two examples, we showcased some delicate interactions
between functor, type equalities and modularity in the context of an
\ml module system. We now see in details how the rules
presented in \cref{ml:module:typing,ml:module:subtyping,ml:module:validity}
produce these behaviors.

\paragraph{Qualified access}
Unqualified module variables are typechecked in a similar manner than
regular variables in the expression language, with the \mysc{ModVar}
typing rule. Qualified access (of the form $X.a$), both for the core and the module language,
need more work.
As with the expression language, the typing environment is simply
a list of declaration. In particular, typing environments do not store paths.
This means that in order to prove $\wt[]{\Env}{p.a}{\tau}$, we must
first verify that the module $p$ typechecks in $\Env$: $\wtm[]{\Env}{p}{\Mm}$.
We then need to verify that the module type $\Mm$ contains a declaration
$(\valm[]{a}{\tau})$.
This is done in the \mysc{QualModVar} rule for the module language.
The rules for the expression language are given
in \cref{ml:module:additions}.

Let us now try to apply these rules to the module $X$ with the following
module type. $X$ contains
a type $t$ and a value $a$ of that type. We note that module type $\Xm$.
\begin{mathpar}
  \renewcommand\added\addedhide
  X:\ \signature{\typeabsm{t};\ \valm{a}{t}}
\end{mathpar}
We wish to typecheck $X.a$. One expected type for this expression is $X.t$.
However, the binding of $v$ in $\Xm$ gives the type $t$, with no mention of
$X$. We need to prefix the type variable $t$ by the access path $X$.
This is done in the rule \mysc{QualModVar} by the substitution
$\substs[]{n_i}{p.n}{\bv[]{\Sm_1}}{\Mm}$ which prefixes all the bound variables of $\Sm_1$, noted $\bv[]{\Sm_1}$, by the path $p$. Note here that we substitute
only by the names declared \emph{before} the variable $a$. Indeed, a variable
or a type can only reference names declared previously in \ml.
To prove that $X.a$ has the type $X.t$,
we can write the following type derivation.
\begin{mathpar}
  \renewcommand\added\addedhide
  \newcommand\Mx{\bindingm[]{X}{\Xm}}
  \inferrule*[Left=QualVar, Right={with $X.t = \subst[]{t}{p.t}{t}$}]
  { \inferrule*[Left=ModVar]
    { \Mx \in \Mx }
    { \wtm{\Mx}{X}{\signature{\typeabsm{t};\ \valm{a}{t}}} }
  }
  { \wt{\Mx}{X.a}{X.t} }
\end{mathpar}

\paragraph{Strengthening}
The strengthening operation, noted $\strengthen{\Mm}{p}$, is defined in
\cref{ml:module:strengthen} and is used in the \mysc{Strength} rule.
It takes a module type $\Mm$ and a path $p$ and returns a module type $\Mm'$
where all the type declarations, abstract or not, have been replaced by
type aliases pointing to the path $p$.
These type aliases are usually called ``manifest types''.
This operator relies on the following idea:
if $p$ is of type $\Mm$, then $p$ is available in the
environment. In order to expose as many type equalities as possible,
it suffices to give $p$ a type where all the type definition point to
definitions available in the environment. This way, we preserve type
equalities even for abstract types. This also mean that type equalities
can be deduced by only looking at the path and the module type. In
particular, we do not need to look at the implementation of $p$,
which is important for the purpose of separate compilation.

\paragraph{Applicative functors}

Let us consider a functor $F$ with the following type. It takes
a module containing a single type $t$ and return a module containing
an abstract type $t'$ and a conversion function.
\begin{mathpar}
  \renewcommand\added\addedhide
  F:\ \functor{X}
  {\signature{\typeabsm{t}}}
  {(\signature{\typeabsm{t'};\valm{make}{\lamtype{X.t}{t'}}})}
\end{mathpar}

If we consider two modules $X_1$ and $X_2$, does $X_1 = X_2$ imply
$F(X_1).t = F(X_2).t$ ? If that is the case, we say that functors are
\emph{applicative}. Otherwise, they are \emph{generative}%
\footnote{SML only supports generative functors. \ocaml originally
  only supported applicative functors, but also supports the generative
  behavior since version 4.03.}.
Here, we consider the applicative behavior of functors.
This is implemented with the last strengthening
rule which ensures that the body of functors is also strengthened.
For example, if $\Mm$ is the type of the functor above, $\Mm/F$ is the following
module type:
\begin{mathpar}
  \renewcommand\added\addedhide
  \functor{X}
  {\signature{\typeabsm{t}}}
  {(\signature{\typem{t'}{F(X).t};\valm{make}{\lamtype{X.t}{t'}}})}
\end{mathpar}

This justifies the presence of application inside paths. Otherwise, such
type manifests inside functors could not be represented.
A more type-theoretic description of generative and applicative functors can be found in \citet{Leroy96}.

\paragraph{Separate compilation}

Separate compilation is an important properties of programming languages.
In fact, almost all so-called ``mainstream'' languages support it.
We can distinguish two aspects of this property:
separate typechecking and separate code generation.
In both cases, it means that in order to process the file
(either to type check it or to transform it into another representation),
we only need to look at the type of its dependencies, not their
implementation.

It turns out that the \ml module system with manifest types lends itself
very well to separate typechecking \citep{Leroy94}. Indeed, let us consider
a program as a list of modules. Each module represents a compilation
unit (\ie a file). Since module bindings in the typing environment only
contains module types, and not the actual module, typechecking a file only
needs the module type of the previous files, which ensure that we can
typecheck each file separately, as long as all its dependencies
have been typechecked before. This is expressed more formally
in \cref{ml:separation}.

\begin{theorem}[Separate Typechecking]\label{ml:separation}
  Given a list of module declarations that form a typed program, there exists
  an order such that each module can be typechecked with only knowledge
  of the type of the previous modules.

  More formally,
  given a list of $n$ declarations $\dm_i$ and a signature $\Sm$ such that
  \[\renewcommand\added\addedhide
    \wtm{}{(\dm_1;\dots;\dm_n)}{\Sm}
  \]
  then there exists $n$ definitions $\Dm_i$ and a permutation $\pi$ such that
  \begin{align*}
    \forall i < n,\ &
    \wtm[]{\Dm_{1};\dots;\Dm_{i}}{\dm_{i+1}}{\Dm_{i+1}}&
    \submod[]{}{\Dm_{\pi(1)};\dots;\Dm_{\pi(n)}}{\Sm}
  \end{align*}
\end{theorem}
\begin{proof}
  It is always possible to reorder declarations in a signature using the
  \mysc{SubStruct} rule. This means we can choose the appropriate permutation of
  definitions that matches the order of declarations. The rest follows by definition
  of the typing relation.
\end{proof}

\begin{figure}[!p]
  \mprset{sep=1.5em}
  \renewcommand\added\addedhide
  \renewcommand\addedrule\addedhide
  \input{theory/module/typing}\vspace{-3mm}
  \caption{Module typing rules -- $\wtm[]{\Env}{m}{\Mm}$}
  \label{ml:module:typing}
% \end{figure}

% \begin{figure}[hbt]
  \renewcommand\added\addedhide
  \renewcommand\addedrule\addedhide
  \input{theory/module/subtyping}\vspace{-3mm}
  \caption{Module subtyping rules -- $\submod[]{\Env}{\Mm}{\Mm'}$}
  \label{ml:module:subtyping}
\end{figure}

\begin{figure}[p]
  \mprset{sep=1.5em}
  \renewcommand\added\addedhide
  \renewcommand\addedrule\addedhide
  \input{theory/module/validity}\vspace{-3mm}
  \caption{Module type validity rules -- $\wfm[]{\Env}{\Mm}$}
  \label{ml:module:validity}
% \end{figure}

% \begin{figure}[hbt]
  \renewcommand\added\addedhide
  \renewcommand\addedrule\addedhide
  \input{theory/module/additions}\vspace{-3mm}
  \caption{Additional typing rules for the expression language}
  \label{ml:module:additions}
% \end{figure}

% \begin{figure}[hbt]
  \renewcommand\added\addedhide
  \renewcommand\addedrule\addedhide
  \input{theory/module/strengthen}\vspace{-3mm}
  \caption{Module strengthening operation -- $\strengthen{\Mm}{p}$}
  \label{ml:module:strengthen}
  % \vspace{3mm}
\end{figure}

% \subsubsection{Inference}
% \label{sec:ml:inference}

% Full inference is one of the greatest strength of the \ml programming
% language.
% While we do not address inference formally in this thesis, here are
% some remarks.
% Inference is of course decidable for the core language using the
% well known $\mathcal W$ algorithm.
% It is ``efficient'', which means here that it is fast for
% usual programs, though pathological cases can be constructed.
% %
% The typechecking rules for modules are not syntax directed. The \mysc{Strength}
% rule, in particular, is free floating. \citet{Leroy94} presents how to
% turn this into a syntax-directed type system, which allows inference as
% long as functor arguments are annotated.

\subsection{Semantics}
\label{sec:ml:semantics}

We now define the semantics of our \ml language. We use a rule-based
big step semantics with traces. Traces allows us to reason
about execution order in a way that is compatible with modules, as
we will see in \cref{sec:ml:sem:traces}.

We note $v$ for values in the expression language and $\vm$ for values
in the module language. Values are defined in \cref{ml:values}.
Values in the expression language can be either constants or lambdas.
Module values are either structures, which are list of bindings of
values, or functors.
We note $\envr$ the execution environment. Execution environments
are a list of value bindings. Note here that the execution environment
is not mutable, since reference cells are not in the language. We
note the concatenation of environment $+$. Environment access is noted
$\bdrin{x}{v}{\envr}$ where $x$ has value $v$ in $\envr$.
The same notation is also used for structures.
Traces are lists of messages. For now, we consider messages
that are values and are emitted with a $\print$ operation.
The empty trace is noted $\niltr$. Concatenation of traces is noted $\ctr$.

Given an expression $e$ (resp. a module $m$), an execution
environment $\envr$, a value $v$ (resp. $\vm$) and a trace $\trace$,
\[\renewcommand\added\addedhide
  \redto{\envr}{e}{v}{\trace}
\]
means that $e$ reduces to $v$ in $\envr$ and prints $\trace$.
The reduction rules are given in \cref{ml:semantics}.
The rules for the expression language are fairly traditional.
Variables and paths must be resolved using the \mysc{Var} and \mysc{QualVar}
rules. Applications are done in two steps: first, we reduce both
the function and the argument with the \mysc{App} rule,
then we apply the appropriate reduction
rule for the application: \mysc{Beta} for lambda expressions, \mysc{Y} for
fixpoints and \mysc{Delta} for constants.
The $\delta$ operation gives meaning to application of a constant
to a value. $\appconst[]{c}{v} = v', \trace$ means that $c$ applied to $v$
returns $v'$ and emits the trace $\trace$. Let bindings
are treated in a similar manner than lambda expressions: the left
hand side is executed, added to the environment, then the right hand side
is executed.

The module language has similar rules for identifiers and application.
In this case, the \mysc{Beta} and \mysc{App} rule have been combined in
\mysc{ModBeta}. Additional rules for declarations are also present.
Type declarations are ignored (\mysc{TypeDecl}). Values and module
declarations (\mysc{ValDecl} and \mysc{ModDecl}) are treated similarly
to let bindings: the body of the binding is executed, added to the environment
and then the rest of the structure is executed.

\subsubsection{Traces and Printing}
\label{sec:ml:sem:traces}

% \dictum[
% Brian Kernighan, \textit{Unix for Beginners} (1979)
% ]{
%   The most effective debugging tool is still careful thought, coupled with judiciously placed print statements.
% }

Traces allow us to visualize the execution order of programs. In
particular, if we prove that code transformation preserves traces,
it ensures that the execution order is preserved.
Traces allow us to reason about execution without introducing
references and other side-effecting operations in our language, which
would make the presentation significantly more complex.

One example of operation using traces is the $\print$ constant.
Typing and semantics of $\print$ are provided in \cref{ml:print}.
$\print$ accepts any value, prints it, and returns it. From a typing
point of view, $\print$ has the same type as the identity: a polymorphic function
which returns its input. We make use of the fact
that the \mysc{Const} typing rule also uses the instanciation for type schemes.
The semantics of $\print$ is provided via the \mysc{Delta} rule: it returns its argument
directly but also emits a trace containing the given argument.

\begin{figure}[h]
  \mprset{sep=1.5em}
  \renewcommand\added\addedhide
  \renewcommand\addedrule\addedhide
  \begin{mathpar}
    \inferrule[PrintTy]
    {}
    { \Typeof(\print) = \forall\alpha.(\lamtype{\alpha}{\alpha}) }
    \and
    \inferrule[PrintExec]
    {}
    { \appconst{\print}{v} = v, \singltr{v} }
  \end{mathpar}
  \vspace{-6mm}
  \caption{Typing and execution rules for $\print$}
  \label{ml:print}
\end{figure}

\newcommand\mye{
  \letin
  {x}
  {\app{\print}{3}}
  {\app{\print}{(x+1)}}
}

We now present an example using $\print$.
We assume the existence of the type
$\intty$, a set of constant corresponding to the integers
and an associated operation $+$.
We wish to type and execute the expression $e$ defined as $\mye$

Let us first show that $e$ is of type $\intty$. The type derivation is provided
in \cref{ml:print:ex1:typing}. The typing derivation is fairly direct:
we use the \mysc{Const} rule to type $\print$ as $\lamtype{\intty}{\intty}$
and apply it to integers with the rule \mysc{App}.
We can now look at the execution of $e$, which returns $4$
with a trace $\singltr{3;4}$. The execution derivation is shown in
\cref{ml:print:ex1:execution}. The first step is to decompose the let-binding.
We first reduce $\app{\print}{3}$, which can be directly done with
the \mysc{Delta} rule. This gives us $3$ with a trace $\singltr{3}$.
We then reduce $\app{\print}{(x+1)}$ in
the environment where $x$ is associated to $3$.
Before resolving the application of $\print$ with the \mysc{Delta} rule,
we need to reduce its argument with the $\mysc{App}$ rule.
We obtain $4$ with a trace $\singltr{4}$.
We return the result of the right hand side of the left and the concatenation
of both traces by usage of the \mysc{LetIn} rule,
which gives us $4$ with a trace $\singltr{3;4}$.

\begin{ex}[!ht]
  \mprset{sep=1em}
  \renewcommand\added\addedhide
  \newcommand\intfy{\lamtype{\intty}{\intty}}
  \vspace{-2mm}
  \begin{mathpar}
    \inferrule*[Left=LetIn, Right=LetIn]
    { \inferrule*[Left=App]
      { \inferrule*[Left=Const]
        {\instanceof{\intfy}{\Typeof(\print)}}
        {\wt{}{\print}{\intfy}} \\
        \inferrule*{\Typeof(3)=\intty}{\wt{}{3}{\intty}}
      }
      { \wt{}{\app{\print}{3}}{\intty} } \\
      \inferrule*[Right=App]
      { \inferrule*
        {\vdots}
        {\wt{}{\print}{\intfy}} \\
        \inferrule*{\vdots}{ \wt{}{x+1}{\intty} }
      }
      {\wt{\binding{x}{\intty}}{\app{\print}{(x+1)}}{\intty} }
    }
    { \wt{}{\mye}{\intty} }
  \end{mathpar}
  \vspace{-2mm}
  \caption{Typing derivation for $e$ -- $\wt{}{e}{\intty}$}
  \label[ex]{ml:print:ex1:typing}
  
  \begin{mathpar}
    \inferrule*[Left=Let, Right=Let]
    { \inferrule*[Left=Delta]
      { \appconst{\print}{3} = 3, \singltr{3} }
      { \redto{}{\app{\print}{3}}{3}{\singltr{3}} } \\
      \inferrule*[Right=App]
      { \inferrule*
        { }
        { \redto{}{\print}{\print}{\niltr} } \\
        \inferrule*
        { \vdots}
        { \redto{\bdr{x}{3}}{x+1}{4}{\niltr} } \\
        \inferrule*[Right=Delta]
        { \appconst{\print}{4} = 4, \singltr{4} }
        { \redto{}{\app{\print}{4}}{4}{\singltr{4}} }
      }
      { \redto{\bdr{x}{3}}{\app{\print}{(x+1)}}{4}{\singltr{4}} }
    }
    { \redto{}{\mye}{4}{\singltr{3;4}} }
  \end{mathpar}
  \vspace{-2mm}
  \caption{Execution derivation for $e$ -- $\redto{}{e}{4}{\singltr{3;4}}$}
  \label[ex]{ml:print:ex1:execution}
\end{ex}

\subsubsection{Modules}

We now present an example of reduction involving modules. Our example
program $\Pm$ is presented in \cref{ml:mod:ex}. It consists of
two declarations: a module declaration $X$ which contains a single
declaration $a$, and the return value of the program, which is equal to
$X.a$. It is fairly easy to see that the program $\Pm$ return a value
of type $\intty$, hence we focus on the execution of $\Pm$, which
is presented in \cref{ml:mod:ex:execution}.
The derivation is slightly simplified for clarity. In particular, rules
such as \mysc{EmptyStruct} are elided.
The first step is to apply the \mysc{Program} and \mysc{ModuleDecl}
rules in order to execute the content of each declaration.
The declaration of $X$, on the left side, can be reduced by
first applying the \mysc{Struct} rule in order to extract the
content of the module structure, then \mysc{ValDecl}, to reduce
the declaration of $a$. These reductions give us the structure
value $\bdr{a}{3}$.
We now execute the declaration of $\ret$. According to the \mysc{ModuleDecl}
rule, we must do so in a new environment containing $X$: $\bdr{X}{\bdr{a}{3}}$.
In order to reduce $X.a$, we must use the \mysc{QualModVar} rule, which reduces
qualified variables. This means we first reduce $X$, which according
to the environment gives us $\bdr{a}{3}$, noted $V$.
We then look up $a$ in $V$, which returns $3$.
To return, we first compose the resulting structure value from both
declaration: $\bdr{X}{\bdr{a}{3}}+\bdr{\ret}{3}$.
We then lookup $\ret$ in this structure, which gives us $3$.

\begin{ex}[h]
  % \small
  \renewcommand\added\addedhide
  \mprset{sep=1.7em}
  \setlength{\jot}{0pt}% Remove interline space in align*
  \begin{subfigure}{1\linewidth}
    \vspace{-6mm}
    \begin{align*}
      &\mathtt{prog}\\
      &\enspace\modulem{X}{\struct{\letm{a}{3}}}\\
      &\enspace\letm[]{\ret}{X.a}\\
      &\dend
    \end{align*}
    \vspace{-6mm}
    \caption{The program $\Pm$}
    \label[ex]{ml:mod:ex}
  \end{subfigure}

  \begin{subfigure}{1\linewidth}
    \begin{mathpar}
      \inferrule*[Left=Program,Right=Program]
      { \inferrule*[Left=ModuleDecl,Right=ModuleDecl]
        { \inferrule*[Left=Struct]
          { \inferrule*[Left=ValDecl]
            { \inferrule*{}{\redto{}{3}{3}{\niltr}} }
            { \redto{}{\letm{a}{3}}{\bdr{a}{3}}{\niltr} }
          }
          { \redto{}
            {\left(\begin{aligned}
                  &\dstruct\\
                  &\enspace\letm{a}{3}\\
                  &\dend\\
                \end{aligned}\right)}
            {\bdr{a}{3}}{\niltr} } \\
          \inferrule*[Right=ValDecl]
          { \inferrule*[Right=QualModVar]
            { \inferrule*[Left=ModVar]
              { \bdrin{X}{V}{\envr} }
              { \redto{\envr}{X}{V\equiv\bdr{a}{3}}{\niltr} } \\
              \bdrin{a}{3}{V} }
            { \redto{\bdr{X}{\bdr{a}{3}}}{X.a}{3}{\niltr} }
          }
          { \redto{\bdr{X}{\bdr{a}{3}}}
            {\letm[]{\ret}{X.a}}{\bdr{\ret}{3}}{\niltr} }
        }
        { \redto{}
          {\left(\begin{aligned}
                &\modulem{X}{\struct{\letm{a}{3}}}\\
                &\letm[]{\ret}{X.a}
              \end{aligned}\right)}
          {\bdr{X}{\bdr{a}{3}}+\bdr{\ret}{3}}{\niltr}
        }
      }
      { \redto{}{\Pm}{3}{\niltr} }
    \end{mathpar}
    \vspace{-6mm}
    \caption{Execution of $\Pm$}
    \label[ex]{ml:mod:ex:execution}
  \end{subfigure}
  \caption{Example of execution with modules}
\end{ex}

% \paragraph{Why big steps?}

% One might wonder why use a big step semantics with traces, instead of a
% small step semantics. Indeed, small step usually make proofs easier,
% especially for simulations which we will use on \eliom later.
% A first remark is that modules are not stable by substitution since
% $(\struct{\dots}).x$ is not valid (and is problematic to type). Hence
% we need to use a semantics with environments and closures.
% %
% Let us now consider doing one step deep inside a structure
% using a small step semantics.
% The previously evaluated declarations in the structure should
% be available in the local environment which mean
% we would need to rebuild environments as we explore a context to execute a
% small step.
% We would also need to manipulate partially evaluated structures as we execute
% declarations.
% Furthermore, typing preservation for small steps would be difficult to
% express in the presence of abstract types.
% While this is all possible, big steps semantics with environments is,
% by comparison, fairly straightforward.

% \section{Properties}
% \label{sec:ml:prop}

% Several properties have been stated for the expression language and the module language
% independently in \citet{wright1994syntactic} and \citet{Leroy94, Leroy95} respectively.
% We will reformulate two properties that are useful for \eliom in the context
% of our formalization of \ml.

\begin{figure}[!p]
  \mprset{sep=1.5em}
  \renewcommand\added\addedhide
  \renewcommand\addedrule\addedhide
  \input{theory/ml/values}
  \caption{\ml values}
  \label{ml:values}
  \input{theory/ml/semantics}\vspace{-3mm}
  \caption{Big step semantics -- $\redto{\envr}{e}{v}{\trace}$}
  \label{ml:semantics}
\end{figure}

\subsubsection{Notes on Soundness}
\label{ml:soundness}

Soundness properties, which correspond to the often misquoted
``Well typed programs cannot go wrong.'', have been proven for many variants
of the \ml language.
Unfortunately, stating and proving the soundness property for big step semantics
and \ml modules requires a fairly large amount of machinery which
we do not attempt to provide here.
Instead, we give pointers to various relevant work containing such proofs.

Soundness for a small step semantics of our expression language is
provided by \citet{wright1994syntactic}.
At a larger scale, \citet{OCamlLight} proves
the soundness of a small step semantics for a very large portion of the
\ocaml expression language using the Locally Nameless Coq framework
\citep{aydemir-popl-08}.
Soundness of a big step semantics has been proved and mechanized for several
richer languages
\citep{DBLP:conf/popl/AminR17,Tofte88,DBLP:conf/esop/OwensMKT16,DBLP:conf/popl/LeeCH07,garrigue2009certified}.

Unfortunately, as far as we are aware, soundness of Leroy's module language
with higher order applicative functors has not be proved directly and is a fairly
delicate subject.
The most recent work of interest is \cite{FingModules}, which presents
an elaboration scheme from \ml modules, including applicative \ocaml-style
modules, into System $F_\omega$. Soundness then relies on soundness of
the elaboration (provided in the article) and soundness of System $F_\omega$.
In this work, the applicative/generative behavior of functors is decided
depending on its purity, which is much more precise than what
is done in \ocaml.
% Notes that our language does not contain side-effects, which means that our
% language has a chance to be sound. The same cannot be said about
% \ocaml in general.

% The traditional technique to easily prove soundness of big step semantics
% is to add a notion of ``step counter'' that will decrease during the
% execution (on each function call, for example). When the counter reaches zero,
% a ``time out'' value is returned and is propagated. The soundness
% property can then be expressed in the following way: If $e$ is of type $\tau$,
% either there exists a counter $c$ such that $eval(e, c) = v$ and $v$ is of type
% $\tau$, or for any counter $c$, $eval(e, c) = Timeout$.

% \begin{theorem}[Type soundness for expressions]\label{ml:soudness}
%   Given $e$ an \ml expression, $\tau$ a type such that $\wt[]{}{e}{\tau}$.
%   Then either the execution of $e$ does not terminate,
%   or there exists a value $v$ and a trace $\trace$ such that
%   $ \renewcommand\added\addedhide
%     \redto{}{e}{v}{\trace}
%   $
% \end{theorem}
% \begin{proof}
%   See \cite{wright1994syntactic},
%   \cite{DBLP:conf/popl/AminR17},
%   \cite{DBLP:conf/popl/LeeCH07},
%   \cite{DBLP:conf/esop/OwensMKT16},
%   \cite{Tofte88}.
%   \TODO{This is actually broken}
% \end{proof}

% \input{theory/ml/soa}

\clearpage
%%% Local Variables:
%%% mode: latex
%%% TeX-master: "../main"
%%% End:

\section{Semantics preservation}\label{sec:simulation:proof}

We now prove \cref{thm:simulation} which states that
given an \etiny program $\Pm$, the trace of its execution is the same
as the concatenation of the traces of $\compile{s}{\Pm}$ and
$\compile{c}{\Pm}$. More formally:

\begin{theorem}[Compilation preserves semantics]
  Given sets of constants where converters are well-behaved,
  given an \etiny program $\Pm$ respecting the slicability hypothesis and
  such that
  $ \redp{\emptyr}{\Pm}{v}{\trace} $
  then
  \begin{align*}
    \redmls{\emptyr}{\compile{s}{\Pm}}{()}{\envfrag}{\envinj}{\trace_s}&&
    \redmlc{\emptyr}{\compile{c}{\Pm}}{v}{\envfrag}{\envfrag'}{\envinj}{\envg}{\trace_c}&&
    \trace = \trace_s\ctr\trace_c
  \end{align*}
\end{theorem}

\subsection{Hoisting}\label{sec:sim:hoisting}

In \cref{sec:compilation:sections}, we mentioned that a useful property
of injections and fragments is that they can be partially lifted
outside sections.
This property can be used to simplify the simulation proofs.
We consider the code transformation that hoists the content of injections
out of fragments, client declarations and mixed functors in a way
that preserve semantics. This transformation can be decomposed
in two parts.

\paragraph{Injections}

We decompose injections inside fragments and client declarations into simpler
components.
For example, the \etiny piece of code presented in \cref{ex:hoistfrag1}
is decomposed in \cref{ex:hoistfrag2} by moving out the application
of the converter and leaving only a call to the $\serial$ converter.
All injections using a converter than is not $\serial$ nor $\fragment$
can be decomposed in such a way.

Since injections
can only be used on variables or constants and that no server bindings can be
introduced inside a fragment, scoping is preserved.
Furthermore, by definition of converters and their client and server
components, this transformation preserves typing.
It also preserves the dynamic semantics as long as the order
of hoisting correspond to the order of evaluation. This can be
seen by inspecting the reduction relation for server code under client
contexts $\pred{}_{c/s}$.
Finally, it trivially preserves the semantics of the compiled program since
it corresponds exactly to how converters are decomposed during compilation.

\begin{ex}[ht]
  \begin{subfigure}[t]{0.45\linewidth}
    \begin{align*}
      &\letin{a}{1 + 2}{}\\
      &\cv{3 + \injf{a}{\intty}}
    \end{align*}\vspace{-4mm}
    \caption{Fragment with injections}
    \label{ex:hoistfrag1}
  \end{subfigure}\hfill
  \begin{subfigure}[t]{0.45\linewidth}
    \begin{align*}
      &\letin{a}{1 + 2}{}\\
      &\letin{a'}{\app{\intty^s}{a}}\\
      &\cv{3 + \app{\intty^c}{\injf{a'}{\serial}}}
    \end{align*}\vspace{-4mm}
    \caption{Fragment with hoisted injections}
    \label{ex:hoistfrag2}
  \end{subfigure}
  \caption{Hoisting on fragments}
\end{ex}

This allows us to assume that reduction of server code in client context
only uses variable lookup and never leads to any evaluation.
In particular, this will avoid having to deal with the case
of fragments being executed inside the reduction of another fragment (to
see why this could happen, consider the case of a converter
of type
$\forall\alpha_c.\conv{(\lamtype{unit}{\cvtype{\alpha_c}})}{\alpha_c}$).

In the rest of this section, we assume that reductions of the \etiny rule
\mysc{Fragment} are always of the following shape:
\begin{mathpar}
  \inferrule
  { \redel[c/s]{\envr_c}{e}{\noinj{e}}{\emptym}{\niltr} }
  { \reds{\envr_s}{\cv{e}}{\rf{r}}{(\bind{r}{\noinj{e}})}{\niltr} }
  \inferrule{}{
    \text{ where }
    \noinj{e} = \substi[]{\injf{x_i}{f_i}}{\envr_s(x_i)}{e}
    \text{ and } f_i\in\{\serial,\fragment\}
  }
\end{mathpar}
and that reductions of the \etiny rule \mysc{ClientDecl}
are always of the following shape:
\begin{mathpar}
  \inferrule
  { }
  { \redel[c/s]{\envr_s}{\dm_c}{\noinj{\dm_c}}{\emptym}{\niltr} }
  \inferrule{}{
    \text{ where }
    \noinj{\dm_c} = \substi[]{\injf{x_i}{f_i}}{\envr_s(x_i)}{\dm_c}
    \text{ and } f_i\in\{\serial,\fragment\}
  }
\end{mathpar}

\paragraph{Injections inside mixed modules}

We also hoist injections completely out of mixed contexts to the outer
englobing scope.
For example in the functor presented in \cref{ex:hoistmixed1},
we can lift the injection out of the functor, as show in
\cref{ex:hoistmixed2}.
This is valid since injections can only reference
content outside of the functor, by typing.
Semantics is similarly preserved since injections inside functors
are reduced immediately when encountering a functor, as per rule
\mysc{ModClosure} in \cref{eliom:semantics}.

This allows us to assume that the reduction of a mixed module
will never lead to the reduction of an injection.

\begin{ex}[ht]
  \begin{subfigure}[t]{0.4\linewidth}
    \begin{align*}
      &\letm[s]{x}{\dots}\\
      &\modulem[\sideX]{F{(X:\Mm)}}{\dstruct}\\
      &\quad\letm[c]{y}{\injf{x}{f}}\\
      &\dend
    \end{align*}\vspace{-4mm}
    \caption{Mixed functor with injections}
    \label{ex:hoistmixed1}
  \end{subfigure}\hfill
  \begin{subfigure}[t]{0.5\linewidth}
    \begin{align*}
      &\letm[s]{x}{\dots}\\
      &\letm[c]{y'}{\injf{x}{f}}\\
      &\modulem[\sideX]{F{(X:\Mm)}}{\dstruct}\\
      &\quad\letm[c]{y}{y'}\\
      &\dend
    \end{align*}\vspace{-4mm}
    \caption{Mixed functor with hoisted injections}
    \label{ex:hoistmixed2}
  \end{subfigure}
  \caption{Hoisting on mixed modules}
\end{ex}

\subsection{Preliminaries}

Let us start with some naming conventions.
Identifiers with a hat, such
as $\co\envg$, are related to the compiled semantics. For example,
while the server environment for the interpreted semantics
is noted $\envr_s$, the environment for the execution of the target
language \ocsis is noted $\co\envr_s$.
This naming convention is only for ease of reading and does not apply
a formal relation between the objects with and without hats, unless
indicated explicitly.

\subsubsection{Remarks about global environments}

Let us make some preliminary remarks about global environments in the \etiny
client generated programs and in \ocsic.

Given a global environment $\envg$ resulting of $\pred{}_c$,
it contains only two kinds of references:
\begin{itemize}
\item Closure fragments, noted $\rf{f}$, which come from the
  execution of $\pbind\ \mathtt{env}$.
  The associated value is always a environment (\ie a signature).
\item Fragment values, noted $\rf{r}$, which come from the execution of
  $\pbind\ \mathtt{with}$.
\end{itemize}
In the rest of this section, we consider that we can always
decompose global environments $\envg$ in two parts:
a fragment value environment $\envg_r$ containing all the references
$\rf{r}$ that were produced by $\pbind\ \mathtt{with}$ and
a fragment closure environment $\envg_f$ containing
only binding of the form $\bdr{\rf{f}}{\envr}$ that were produced by
$\pbind\ \mathtt{env}$.
\\

Similarly, given a global environment $\co\envg$ used in \ocsic.
There are only three kind of references:
\begin{itemize}
\item Closure fragments, noted $\rf{f}$, which come from the slicing of syntactic
  fragments in the source program.
  The associated value is always a closure.
\item Fragment values, noted $\rf{r}$ come from the execution of
  fragments in the fragment queue.
\item Injections, noted $\rf{x}$. The associated values
  must be serializable, and hence
  can only be references or constants in $\Const_\base$.
\end{itemize}
In the rest of this section, we consider that we can always decompose global
compiled environment $\co\envg$ into
a fragment closure environment $\co\envg_f$, a fragment value environment
$\co\envg_r$ and an injection environment $\envinj$.

% Furthermore, only references associated with fragment values appear
% in \eliom programs, where they are the result of a $\bindg$.
% As a syntactic convenience, we will note $\envg\co\subset\co\envg$
% to signify that we can decompose $\co\envg$ in
% $\co\envg_f\cup\co\envg_r\cup\envinj$ and $\envg\equivc{\co\envg}\co\envg_r$.

\subsubsection{Client equivalence}

\begin{definition}[Client values equivalence]
  Given $v$ an \eliom client value, $v'$ an \ocsic value
  and $\envinj$ an environment of references,
  $v$ and $v'$ are equivalent under $\envinj$,
  noted $v \equivc{\envinj} v'$, if and only if they are equals after
  substitution by $\envinj$: $\substenv{\envinj}{v} = \substenv{\envinj}{v'}$.

  We extend this notation to environments and traces.
  % \begin{align*}
  %   c &\equivc c
  %   &\text{ for all } c\in\Const_c\cup\Const_\base\\
  %   \closure{x}{\envr}{e}
  %     &\equiv\closure{x'}{\envr'}{e'}
  %   &\text{ where for all }
  %     \envg,v,v'\text{ such that } v_x\equivc v'_x\text{ then }\\
  %     &&\redc{\envr\cup\bdr{x}{v_x}}{e}{v}{\trace}{\envg}{\envg} \land
  %     \redmlc{\envr\cup\bdr{x'}{v'_x}}{e'}{v'}{\envg}{\envg}{\trace}
  %     \implies
  %     v\equivc v'
  % \end{align*}
\end{definition}

\begin{definition}[Global environment equivalence]
  We say that an \etiny global environment
  $\envg = \envg_f\cup\envg_r$
  and an \ocsic global environment
  $\co\envg = \co\envg_f\cup\co\envg_r\cup\envinj$
  are synchronized if and only if the following conditions hold.

  \begin{itemize}[leftmargin=*]
  \item The reference environments are equivalent:
    $\envg_r\equivc{\envinj}\co\envg_r$
  \item The domains of $\envg_f$ and $\co\envg_f$ coincides, and:
    \begin{itemize}
    \item
    For each $\rf{f}$ in these environments such that
    $\bdrin{\rf{f}}{\envr}{\envg_f}$ and that
    $\bdrin{\rf{f}}{\closure{x_0\dots x_n}{\co{\envr}}{e}}{\co\envg_f}$, then
    the following property must hold.

    We must have that $\envr\equivc{\envinj}\co\envr$ and that
    for all $v_0,\dots,v_n$, $\co v_0,\dots,\co v_n$ such that
    for all $i$, $v_i\equivc{\envinj}\co{v}_i$;
    then:
    \begin{align*}
      \redc{\envr}{\subst[]{x_i}{v_i}{e}_i}{v}
      {\trace}{\envg}{\envg}
      \implies
      \redmlce{}
      {\app{\closure{\overline{x}_i}{\co{\envr}}{e}}{\co{v}_0\dots \co{v}_n}}
      {\co{v}}{\co\envg}{\co\envg}{\co{\trace}}
    \end{align*}
    with $v\equivc{\envinj}\co{v}$ and $\trace\equivc{\envinj}\co{\trace}$

    \item
    For each $\rf{F}$ in these environments such that
    $\bdrin{\rf{F}}{\envr}{\envg_f}$ and that
    $\bdrin{\rf{F}}{\functorenvi{\co\envr}{Y}{\Mm}{\mm}}
    {\co\envg_f}$, we have $\envr\equivc{\envinj}\co\envr$.
    \end{itemize}
  \end{itemize}
\end{definition}

\begin{definition}[Fragment closure environment]
  We consider that $\co\envg_f$ is a fragment closure environment for
  the \etiny server expression $e_s$, noted $FCE(\co\envg_f,e_s)$, if
  for each $\bdr{\rf{f}}{\closure{\overline{x}_i}{\co\envr}{e'}}$ in
  $\co\envg_f$, for each $\cv{e}_{\rf{f}}$ in $\frags{e_s}$
  we have $e' = \subst[]{\injf{x_i}{f_i}}{x_i}{e}_i$.

\end{definition}
\begin{definition}[Functor closure environment]
  We consider that $\co\envg_f$ is a functor closure environment for
  the \etiny module expression $\mm$, noted $FCE(\co\envg_f,\mm)$, if
  for each $\bdr{\rf{F}}{\functorenvi{\co\envr}{Y}{\Mm}{\co\sm}}$ in
  $\co\envg_f$, for each $(\struct{\sm}_{\rf{F}})$ in $\mm_s$
  we have $\co\sm = \compile{c}{\sm}$. Additionally, we require
  that $\co\envg_f$ be a fragment closure environment
  for each expression contained in $\sm$.
\end{definition}

In the rest of this section, we use the same notation for both properties.
We extend this notation to server declarations, server values
(by looking under closures) and server environments.

\begin{lemma}[Reduction up to equivalence]
  \label{redequivc}\label{redequivmodc}
  Given $\envr$, $\co\envr$,
  $\envg = \envg_f\cup\envg_r$,
  $\co\envg = \co\envg_f\cup\co\envg_r\cup\envinj$,
  $e$ and $\co e$
  such that:
  \begin{align*}
    \envr&\equivc{\envinj}\co\envr
    &\envg&\equivc{\envinj}\co\envg
    &e[\envinj] &= \co e[\envinj]
    &\redc{\envr}{e}{v}{\trace}{\envg}{\envg}
  \end{align*}
  Then we have:
  \begin{align*}
    &\redmlce{\co\envr}{\co{e}}{\co{v}}{\co\envg}{\co\envg}{\co\trace}
    &v&\equivc{\envinj}\co v
    &\trace&\equivc{\envinj}\co\trace
  \end{align*}
\end{lemma}
\begin{proof}
  The only difference between \etiny client expressions and
  \ocsic expressions are the presence of extra references for injections
  in \ocsic. Indeed, syntactic injections have been removed either
  by the server execution or by compilation and $\bindg$ constructs
  are only accessible at the module level. Since we assume
  that the original expression $e$ and the compiled expression $\co e$
  are the same up to the injection environment $\envinj$, we can
  trivially mimic the execution of $e$ in $\co e$ by induction.
\end{proof}

\subsubsection{Server equivalence}

\newcommand\substfrag[1]
{\substi[]{\cv{e_i}_{\rf{f}}}{\pfragment{\rf{f}}{\overline{x_{i,j}}}}{#1}}

% \begin{definition}[Compilation of server expressions]
%   In order to simplify the presentation, we extend the server
%   compilation relation on expressions.
%   \begin{align*}
%     \compile{s}{e}
%     &= \substi[]{\cv{e_i}_{\rf{f}}}{\pfragment{\rf{f}}{\overline{x_{i,j}}}}{e}
%     &\text{ where $\overline{x_{i,j}} = \injs{e_i}$}
%   \end{align*}

%   We also extend this notation values and environments. In this case,
%   substitution should be also done inside closures.
% \end{definition}

\begin{definition}[Server value equivalence]
  Given $v$ an \eliom server value, $\co v$ an \ocsis value.
  We say they are equivalent, noted $v\equivs{}\co v$ if
  and only if
  \begin{align*}
    \substfrag{v} &= \co v
    &\text{ where $\overline{x_{i,j}} = \injs{e_i}$}
  \end{align*}
  We extend this notation to environments and traces.
\end{definition}

\subsection{Server expressions and structures}

We first look at server expressions and structures.
By definition of the server reduction relation for \etiny,
the emitted program is a series of binds.

\begin{lemma}[Server expressions are simulable]
  \label{simulation:serverexpr}
  We consider an \etiny server expression $e$;
  the \etiny environments $\envr_s$, $\envr_c$ and
  $\envg = \envg_f\cup\envg_r$;
  the target environment $\co\envr_s$, $\co\envr_c$ and
  $\co\envg = \co\envg_f\cup\co\envg_r\cup\envinj$.

  If the expression $e$ has valid server and client executions:
  \begin{align*}
    \redel[s]{\envr_s}{e}{v}{\clientprog}{\trace_s}
    &&\redc{\envr_c}{\clientprog}{\emptyr}{\trace_c}{\envg}{\envg'}
  \end{align*}
  and the following invariants hold:
  \begin{align*}
    \co\envr_c &\equivc{\envinj}\envr_c&
    \co\envr_s &\equivs{}\envr_s&
    \co\envg &\equivc{}\envg&
    &\fragenv{\co\envg_f,e}&
    &\fragenv{\co\envg_f,\envr_s}
  \end{align*}

  Then $\co e = \substfrag{e}$ has an equivalent execution.
  \begin{mathpar}
    \inferrule
    { }
    { \redmls{\co\envr_s}{\co{e}}{\co v}
      {\envfrag_\bullet}{\emptyr}{\co\trace_s}}
    \and
    \inferrule
    { }
    { \redmlc{\co\envr_c}{\pexec}{\emptym}
      {\envfrag_\bullet\cfrag\tokend}{\emptyfrag}
      {\co\envg}{\co\envg'}{\co\trace_c}
    }
  \end{mathpar}
  with the following invariants:
  \begin{align*}
    \co\envg'&\equivc{}\envg'&
    &\fragenv{\co\envg_f',v}&
    % \co\envg&\subset\co\envg'&
    \co v&\equivs{} v&
    \co\trace_s&\equivs{}\trace_s&
    \co\trace_c&\equivc{\envinj}\trace_c
  \end{align*}
\end{lemma}
\begin{proof}
  We consider an expression $e$;
  the \etiny environments $\envr_s$, $\envr_c$ and
  $\envg = \envg_f\cup\envg_r$;
  the target environment $\co\envr_s$, $\co\envr_c$ and
  $\co\envg = \co\envg_f\cup\co\envg_r\cup\envinj$.
  such that
  \begin{align*}
    \co\envg&\equivc{}\envg&
    \co\envr_c &\equivc{\envinj}\envr_c&
    \co\envr_s &\equivs{}\envr_s&
    &\fragenv{\co\envg_f,e}&
    &\fragenv{\co\envg_f,\envr_s}
  \end{align*}

  We will proceed by induction over the executions of $e$ and $\clientprog$.
  The only case of interest is when the server expression is a fragment.

\begin{subproof}[Case $\cv{e}_{\rf{f}}$]\item
  We assume that the following executions hold:
  \begin{mathpar}
    \inferrule
    { \bdrin{x_i}{v_i}{\envr_s} }
    { \redx{\envr_s}{\cv{e}_{\rf{f}}}{\rf{r}}
      {\bindw{r}{f}{\noinj{e}}}{\niltr} }
    \and
    \inferrule
    { \bdrin{\rf{f}}{\envr}{\envg} \\
      \redc{\envr}{\noinj{e}}{v_c}{\trace_c}{\envg}{\envg} }
    { \redc{\envr_c}{(\bindw{r}{f}{\noinj{e}})}{\emptyr}
      {\trace_c}{\envg}{\envg'}
    }
  \end{mathpar}
  where
  $\noinj{e} =
  \substi[]{\injf{x_i}{f_i}}{\injval{v_i}}{e}$ and
  $\envg' = \envg\cup\bdr{\rf{r}}{v_c}$.
  We have $\co e$ equal to $\pfragment{\rf{f}}{x_1\dots x_n}$.

  We first consider the execution of $\co e$.
  We can easily
  construct the following execution.
  \begin{mathpar}
    \inferrule
    { \bdrin{x_i}{\co{v}_i}{\co\envr_s} }
    { \redmls{\co\envr_s}
      {\pfragment{\rf{f}}{x_1\dots x_n}}
      {\rf{r}}
      {\tokfrag{r}
        {\rf{f}\ \injval{\co v_1}\dots\injval{\co v_n}}}
      {\emptyr}
      {\niltr}
    }
  \end{mathpar}

  By hypothesis, for each $i$, $v_i\equivs{\co\envg}\co{v}_i$.
  This gives us that
  $\injval{v_i}\equivc{\co\envg}\injval{\co{v}_i}$.
  We trivially have that
  $\rf{r}\equivs{\co\envg}\rf{r}$
  \\

  Let us now look at the client execution.
  By client execution of $\clientprog$, $\bdrin{\rf{f}}{\envr}{\envg}$.
  Since $\envg\equivc{\envinj}\co\envg$, we have
  $\bdr{\rf{f}}{\closure{x_0\dots x_n}{\co{\envr}}{e'}}\in\co\envg$
  and $\envr\equivc{\co\envg}\co\envr$.
  Furthermore, since $\fragenv{\co\envg_f,\cv{e}_{\rf{f}}}$, we
  know that that $e' = \substi[]{\injf{x_i}{f_i}}{x_i}{e}$.
  We have by hypothesis that
  $\redc{\envr_c}{\noinj{e}}{v_c}{\trace_c}{\envg}{\envg}$.
  Since $\noinj{e} = \subst[]{x_i}{\injval{v_i}}{e'}$
  and since for all $i$, $\injval{v}_i\equivc{\co\envg}\injval{\co{v}}_i$,
  we can use \cref{redequivc} to build the following reduction:
  \begin{mathpar}
    \inferrule
    { \inferrule
      { \inferrule
        { \redmlce{\co\envr_c\cup\bdr{x_i}{\co{v}_i}_i}
          {e'}{\co{v}}
          {\co\envg}{\co\envg}{\co\trace_c} }
        { \redmlce{\co\envr_c}
          { \closure{x_1\dots x_n}{\co{\envr}}{e'}
            \ \injval{\co v_1}\dots\injval{\co v_n} }
          { \co{v} }
          {\co\envg}{\co\envg}
          {\co\trace_c}
        }
      }
      { \redmlce{\co\envr_c}
        { \rf{f}\ \injval{\co v_1}\dots\injval{\co v_n} }
        { \co{v} }
        {\co\envg}{\co\envg}
        {\co\trace_c}
      }
    }
    { \redmlc{\co\envr_c}
      { \pexec }{\emptym}
      { \tokfrag{r}
        {\rf{f}\ \injval{\co v_1}\dots\injval{\co v_n}}
      }{\emptyfrag}
      {\co\envg}{\co\envg'}
      {\co\trace_c}
    }
  \end{mathpar}
  Where $\co\envg' = \co\envg\cup\bdr{\rf{r}}{\co{v}}$.
  By \cref{redequivc}, we have
  that $v\equivc{\co\envg}\co{v}$ and
  $\trace_c\equivc{\co\envg}\co{\trace}_c$.
  The only part that is changed in $\envg'$ and $\co\envg'$ is
  the fragment reference environment, hence we easily have that
  $\co\envg'\equivc{}\envg'$, which concludes.
\end{subproof}

\begin{subproof}[Other cases]\item
  In other cases, we first note that references manipulated
  inside server code can only fragment references $\rf{r}$.
  By hypothesis, the same references are considered before and
  after compilation. Since the fragment closure environment
  hypothesis ranges over all server expressions, including the one
  in closures, it is easy to preserve it during execution.
  The rest is a very simple induction.
\end{subproof}
\end{proof}

\begin{corollary}[Server module declarations are simulable]
  \label{redequivmods}
  \label{simulation:servermod}
  We consider an \etiny server declaration $\dm_s$;
  the \etiny environments $\envr_s$, $\envr_c$ and
  $\envg = \envg_f\cup\envg_r$;
  the target environment $\co\envr_s$, $\co\envr_c$ and
  $\co\envg = \co\envg_f\cup\co\envg_r\cup\envinj$.

  If the expression $e$ has valid server and client executions:
  \begin{align*}
    \redel[s]{\envr_s}{\dm}{\vm}{\clientprog}{\trace_s}
    &&\redc{\envr_c}{\clientprog}{\emptyr}{\trace_c}{\envg}{\envg'}
  \end{align*}
  and the following invariants hold:
  \begin{align*}
    \co\envr_c &\equivc{\envinj}\envr_c&
    \co\envr_s &\equivs{}\envr_s&
    \co\envg &\equivc{}\envg&
    &\fragenv{\co\envg_f,\dm}&
    &\fragenv{\co\envg_f,\envr_s}
  \end{align*}

  Then $\co\dm = \substfrag\dm$ have an equivalent execution.
  \begin{mathpar}
    \inferrule
    { }
    { \redmls{\co\envr_s}{\co\dm}{\co \vm}
      {\envfrag_\bullet}{\emptyr}{\co\trace_s}}
    \and
    \inferrule
    { }
    { \redmlc{\co\envr_c}{\pexec}{\emptym}
      {\envfrag_\bullet\cfrag\tokend}{\emptyfrag}
      {\co\envg}{\co\envg'}{\co\trace_c}
    }
  \end{mathpar}
  with the following invariants:
  \begin{align*}
    \co\envg'&\equivc{}\envg'&
    &\fragenv{\co\envg_f',\vm}&
    % \co\envg&\subset\co\envg'&
    \co\vm&\equivs{}\vm&
    \co\trace_s&\equivs{}\trace_s&
    \co\trace_c&\equivc{\envinj'}\trace_c
  \end{align*}
\end{corollary}

\subsection{Mixed structures}

\begin{lemma}[Structures are simulable]\label{lemma:simulation:struct}
  We consider a slicable structure $\sm$;
  the \etiny environments $\envr_s$, $\envr_c$ and
  $\envg = \envg_f\cup\envg_r$;
  the target environment $\co\envr_s$, $\co\envr_c$ and
  $\co\envg = \co\envg_f\cup\co\envg_r\cup\envinj$.

  If the structure has valid server and client executions:
  \begin{align*}
    \redx{\envr_s}{\sm}{\vm_s}{\clientprog}{\trace_s}
    &&\redc{\envr_c}{\clientprog}{\vm_c}{\trace_c}{\envg}{\envg'}
  \end{align*}
  and the following invariants hold:
  \begin{align*}
    \co\envr_c &\equivc{\envinj}\envr_c&
    \co\envr_s &\equivs{}\envr_s&
    \co\envg &\equivc{}\envg&
    &\fragenv{\co\envg_f,\sm}&
    &\fragenv{\co\envg_f,\envr_s}
  \end{align*}
  then for any $\envfrag$,
  the compiled structures have equivalent executions
  \begin{align*}
    \redmls{\co\envr_s}{\compile{s}{\sm}}{\co\vm_s}
    {\envfrag_\bullet}{\envinj_\bullet}{\co\trace_s}
    &&\redmlc{\co\envr_c}{\compile{c}{\sm}}{\co\vm_c}
       {\envfrag_\bullet\cfrag\envfrag}{\envfrag'}
       {\envinj_\bullet\cup\co\envg}{\co{\envg}'}{\co\trace_c}
  \end{align*}
  with the following invariants:
  \begin{align*}
    \co\envg' &\equivc{}\envg'&
    \co\vm_s&\equivs{}\vm_s&
    \co\trace_s&\equivs{}\trace_s\\
    \fragenv{&\co\envg_f',\vm_s}&
    \co\vm_c&\equivc{\co\envg'}\vm_c&
    \co\trace_c&\equivc{\co\envg'}\trace_c
  \end{align*}
\end{lemma}
\begin{proof}
  \setlength{\jot}{0.5pt}

  We consider a slicable structure $\sm$;
  the \etiny environments $\envr_s$, $\envr_c$ and
  $\envg = \envg_f\cup\envg_r$;
  the target environment $\co\envr_s$, $\co\envr_c$ and and
  $\co\envg = \co\envg_f\cup\co\envg_r\cup\envinj$.
  such that
  \begin{align*}
    \co\envr_c &\equivc{\co\envg}\envr_c&
    \co\envr_s &\equivs{}\envr_s&
    \co\envg &\equivc{}\envg&
    &\fragenv{\co\envg_f,\sm}&
    &\fragenv{\co\envg_f,\envr_s}
  \end{align*}

  We will now proceed by induction over the execution of $\sm$.

% \begin{subproof}[Case $\emptym$]\item
%   \TODO{}
% \end{subproof}

\begin{subproof}[Case $\sm = \dm_b;\sm'$ -- Base declaration]\item
  We assume that the following executions hold:
  \begin{mathpar}
    \mprset{sep=1.5em}
    \inferrule*
    { \redel[\base]{\envr_s}{\dm_\base}{\vm_s}{\emptym}{\trace_s} \\
      \redx{\envr_s+\vm_s}{\sm'}{\vm'_s}{\clientprog}{\trace'_s}
    }
    { \redx{\envr_s}
      {\dm_\base;\sm'}{\vm_s+\vm'_s}{(\dm_\base;\clientprog)}{\trace_s\ctr\trace'_s}
    }
    \and
    \inferrule*
    { \redc{\envr_c}{\dm_b}{\vm_c}{\trace_c}{\envg}{\envg} \\
      \redc{\envr_c+\vm_c}{\clientprog}{\vm'_c}{\trace'_c}{\envg}{\envg'} }
    { \redc{\envr_c}{\dm_b;\clientprog}{\vm_c+\vm'_c}
      {\trace_c\ctr\trace'_c}{\envg}{\envg'}
    }
  \end{mathpar}

  Let us consider the executions of $\dm_\base$.
  By definition of base, it contains neither injections nor fragments.
  By \cref{prop:execbase},
  $\redel[\base]{\envr}{\dm_\base}{\vm_s}{\emptym}{\trace_s}$
  and $\redc{\envr_c}{\dm_b}{\vm_b}{\trace_c}{\envg}{\envg}$ both
  correspond to $\redel[]{\envr_s}{\dm_\base}{\vm_s}{\emptym}{\trace_s}$
  and $\redel[]{\envr_c}{\dm_\base}{\vm_c}{\emptym}{\trace_c}$ respectively.
  By definition, base fragments can't be present, hence
  we also have $\fragenv{\co\envg_f,\vm_s}$

  Additionally, the compilation functions are the identity on base, which mean
  that $\compile{s}{\dm_b}$ and $\compile{c}{\dm_b}$ contains only \ml constructs.
  The reduction relation over \ocsis and \ocsic coincide with the \ml one on the
  \ml fragment of the language. Hence, for any $\envfrag$, we have
  $\redmls{\envr_s}{\compile{s}{\dm_b}}{\co\vm_s}
  {\emptyfrag}{\emptyr}{\co\trace_s}$
  and
  $\redmlc{\envr_c}{\compile{c}{\dm_b}}{\co\vm_c}
  {\envfrag}{\envfrag}{\co\envg}{\co\envg}{\co\trace_c}$
  with
  \begin{align*}
    \co\vm_s&\equivs{}\vm_s&
    \co\vm_c&\equivc{\co\envg}\vm_c&
    \co\trace_s&\equivs{}\trace_s&
    \co\trace_c&\equivc{\co\envg}\trace_c
  \end{align*}

  Let us consider the execution of $\sm'$ and $\clientprog$.
  We easily have the following properties:
  \begin{align*}
    \co\envr_c+\co\vm_c &\equivc{\envinj}\envr_c+\vm_c&
    \co\envr_s+\co\vm_s &\equivs{}\envr_s+\vm_s&
    \co\envg &\equivc{}\envg&
    &\fragenv{\co\envg_f,\sm}&
    &\fragenv{\co\envg_f,\envr_s+\vm_s}
  \end{align*}
  hence, by induction on the execution of $\sm'$ and $\clientprog'$, we have
  $\redmls{\co\envr_s+\co\vm_s}{\compile{s}{\sm'}}{\co\vm'_s}
  {\envfrag_\bullet}{\envinj_\bullet}{\co\trace'_s}$
  and
  $\redmlc{\co\envr_c+\co\vm_c}{\compile{c}{\sm'}}{\co\vm'_c}
  {\envfrag_\bullet \cfrag \envfrag}{\envfrag}
  {\co\envg\cup\envinj_\bullet}{\co\envg'}{\co\trace'_c}$
  for any $\envfrag$,
  with
  \begin{align*}
    \co\envg' &\equivc{}\envg'&
    \co\vm'_s&\equivs{}\vm'_s&
    \co\trace_s&\equivs{}\trace_s\\
    \fragenv{&\co\envg_f',\vm'_s}&
    \co\vm'_c&\equivc{\co\envg'}\vm'_c&
    \co\trace'_c&\equivc{\co\envg'}\trace'_c
  \end{align*}

  We can then build the following derivations:
  \begin{mathpar}
    \mprset{sep=1.5em}
    \inferrule*
    {
      \redmls{\co\envr_s}{\compile{s}{\dm_b}}{\co\vm_s}
      {\emptyfrag}{\emptyr}{\trace_s} \\
      \redmls{\co\envr_s+\co\vm_s}{\compile{s}{\sm'}}{\co\vm'_s}
      {\envfrag_\bullet}{\envinj_\bullet}{\co\trace'_s}
    }
    { \redmls{\co\envr_s}
      {\compile{s}{\dm_\base;\sm'}}{\co\vm_s+\co\vm'_s}
      {\envfrag_\bullet}{\envinj_\bullet}{\co\trace_s\ctr\co\trace'_s}
    }
    \and
    \inferrule*
    { \redmlc{\co\envr_c}{\compile{c}{\dm_b}}{\co\vm_c}
      {\envfrag_\bullet \cfrag \envfrag}{\envfrag_\bullet \cfrag \envfrag}
      {\co\envg\cup\envinj_\bullet}{\co\envg\cup\envinj_\bullet}
      {\co\trace_c} \quad
      \redmlc{\co\envr_c+\co\vm_c}{\compile{c}{\sm'}}{\co\vm'_c}
      {\envfrag_\bullet \cfrag \envfrag}{\envfrag}
      {\co\envg\cup\envinj_\bullet}{\co\envg'}{\co\trace'_c}
    }
    { \redmlc{\envr_c}{\compile{c}{\dm_b;\sm}}{\co\vm_c+\co\vm'_c}
      {\envfrag_\bullet \cfrag \envfrag}{\envfrag}
      {\co\envg\cup\envinj_\bullet}{\co\envg'}{\co\trace_c\ctr\co\trace'_c}
    }
  \end{mathpar}
  and the following invariants are easily verified:
  \begin{align*}
    \co\envg' &\equivc{}\envg'&
    \co\vm_s+\co\vm'_s&\equivs{}\vm_s+\vm'_s&
    \co\trace_s\ctr\co\trace'_s&\equivs{}\trace_s\ctr\trace'_s\\
    \fragenv{&\co\envg_f',\vm_s+\vm'_s}&
    \co\vm_c+\co\vm'_c&\equivc{\co\envg'}\vm_s+\vm'_c&
    \co\trace_c\ctr\co\trace'_c&\equivc{\co\envg'}\trace_c\ctr\trace'_c
    \qedhere
  \end{align*}
\end{subproof}

\begin{subproof}[Case $\sm = \dm_s;\sm'$ -- Server declaration]\item
  We assume that the following executions hold:
  \begin{mathpar}
    \mprset{sep=1.5em}
    \inferrule*
    { \redel[s]{\envr_s}{\dm_s}{\vm_s}{\clientprog}{\trace_s} \\
      \redx{\envr_s+\vm_s}{\sm'}{\vm'_s}{\clientprog'}{\trace'_s}
    }
    { \redx{\envr_s}
      {\dm_s;\sm'}{\vm_s+\vm'_s}
      {(\bindenv{f_i})_i;\clientprog;\clientprog'}{\trace_s\ctr\trace'_s}
    }
    \and
    \inferrule*
    { \redc{\envr_c}{(\bindenv{f_i})_i}
      {\{\}}{\niltr}{\envg}{\envg'} \\
      \redc{\envr_c}{\clientprog}
      {\vm_c}{\trace_c}{\envg'}{\envg''} \\
      \redc{\envr_c}{\clientprog'}
      {\vm'_c}{\trace'_c}{\envg''}{\envg'''}
    }
    { \redc{\envr_c}{(\bindenv{f_i})_i;\clientprog;\clientprog'}
      {\vm_c+\vm'_c}{\trace_c\ctr\trace'_c}{\envg}{\envg''} }
  \end{mathpar}

  Let us note $\cv{e_i}_{\rf{f_i}}$ the fragments syntactically present
  in $\dm_s$.
  let us note $\cv{e_j}_{\rf{f_j}}$ the fragments \emph{executed} during the reduction of
  $\dm_s$ and $\rf{r_j}$ the associated fresh variables.

  We have the following compilations:
  \begin{align*}
    \compile{s}{\dm_s;\sm'}
    &= \substi[]{\cv{e_i}_{\rf{f_i}}}{\pfragment{\rf{f}_i}
      {\overline{x_{i,k}}}}{\compile{s}{\dm_s}};
      \pend;\compile{s}{\sm'}\\
    \compile{c}{\dm_s;\sm'}
    &=\left( \bindg \rf{f}_i =
      \lam{\overline{x_{i,k}}}
      {\subst[]
      {\injf{x_{i,k}}{f_{i,k}}}
      {x_{i,k}}{e_i}};
      \right)_i;
      \pexec;\compile{c}{\sm'}
  \end{align*}

  After hoisting, converters can only be the $\serial$ or $\fragment$. Its
  server and client parts are the identity, hence we simply omit them.
  We also note that $\compile{s}{\dm_s}$ differs with $\dm_s$ only
  on type annotations and type declarations which are ignored by reduction
  relations. We note
  $\co\dm_s = \substi[]{\cv{e_i}}{\pfragment{\rf{f}_i}
    {\overline{x_{i,k}}}}{\compile{s}{\dm_s}}$.
  \\

  Let us consider the reduction of
  $(\bind{\rf{f}_i}{\lam{\overline{x}_i}{\co{e_i}}})_i$.
  Let us note $e'_i = \subst[]{\injf{x_{i,j}}{f_{i,j}}}{x_{i,j}}{e_i}$.
  For any queue $\envfrag$, we have the following reduction:
  \begin{mathpar}
    \inferrule
    { \forall i,\ %
      \inferrule*
      { \redmlc{\co\envr_c}
        {\lam{x_1\dots x_m}{e'_i}}
        {\closure{\overline{x}_i}{\co\envr_c}{e'_i}}
        {\envfrag}{\envfrag}{\co\envg_i}{\co\envg_i}{\niltr}
      }
      { \redmlc{\co\envr_c}
        {\bind{\rf{f}_i}{\lam{\overline{x}_i}{e'_i}}}
        {\emptyr}
        {\envfrag}{\envfrag}{\co\envg_i}{\co\envg_{i+1}}{\niltr}
      }
    }
    { \redmlc{\co\envr_c}
      {(\bind{\rf{f}_i}{\lam{\overline{x}_i}{e'_i}})_i}{\emptyr}
      {\envfrag}{\envfrag}{\co\envg_1}{\co\envg_{n+1}}{\niltr}
    }
  \end{mathpar}
  where $\co\envg_1 = \co\envg$ and
  $\co\envg_{i+1} = \co\envg_i\cup
  \bdr{\rf{f_i}}{\closure{x_1\dots x_m}{\co{\envr_c}}{e'_i}}$.
  Let $\envg_{bind}$ be
  $\bdr{\rf{f_i}}{\closure{x_1\dots x_m}{\co{\envr_c}}{e'_i}}_i$.
  We note $\co\envg' = \co\envg_{n+1} = \co\envg \cup \envg_{bind}$
  and
  $\co\envg'_f = \co\envg_f\cup \envg_{bind}$.
  \\

  Since
  $\redc{\envr_c}{(\bindenv{f_i})_i}{\{\}}{\niltr}{\envg}{\envg'}$,
  we have $\envg' = \envg\cup\bdr{\rf{f_i}}{\envr_c}_i$.
  and $\envr_c\equivc{}\co\envr_c$, we have that
  $\envg'\equivc{}\co\envg'$.
  Furthermore, given one of the $\rf{f_i}$ in $\envg_{bind}$, each fragment
  annotated with this $\rf{f_i}$ syntactically appear in $\dm_s$ by
  uniqueness of the annotation function.
  This also holds inside functors, since
  each $\rf{f_i}$ will be prefixed by a unique module reference.
  Hence $\fragenv{\co\envg'_f,\dm_s}$
  and $\fragenv{\co\envg'_f,\envr_s}$.
  \\

  We now have all the ingredients to uses \cref{redequivmods} on
  the execution of $\dm_s$ and $\clientprog$.
  This gives us the following reductions:
  \begin{mathpar}
    \inferrule
    { }
    { \redmls{\co\envr_s}{\co\dm_s}{\co \vm_s}
      {\envfrag_\bullet}{\emptyr}{\co\trace_s}}
    \and
    \inferrule
    { }
    { \redmlc{\co\envr_c}{\pexec}{\emptym}
      {\envfrag_\bullet\cfrag\tokend}{\emptyfrag}
      {\co\envg'}{\co\envg''}{\co\trace_c}
    }
  \end{mathpar}
  with the following invariants:
  \begin{align*}
    \co\envg''&\equivc{}\envg''&
    &\fragenv{\co\envg_f'',\vm}&
    % \co\envg&\subset\co\envg'&
    \co\vm&\equivs{}\vm&
    \co\trace_s&\equivs{}\trace_s&
    \co\trace_c&\equivc{\envinj}\trace_c
  \end{align*}
  We remark that $\envinj'' = \envinj$ since no injection
  took place during a server section and that $\co\envg_f'' = \co\envg_f'$,
  by definition of the reduction for $\dexec$.
  \\

  We now consider the execution of $\sm'$.
  The following invariants holds:
  \begin{align*}
    \co\envr_c &\equivc{\envinj}\envr_c&
    \co\envr_s+\co\vm_s &\equivs{}\envr_s+\vm_s&
    \co\envg'' &\equivc{}\envg''&
    &\fragenv{\co\envg''_f,\sm'}&
    &\fragenv{\co\envg''_f,\envr_s'+\vm_s}
  \end{align*}
  By induction on the execution of $\sm'$ and $\clientprog'$, we have
  $\redmls{\co\envr_s+\co\vm_s}{\compile{s}{\sm'}}{\co\vm'_s}
  {\envfrag'_\bullet}{\envinj'_\bullet}{\co\trace'_c}$
  and
  $\redmlc{\co\envr_c}{\compile{c}{\sm'}}{\co\vm'_c}
  {\envfrag'_\bullet \cfrag \envfrag'}{\envfrag'}
  {\co\envg''\cup\envinj'_\bullet}{\co\envg'''}{\co\trace'_c}$
  where%
  \begin{align*}
    \co\envg''' &\equivc{}\envg'''&
    \co\vm'_s&\equivs{}\vm'_s&
    \co\trace'_s&\equivs{}\trace'_s\\
    \fragenv{&\co\envg_f''',\vm'_s}&
    \co\vm'_c&\equivc{\co\envg'}\vm'_c&
    \co\trace'_c&\equivc{\co\envg'}\trace'_c
  \end{align*}

  Finally, we can construct the following executions:
  \begin{mathpar}
    \mprset{sep=1.5em}
    \inferrule
    { \redmls{\co{\envr_s}}
      {\co\dm_s}{\co\vm_s}
      {\envfrag_\bullet}
      {\emptyr}{\co\trace_s}
      \\
      \redmls{\envr_s+\vm_s}{\compile{s}{\sm'}}{\co\vm'_s}
      {\envfrag'_\bullet}{\envinj'_\bullet}{\co\trace'_s}
    }
    { \redmls{\co\envr_s}
      {\co\dm_s;\pend;\compile{s}{\sm'}}
      {\co\vm_s+\co\vm'_s}
      {\envfrag_\bullet\cfrag\tokend\cfrag\envfrag'_\bullet}
      {\envinj'_\bullet}{\co\trace_s\ctr\co\trace'_s}
    }
    \and
    \inferrule
    { \redmlc{\co\envr_c}
      {\co\clientprog}{\emptyr}
      {\envfrag}{\envfrag}
      {\co\envg\cup\envinj'_\bullet}{\co\envg'\cup\envinj'_\bullet}{\niltr}
      \and
      \inferrule*
      { \redmlc{\co\envr_c}{\compile{c}{\sm'}}{\vm'_c}
        {\envfrag'_\bullet\cfrag\envfrag'}{\envfrag'}
        {\co\envg''\cup\envinj'_\bullet}{\co\envg'''}{\co\trace'_c}
      }
      { \redmlc{\co\envr_c}{\pexec;\compile{c}{\sm'}}{\co\vm'_c}
        {\envfrag}{\envfrag'}
        {\co\envg'\cup\envinj'_\bullet}{\co\envg'''}{\co\trace_c\ctr\co\trace'_c}
      }
    }
    { \redmlc{\co\envr_c}
      {\co\clientprog;\pexec;\compile{c}{\sm'}}
      {\co\vm'_c}
      {\envfrag}{\envfrag'}
      {\co\envg\cup\envinj'_\bullet}{\co\envg'''}{\co\trace_c\ctr\co\trace'_c}
    }
  \end{mathpar}
  where
  $\envfrag = \envfrag_\bullet\cfrag\dend\cfrag
  \envfrag'_\bullet\cfrag\envfrag'$ and
  $\co\clientprog = (\bind{\rf{f}_i}{\lam{\overline{x}_i}{e'_i}})_i$.
  We verify the following invariants:
  \begin{align*}
    \co\envg''' &\equivc{}\envg'''&
    \co\vm_s+\co\vm'_s&\equivs{}\vm_s+\vm'_s&
    \co\trace_s+\co\trace'_s&\equivs{}\trace_s+\trace'_s\\
    \fragenv{&\co\envg_f''',\vm_s+\vm'_s}&
    \co\vm'_c&\equivc{\co\envg'''}\vm'_c&
    \co\trace_c+\co\trace'_c&\equivc{\co\envg'''}\trace_c+\trace'_c
    \qedhere
  \end{align*}
\end{subproof}

\begin{subproof}[Case $\sm = \dm_c;\sm'$ -- Client declaration]\item
  \setlength{\jot}{0pt}% Remove interline space in align*
  \newcommand\compdmc{\co\dm_c}%
  We assume that the following executions hold:
  \begin{mathpar}
    \mprset{sep=1.5em}
    \inferrule*
    { \redel[c/s]{\envr_s}{\dm_c}{\noinj{\dm_c}}{\emptym}{\niltr} \\
      \redx{\envr_s}{\sm'}{\vm'_s}{\clientprog'}{\trace'_s}
    }
    { \redx{\envr_s}
      {\dm_c;\sm'}{\vm'_s}
      {(\noinj{\dm_c};\clientprog')}{\trace'_s}
    }
    \and
    \inferrule*
    { % \redc{\envr_c}{\clientprog}{\emptyr}{\trace_c}{\envg}{\envg'} \\
      \redc{\envr_c}{\noinj{\dm_c}}{\vm_c}{\trace_c}{\envg}{\envg} \\
      \redc{\envr_c+\vm_c}{\clientprog'}{\vm'_c}{\trace'_c}{\envg}{\envg'}
    }
    { \redc{\envr_c}{\noinj{\dm_c};\clientprog'}{\vm_c+\vm_c'}
      {\trace_c\ctr\trace'_c}{\envg}{\envg'}
    }
  \end{mathpar}

  Let us note  $\injf{x_i}{f_i}$ the injections in $\dm_c$ and
  $\rf{x}_i$ the associated fresh variables.
  Since hoisting has been applied, all the $f_i$ are either $\serial$
  or $\fragment$. Furthermore, no fragments are executed due to the
  execution of injections and
  $\noinj{\dm_c} = \substi[]{\injf{x_i}{f_i}}{\envr_s(x_i)}{\dm_c}$.

  We have the following compilations:
  \begin{align*}
    \compile{s}{\dm_c;\sm}
    &=\left(\pinj{\rf{x}_i}{x_i};\right)_i
      \pend;
      \compile{s}{\sm}\\
    \compile{c}{\dm_c;\sm}
    &=\pexec;
      \subst[]{\injf{x_i}{f_i}}{\rf{x}_i}{\dm_c}_i;
      \compile{c}{\sm}
  \end{align*}

  In the rest of this proof, we note $\compdmc =
  \subst[]{\injf{x_i}{f_i}}{\rf{x}_i}{\dm_c}_i$.

  We consider the server reduction $\dm_c\pred{}_{c/s}\noinj{\dm_c}$.
  We know that
  $\noinj{\dm_c} = \substi[]{\injf{x_i}{f_i}}{\injval{\envr_s(x_i)}}{\dm_c}$.
  Let us note $v_i = \envr_s(x_i)$.
  We can build the following $\ocsis$ reduction:
  % We can decompose $\clientprog$ in $\clientprog_1;\dots;\clientprog_n$,
  % where each $\clientprog_i$ is the result of the evaluation
  % of the injection $\injf{x_i}{f_i}$.
  % We have for each injection the following two evaluations
  % with $\envg_1 = \envg$ and $\envg_n = \envg'$
  % \begin{mathpar}
  %   \inferrule
  %   { \reds{\envr_s}{\app{f_i^s}{x_i}}{v_i}{\clientprog_i}{\trace_i} }
  %   { \redel[c/s]{\envr_s}
  %     {\injf{x_i}{f_i}}{\app{f_i^c}{\injval{v_i}}}{\clientprog_i}{\trace_i} }
  %   \and
  %   \inferrule
  %   { }
  %   { \redc{\envr_c}{\clientprog_i}{\emptyr}{\trace^c_i}{\envg_i}{\envg_{i+1}} }
  % \end{mathpar}
  % %
  % We can apply \cref{simulation:serverexpr} on each injection, which gives us
  % for each $i$:
  % \TODO{Make sure that all the conditions hold}
  % \begin{align*}
  %   \clientprog_i &= \left( \bind{y_{j_i}}{e_{j_i}}\right)_{j_i}&
  %   {\envfrag_\bullet}_i &= \tokfrag{y_{j_i}}{\rf{f_{j_i}}\ a_{j_i}}_{j_i}&
  %   \envg^f_i &= \bdr{\rf{f_{j_i}}}{f_{j_i}}_{j_i}&
  %   \envinj_i &= \bdr{\rf{x}_i}{\injval{v_i}}
  % \end{align*}\vspace{-8mm}
  % \begin{align*}
  %   \overline{\envg}_{i+1} \setminus \overline{\envg}_i
  %    & = \envg_{i+1} \setminus \envg_i = \bdr{\rf{y_{j_i}}}{v_{j_i}}&
  %   \envinj_i \cup \envg^f_i \cup \envg_i &\subset \overline{\envg}_i
  % \end{align*}
  \begin{mathpar}
    \inferrule
    { \forall i.\ %
      \inferrule*
      { \bdrin{x_i}{\co v_i}{\co\envr_s} }
      { \redmls{\co\envr_s}
        {\pinj{\rf{x_i}}{x_i}}{\emptym}
        {\emptyfrag}{\bdr{\rf{x_i}}{\injval{\co v_i}}}{\niltr} }
    }
    { \redmls{\co\envr_s}
      {(\pinj{\rf{x_i}}{x_i};)_i;\pend}{\emptym}
      {\tokend}{\bdr{\rf{x_i}}{\injval{\co v_i}}_i}{\niltr}
    }
    % \hfill
    % \forall\envfrag\exists\overline{\envg}'
    % \inferrule
    % { \redmlc{\envr_c}{\pexec}{\emptym}
    %   {\envfrag}{\envfrag'}
    %   {\overline{\envg}_{i+1}}{\overline{\envg}'}{\trace_i^c}
    % }
    % { \redmlc{\envr_c}{\pexec}{\emptym}
    %   {{\envfrag_\bullet}_i\cfrag\envfrag}{\envfrag'}
    %   {\overline{\envg}_i}{\overline{\envg}'}{\trace_i^c}
    % }
  \end{mathpar}
  Since $\co\envr_s\equivs{}\envr_s$, we also have
  that $\co v_i\equivs{}v_i$ and
  $\injval{\co v_i}\equivc{\envinj}\injval{v_i}$ for each $i$.
  We note $\envinj_\bullet = \bdr{\rf{x_i}}{\injval{\co v_i}}_i$.
  By definition of the slicing relation, the $\rf{x_i}$ are fresh, hence
  they are not bound in $\co\envg$. We can thus construct
  the global environment $\co\envg' = \co\envg\cup\envinj_\bullet$.
  Since we only extend the part with injection references, we
  still have that $\envg\equivc{}\co\envg'$.
  % Le us note
  % $\envfrag_\bullet = {\envfrag_\bullet}_1\cfrag\dots\cfrag{\envfrag_\bullet}_n$ and
  % $\envinj = \cup_i\envinj_i$.
  % Let $\overline{\envg}$ be $\envg\cup\envinj\cup(\cup_i\envg^f_i)$.
  % We can now build the following execution by repeatedly using the
  % $\dexec$ reduction for each injection.
  % \begin{mathpar}
  %   \inferrule
  %   { \vdots }
  %   { \redmlc{\envr_c}{\pexec}{\emptym}
  %     {{\envfrag_\bullet}\cfrag\tokend\cfrag\envfrag}{\envfrag}
  %     {\overline{\envg}}{\overline{\envg}'}{\ctr_i\trace_i^c}
  %   }
  % \end{mathpar}
  % By inspection of the
  % execution of $\clientprog = \clientprog_1;\dots;\clientprog_n$,
  % we have that $\trace_c = \ctr_i\trace_i^c$.
  % By the various equalities above, we have that
  % $\overline{\envg}'\setminus\overline{\envg} = \envg'\setminus\envg =
  % \bdr{\rf{r_j}}{v_j}_j$.
  \\

  We now consider the client reduction
  $\redc{\envr_c}{\noinj{\dm_c}}{\vm_c}{\trace_c}{\envg}{\envg}$.
  We know that
  $\noinj{\dm_c}$ is equal to
  $\substi[]{\injf{x_i}{f_i}}{\injval{v_i}}{\dm_c}$,
  hence the reduction tree contains for each $i$ a reduction
  $\redc{}{\injval{v_i}}{\injval{v_i}}{\niltr}{\envg}{\envg}$.
  To obtain a reduction of
  $\compdmc = \subst[]{\injf{x_i}{f_i}}{\rf{x_i}}{\dm_c}_i$,
  we simply substitute each of these subreduction by
  one of the form
  $\redmlc{}{\rf{x_i}}{\injval{\co v_c}}
  {\envfrag}{\envfrag}{\co\envg'}{\co\envg'}{\niltr}$.
  for any queue $\envfrag$.
  By \cref{redequivmodc}, we can build the following reduction:
  \begin{mathpar}
    \inferrule{ }
    { \redmlc{\co\envr_c}{\compdmc}{\co\vm_c}
      {\envfrag}{\envfrag}
      {\co\envg'}{\co\envg'}{\co\trace_c}
    }
  \end{mathpar}
  where $\co\vm_c\equivc{\envinj'}\vm_c$ and
  $\co\trace_c\equivc{\envinj'}\trace_c$, for any queue $\envfrag$.
  \\

  We now consider the execution of $\sm'$.
  We have the following properties:
  \begin{align*}
    \co\envr_c+\co\vm_c &\equivc{\envinj'}\envr_c+\vm_c&
    \co\envr_s &\equivs{}\envr_s&
    \co\envg' &\equivc{}\envg&
    &\fragenv{\co\envg_f,\sm'}&
    &\fragenv{\co\envg_f,\envr_s}
  \end{align*}
  By induction on the execution of $\sm'$ and $\clientprog'$, we have
  $\redmls{\co\envr_c+\co\vm_c}{\compile{c}{\sm'}}{\co\vm'_s}
  {\envfrag'_\bullet}{\envinj'_\bullet}{\co\trace'_c}$
  and
  $\redmlc{\co\envr_s}{\compile{s}{\sm'}}{\co\vm'_c}
  {\envfrag'_\bullet \cfrag \envfrag'}{\envfrag'}
  {\envinj'_\bullet\cup\co\envg'}{\co\envg''}{\co\trace'_c}$
  where%
  \begin{align*}
    \co\envg'' &\equivc{}\envg'&
    \co\vm'_s&\equivs{\envinj''}\vm'_s&
    \co\trace'_s&\equivs{\envinj''}\trace'_s\\
    \fragenv{&\co\envg_f'',\vm'_s}&
    \co\vm'_c&\equivc{\envinj''}\vm'_c&
    \co\trace'_c&\equivc{\envinj''}\trace'_c
  \end{align*}

  Finally, we can build the following derivations:
  \begin{mathpar}
    \mprset{sep=1.1em}
    \inferrule* {
      \forall i,\ %
      \inferrule*{ }
      { \redmls{\co\envr_s}
        {\pinj{\rf{x}_i}{x_i}}{\emptym}
        {\emptyfrag}{\envinj_\bullet}{\niltr}
      } \\
      \inferrule*{ }
      { \redmls{\co\envr_s}
        {\compile{s}{\sm'}}{\co\vm'_s}
        {\envfrag'_\bullet}{\envinj'_\bullet}{\co\trace'_s}
      }
    }
    { \redmls{\co\envr_s}
      {\left(\pinj{\rf{x}_i}{\app{f_i^s}{x_i}};\right)_i
        \pend;\compile{s}{\sm'}}{\co\vm'_s}
      {\tokend\cfrag\envfrag'_\bullet}
      {\envinj_\bullet\cup\envinj'_\bullet}{\co\trace'_s}
    }
    \and
    \inferrule*
    { % \redmlc{\co\envr_c}
      % {\pexec}{\emptym}
      % {\envfrag_\bullet\cfrag\tokend\cfrag\envfrag}{\envfrag}
      % {\co\envg}{\co\envg'}{\co\trace_c}
      \inferrule*
      { \redmlc{\co\envr_c}
        {\compdmc}{\co\vm_c}
        {\envfrag'_\bullet\cfrag\envfrag'}{\envfrag'_\bullet\cfrag\envfrag'}
        {\co\envg'\cup\envinj'_\bullet}
        {\co\envg'\cup\envinj'_\bullet}{\co\trace'_c}
        \quad
        \redmlc{\co\envr_c+\co\vm_c}{\compile{c}{\sm'}}{\co\vm'_c}
        {\envfrag'_\bullet\cfrag\envfrag'}{\envfrag'}
        {\co\envg'\cup\envinj'_\bullet}{\co\envg''}
        {\co\trace'_c}
      }
      { \redmlc{\co\envr_c}
        {\compdmc;\compile{c}{\sm'}}{\co\vm_c+\co\vm'_c}
        {\envfrag'_\bullet\cfrag\envfrag'}
        {\envfrag'}
        {\co\envg'\cup\envinj'_\bullet}{\co\envg''}
        {\co\trace_c\ctr\co\trace'_c}
      }
    }
    { \redmlc{\co\envr_c}
      {\pexec;\compdmc;\compile{c}{\sm'}}{\co\vm_c+\co\vm'_c}
      {\tokend\cfrag\envfrag'_\bullet\cfrag\envfrag'}
      {\envfrag'}
      {\co\envg\cup\envinj_\bullet\cup\envinj'_\bullet}{\co\envg''}
      {\co\trace_c\ctr\co\trace'_c}
    }
  \end{mathpar}

  We verify the following invariants:

  \begin{align*}
    \co\envg'' &\equivc{}\envg'&
    \co\vm_s&\equivs{}\vm_s&
    \co\trace_s&\equivs{}\trace_s\\
    \fragenv{&\co\envg_f'',\vm_s+\vm'_s}&
    \co\vm_c+\co\vm'_c&\equivc{\co\envg''}\vm_c+\vm'_c&
    \co\trace_c+\co\trace'_c&\equivc{\co\envg''}\trace_c+\trace'_c&
    \qedhere
  \end{align*}
\end{subproof}

\begin{subproof}[Case \textnormal{$\modulem[\sideX]{X}{\mm};\sm'$} -- Declaration of a mixed module]\item
  We assume that the following executions hold:
  \begin{mathpar}
    \mprset{sep=1.3em}
    \inferrule*
    { \inferrule*
      { \redxe{\envr_s}{\mm}{\vm_s}{\mm^c}{\clientprog}{\trace_s} }
      { \redx{\envr_s}
        {\modulem[\sideX]{X}{\mm}}{\bdr{X}{\vm_s}}
        {\modulem[]{X}{\mm^c};\clientprog}{\trace_s}
      } \\
      \redx{\envr_s+\bdr{X}{\vm_s}}{\sm'}{\vm'_s}{\clientprog'}{\trace'_s}
    }
    { \redx{\envr_s}
      {\modulem[\sideX]{X}{\mm};\sm'}
      {\bdr{X}{\vm_s}+\vm'_s}
      {(\clientprog;\modulem[]{X}{\mm^c};\clientprog')}{\trace_s\ctr\trace'_s}
    }
    \and
    \inferrule*
    { \redc{\envr_c}{\clientprog}
      {\{\}}{\trace_c}{\envg}{\envg'}
      \quad
      \inferrule*
      { \redc{\envr_c}{\mm^c}{\vm_c}{\trace'_c}{\envg'}{\envg''} }
      { \redc{\envr_c}
        {\modulem[]{X}{\mm^c}}{\bdr{X}{\vm_c}}{\trace'_c}
        {\envg'}{\envg''}
      } \quad
      \redc{\envr_c+\bdr{X}{\vm_c}}{\clientprog'}
      {\vm'_c}{\trace''_c}{\envg''}{\envg'''} }
    { \redc{\envr_c}
      {\clientprog;\modulem[]{X}{\mm^c};\clientprog'}{\bdr{X}{\vm_c}+\vm'_c}
      {\trace_c\ctr\trace'_c\ctr\trace''_c}{\envg}{\envg'''}
    }
  \end{mathpar}

  Let us assumes that we can build the following reductions
  \begin{mathpar}
    \inferrule{ }
    { \redmls{\co\envr_s}
      {\compile{s}{\modulem[\sideX]{X}{\mm}}}{\bdr{X}{\co\vm_s}}
      {\envfrag_\bullet}{\envinj}{\co\trace_s}
    }
    \and
    \inferrule{ }
    { \redmlc{\co\envr_c}
      {\compile{c}{\modulem[\sideX]{X}{\mm}}}{\bdr{X}{\co\vm_c}}
      {\envfrag_\bullet\cfrag\envfrag}{\envfrag}
      {\envinj\cup\co\envg}{\co\envg'}
      {\co\trace_c}
    }
  \end{mathpar}
  for any $\envfrag$, and that the following invariants hold:
  \begin{align*}
    \co\envg' &\equivc{}\envg''&
    \co\vm_s&\equivs{}\vm_s&
    \co\trace_s&\equivs{}\trace_s\\
    \fragenv{&\co\envg_f',\vm_s}&
    \co\vm_c&\equivc{\co\envg''}\vm_c&
    \co\trace_c&\equivc{\co\envg'}\trace_c+\trace'_c
  \end{align*}
  By induction on the execution of $\sm'$ and $\clientprog'$,
  we can build the following reduction:
  $\redmls{\co\envr_s+\bdr{X}{\co\vm_s}}{\compile{s}{\sm'}}{\vm'_s}
  {\envfrag'_\bullet}{\envinj'}{\trace'_s}$
  and
  $\redmlc{\co\envr_c+\bdr{X}{\co\vm_c}}{\compile{c}{\sm'}}{\vm'_c}
  {\envfrag'_\bullet\cfrag\envfrag}{\envfrag}
  {\envinj'\cup\co\envg'}{\co\envg''}{\trace'_c}$,
  which allows us to conclude.

  To build the compiled reduction,
  we will operate by case analysis over $\mm$.

  \begin{subproof}[Subcase \textnormal{$\mm = \struct{\sm}_{\rf{X}}$} -- Declaration of a mixed structure]\item
  We have $\clientprog = \bind{X}{(\struct{\clientprog_0})}$ and
  $\mm^c = \rf{X}$
  with the following reductions:
  \begin{mathpar}
    \inferrule
    { \redx{\envr_s}{\sm}{\vm_s}{\clientprog}{\trace_s} }
    { \redxe{\envr_s}{\struct{\sm}}{\vm_s+\bdr{\dyn}{\rf{X}}}{\rf{X}}
      {\clientprog}{\trace_s} }
    \and
    \inferrule
    { \inferrule
      { }
      { \redc{\envr_c}{\clientprog_0}{\vm_c}{\trace_c}
        {\envg}{\envg'} }
      \\
      \inferrule
      { \bdrin{\rf{X}}{\vm_c}{\envg''} }
      { \redc{\envr_c}{\rf{X}}{\vm_c}{\niltr}{\envg''}{\envg''} }
    }
    { \redc{\envr_c}{(\bind{X}{\struct{\clientprog_0}};\ \modulem[]{X}{\rf{X}})}
      {}
      {\trace_c}{\envg}{\envg''}
    }
  \end{mathpar}
  where
  $\envg'' = \envg'\cup\bdr{\rf{X}}{\vm_c}$.
  \\

  We have the following compilations:
  \begin{align*}
    \compile{s}{\modulem[\sideX]{X}{\struct{\sm}}_{\rf{X}}}
    &= \begin{aligned}
      &\modulem[]{X}{\mathtt{struct}}\\
      &\quad\modulem[]{\dyn}{\rf{X}};\\
      &\quad\compile{s}{\sm}\\
      &\mathtt{end}
  \end{aligned}\\[2mm]
    \compile{c}{\modulem[\sideX]{X}{\struct{\sm}}_{\rf{X}}}
    &= \begin{aligned}
      &\binddyn{\rf{X}}{\struct{\compile{c}{\sm}}};\\
      &\modulem[]{X}{\rf{X}};
    \end{aligned}
  \end{align*}

  By induction on the execution of $\sm$ and $\clientprog$,
  we have
  $\redmls{\co\envr_s}{\compile{s}{\sm}}{\co\vm_s}
  {\envfrag_\bullet}{\envinj_\bullet}{\co\trace_s}$
  and
  $\redmlc{\co\envr_c}{\compile{c}{\sm}}{\co\vm_c}
  {\envfrag_\bullet \cfrag \envfrag}{\envfrag}
  {\co\envg\cup\envinj_\bullet}{\co\envg'}{\co\trace_c}$
  with the following invariants:
  \begin{align*}
    \co\envg' &\equivc{}\envg'&
    \co\vm_s&\equivs{}\vm_s&
    \co\trace_s&\equivs{}\trace_s\\
    \fragenv{&\co\envg_f',\vm_s}&
    \co\vm_c&\equivc{\co\envg'}\vm_c&
    \co\trace_c&\equivc{\co\envg'}\trace_c
  \end{align*}

  We can then build the following executions:
  \begin{mathpar}
    \inferrule
    { \redmls{\co\envr_s}
      {\compile{s}{\sm}}
      {\co\vm_s}
      {\envfrag_\bullet}{\envinj_\bullet}{\co\trace_s}
    }
    { \redmls{\co\envr_s}
      {\compile{s}{\modulem[\sideX]{X}{\struct{\sm}_{\rf{X}}}}}
      {\bdr{X}{\bdr{\dyn}{\rf{X}}+\co\vm_s}}
      {\envfrag_\bullet}{\envinj_\bullet}{\co\trace_s}
    }
    \and
    \inferrule
    { \inferrule
      { \redmlc{\co\envr_c}{\compile{c}{\sm}}{\co\vm_c}
        {\envfrag_\bullet \cfrag \envfrag}{\envfrag}
        {\co\envg\cup\envinj_\bullet}{\co\envg'}{\co\trace_c}
      }
      { \redmlc{\co\envr_c}
        {\modulem[]{X}{\struct{\compile{c}{\sm}}}}
        {\bdr{X}{\co\vm_c}}
        {\envfrag_\bullet \cfrag \envfrag}{\envfrag}
        {\co\envg\cup\envinj_\bullet}{\co\envg'}{\co\trace_c}
      }
    }
    { \redmlc{\co\envr_c}
      {\compile{c}{\modulem[\sideX]{X}{\struct{\sm}}}}
      {\bdr{X}{\co\vm_c}}
      {\envfrag_\bullet \cfrag \envfrag}{\envfrag}
      {\co\envg\cup\envinj_\bullet}{\co\envg''}
      {\co\trace_c}
    }
  \end{mathpar}
  Where $\co\envg'' = \co\envg'\cup\bdr{\rf{X}}{\co\vm_c}$.
  We verify the following invariants:
  \begin{align*}
    \co\envg'' &\equivc{}\envg'&
    \co\vm_s+\bdr{\dyn}{\rf{X}}&\equivs{}\vm_s+\bdr{\dyn}{\rf{X}}&
    \co\trace_s&\equivs{}\trace_s\\
    \fragenv{&\co\envg_f'',\vm_s}&
    \co\vm_c&\equivc{\co\envg'}\vm_c&
    \co\trace_c&\equivc{\co\envg'}\trace_c
  \end{align*}
  which concludes.
  \end{subproof}

  \begin{subproof}[Subcase
      \textnormal{$\mm = \functori{X}{\Mm}{\struct{\sm}_{\rf{F}}}$} -- Declaration of a mixed functor]\item
  In this case, we have
  the client program $\clientprog = \bindenv{F}$ and
  the module expression
  $\mm^c = \functori{X}{\Mm}{\struct{\restrict{c}{\sm}}}$.
  The following reductions hold:
  \begin{mathpar}
    \setlength{\jot}{-1pt}% Remove interline space in align*
    \inferrule
    { }
    { \redx{\envr_s}
      { \left(
          \begin{aligned}
            &\modulem[\sideX]{F(X_i:\Mm_i)_i}{\mathtt{struct}}\\
            &\quad\sm\\
            &\mathtt{end}_{\rf{F}}
          \end{aligned}\right)}
      {\bdr{F}{\vm_s}}
      { \left(\begin{aligned}
            &\bindenv{F}\\
            &\modulem[\sideX]{F(X_i:\Mm_i)_i}{\mathtt{struct}}\\
            &\quad\restrict{c}{\sm}\\
          &\dend
        \end{aligned}\right)}
      {\niltr} }
    \and
    \inferrule
    { }
    { \redc{\envr_c}
      { \left(\begin{aligned}
            &\bindenv{F}\\
            &\modulem[\sideX]{F(X_i:\Mm_i)_i}{\mathtt{struct}}\\
            &\quad\restrict{c}{\sm}\\
            &\dend
        \end{aligned}\right)}
      {\bdr{F}{\vm_c}}
      {\niltr}{\envg}{\envg'} }
  \end{mathpar}
  Where $\envg' = \envg \cup\bdr{\rf{F}}{\envr_c}$ and the following values:
  \begin{align*}
    \vm_s &= \functorenvi{\envr_s}{X}{\Mm}{\struct{\sm}}\\
    \vm_c &= \functorenvi{\envr_c}{X}{\Mm}{\struct{\restrict{c}{\sm}}}
  \end{align*}
  We recall that by hoisting, the body of the functors contains
  no injection, hence we don't need to evaluate server code in the client
  part.

  We have the following compilations:
  \begin{align*}
  \compile{s}{
    \begin{aligned}
      &\modulem[\sideX]{F(X_i:\Mm_i)_i}{\mathtt{struct}}\\
      &\quad\sm\\
      &\mathtt{end}_{\rf{F}}
    \end{aligned}}
   &=
    \begin{aligned}
      &\modulem[]{F(X_i:\compile{s}{\Mm_i})_i}
      {\mathtt{struct}}\\
      &\quad
      \modulem[]{\dyn}{\fragm{\rf{F}}{(\getdyn{X_i})_i}};\\
      &\quad\compile{s}{\sm}\\
      %{\substi[c]{v_i}{X_{i'}.v}{\substi[c]{t_i}{X_{i'}.t}{s'}}}\\
      &\mathtt{end}
    \end{aligned}\\[2mm]
  \compile{c}{
    \begin{aligned}
      &\modulem[\sideX]{F(X_i:\Mm_i)_i}{\mathtt{struct}}\\
      &\quad\sm\\
      &\mathtt{end}_{\rf{F}}
    \end{aligned}}
  &=
    \begin{aligned}
      &\binddyn{\rf{F}(X_i:\compile{c}{\Mm_i})_i}
      {\struct{\compile{c}{\sm}}};\\
      &\modulem[]{F(X_i:\compile{c}{\Mm_i})_i}
      {\struct{\restrict{c}{\sm}}};
    \end{aligned}\\[2mm]
  \end{align*}

  We trivially have the following execution:
  \begin{mathpar}
    \setlength{\jot}{-1pt}% Remove interline space in align*
    \inferrule
    { }
    { \redmls{\co\envr_s}
      { \compile{s}{
          \begin{aligned}
            &\modulem[\sideX]{F(X_i:\Mm_i)_i}{\mathtt{struct}}\\
            &\quad\sm\\
            &\mathtt{end}_{\rf{F}}
          \end{aligned}}}
      {\bdr{F}{\co\vm_s}}
      {\envfrag_\bullet}{\{\}}{\niltr} }
    \and
    \inferrule
    { }
    { \redmlc{\co\envr_c}
      { \compile{c}{
          \begin{aligned}
            &\modulem[\sideX]{F(X_i:\Mm_i)_i}{\mathtt{struct}}\\
            &\quad\sm\\
            &\mathtt{end}_{\rf{F}}
          \end{aligned}}}
      {\bdr{F}{\co\vm_c}}
      {\envfrag}{\envfrag}
      {\co\envg}{\co\envg'}
      {\niltr}
    }
  \end{mathpar}
  Where $\envg' = \envg \cup\bdr{\rf{F}}{\co\vm_{\rf{F}}}$ with the following values:
  \begin{align*}
    \co\vm_s &= \functorenvi{\co\envr_s}{X}{\Mm}{\struct{\compile{s}{\sm}}}\\
    \co\vm_c &= \functorenvi{\co\envr_c}{X}{\Mm}{\struct{\restrict{c}{\sm}}}\\
    \co\vm_{\rf{F}}
    &= \functorenvi{\co\envr_c}{X}{\Mm}{\struct{\compile{c}{\sm}}}\\
  \end{align*}

  We now need to show that the invariants still hold. We easily
  have that $\co\vm_c\equivc{\co\envg}\vm_c$.
  By definition of equivalence over mixed functors, we have
  $\co\vm_s\equivs{}\vm_s$. Indeed, the body of the functor in $\co\vm_s$
  is the server compilation of the body of the mixed
  functor $\vm_s$ and the captured environments corresponds.
  Finally, we have that the body of $\co\vm_{\rf{F}}$ is the
  client compilation of the body of the mixed functors and that the
  capture environment corresponds to $\envg'(\rf{F})$. Thus
  we get that $\fragenv{\co\envg_f',\vm_s}$.
  By definition of the annotation function, the reference $\rf{F}$ could not
  have appeared on a previously executed structure,
  hence we still have that $\fragenv{\co\envg_f',\co\rho_s}$.

  Hence, all the following invariants are respected, which concludes.
  \begin{align*}
    \co\envg' &\equivc{}\envg'&
    \co\vm_s&\equivs{}\vm_s\\
    \fragenv{&\co\envg_f',\vm_s}&
    \co\vm_c&\equivc{\co\envg''}\vm_c&&\qedhere
  \end{align*}

  \end{subproof}

  Otherwise, $\mm$ is a module expression. By definition of slicability,
  $\mm$ does not syntactically contain any structure. In the general
  case, we should proceed by induction over module expressions. We
  will simply present the case of a mixed functor application
  where the functor returns a mixed structure.

  We consider $\mm = F(X_1)\dots(X_n)$.
  We have $\mm^c = F(X_1)\dots(X_n)$
  with the following executions:

  \begin{mathpar}
    \setlength{\jot}{-1pt}
    \inferrule
    { \inferrule
      { \redx{\envr_s}{F}
        {\functorenvi[\sideX]{\envr'_s}{Y}{\Mm}{\struct{\sm}_{\rf{F}}}}
        {\emptym}
        {\niltr} \\
        \redx{\envr_s}{X_i}{\vm_i^s}{\emptym}{\niltr} \\
        \bdrin{\dyn}{\rf{R_i}}{\vm_i^s} \\
        \rf{R}\text{ fresh} \\
        \redx{\envr'_s+\bdr{Y_i}{\vm_i^s}_i}
        {\substi[]{\rf{f_i}}{\rf{R.f_i}}{\sm}}{\vm}{\clientprog_0}{\trace} }
      { \redx{\envr_s}{F(X_1)\dots(X_n)}{\vm_s+\bdr{\dyn}{\rf{R}}}
        { \clientprog = \left(\begin{aligned}
              &\bind{R}{\dstruct}\\
              &\quad(\modulem[]{Y_i}{\rf{R_i}};)_i\\
              &\quad\clientprog_0\\
              &\dend\ \mathtt{with}\ \rf{F}
            \end{aligned}
          \right) }
        {\trace} }
    }
    { \redx{\envr}{\modulem[\sideX]{X}{F(X_1)\dots(X_n)}}{\vm_s}
      {\left(\clientprog;\modulem[]{X}{F(X_1)\dots(X_n)}\right)}
      {\trace_s}
    }
  \end{mathpar}
  \begin{mathpar}
    \setlength{\jot}{-1pt}
    \inferrule
    { \bdrin{\rf{F}}{\envr_{\rf{F}}}{\envg} \\
      \bdrin{\rf{R_i}}{\vm_i^c}{\envg} \\
      \redc{\envr_{\rf{F}}+\bdr{Y_i}{\vm_i^c}_i}
      {\clientprog_0}{\vm_c}{\trace_c}
      {\envg}{\envg'}
    }
    { \redc{\envr_c}
      {\left(\begin{aligned}
            &\bind{R}{\dstruct}\\
            &\quad(\modulem[]{Y_i}{\rf{R_i}};)_i\\
            &\quad\clientprog_0\\
            &\dend\ \mathtt{with}\ \rf{F}
          \end{aligned}
        \right)}
      {\emptym}
      {\trace_c}{\envg}{\envg'\cup\bdr{\rf{R}}{\vm_c}}
    }
    \and
    \inferrule
    { \redc{\envr_s}{F}
      {\functorenvi{\envr'_c}{Y}{\Mm}{\struct{\sm_c}}}
      {\niltr}{\cdots}{\cdots} \\
      \redc{\envr_c}{X_i}{\vm_i}{\niltr}{\cdots}{\cdots} \\
      \redc{\envr'_c+\bdr{Y_i}{\vm_i}_i}{\sm_c}{\vm'_c}{\trace'_c}
      {\envg'\cup\bdr{\rf{R}}{\vm_c}}{\envg''}
    }
    { \redc{\envr_c}{\modulem[]{X}{F(X_1)\dots(X_n)}}{\bdr{X}{\vm'_c}}
      {\trace'_c}{\envg'\cup\bdr{\rf{R}}{\vm_c}}{\envg''} }
  \end{mathpar}
  We note $\vm_F$ the value of $F$ in $\envg_s$, which is
  $\functorenvi{\envr'_s}{Y}{\Mm}{\struct{\sm}}$.

  We have the following compilations:
  \begin{align*}
    \compile{s}{\modulem[\sideX]{X}{F(X_1)\dots(X_n)}}
    &= \begin{aligned}
      &\modulem[]{X}{F(X_1)\dots(X_n)};\\
      &\pend;
    \end{aligned}\\[2mm]
    \compile{c}{\modulem[\sideX]{X}{F(X_1)\dots(X_n)}}
    &= \begin{aligned}
      &\modulem[]{X}{F(X_1)\dots(X_n)};\\
      &\pexec;
    \end{aligned}
  \end{align*}

  Let us now look at the execution of the server application
  $F(X_1)\dots(X_n)$.
  By hypothesis, $\vm_F$ is a mixed functor. By equivalence, we know
  that $\co\envr_s(F) = \co\vm_F$ where $\co\vm_F\equivs{}\vm_F$.
  By definition of the equivalence on server values, $\co\vm_F$
  is of the following shape:
  \begin{align*}
    \co\vm_F
    &= \left(\begin{aligned}
        &\functorenvi{\co\envr'_s}{Y}{\Mm}{\dstruct}\\
        &\quad\modulem[]{\dyn}{\fragm{\rf{F}}{(\getdyn{Y_i})_i}};\\
        &\quad\compile{s}{\sm}\\
        &\dend
      \end{aligned}\right)
  \end{align*}
  where $\co\envr'_s\equivs{\co\envg}\envr'_s$.
  For each $i$ we note $\co\vm_i^s = \co\envr_s(X_i)$.
  By equivalence, we have $\co\vm_i^s \equivs{} \vm_i^s$.

  Furthermore, since $\co\envg\equivc{}\envg$ and by hypothesis, $\rf{F}$ is
  also bound in $\co\envg$. We note $\vm_{\rf{F}}$ the corresponding value.
  Since $\fragenv{\co\envg, \vm_c}$ (via $\envr_s$), then $\co\vm_{\rf{F}}$
  is of the following shape:
  \begin{align*}
    \co\vm_{\rf{F}}
    &= \left(\begin{aligned}
         &\functorenvi{\co\envr_{\rf{F}}}{Y}{\Mm}{\dstruct}\\
         &\quad\compile{c}{\sm}\\
         &\dend
       \end{aligned}\right)
  \end{align*}
  where $\co\envr_{\rf{F}}\equivc{\co\envg}\envr_{\rf{F}} = \envg(\rf{F})$.
  Additionally, for each $i$ we note $\vm_i^c = \co\envg(\rf{R_i})$.
  By equivalence, we have $\co\vm_i^c \equivs{} \vm_i^c$.

  We can now proceed by induction on $\sm$ and $\clientprog_0$ in the
  environment $\co\envg$,
  $\co\envr_c\cup\bdr{Y_i}{\rf{R_i}}$ and
  $\co\envr_s\cup\bdr{\dyn}{\rf{R}}\cup\bdr{Y_i}{\co\vm_i^s}_i$.
  We obtain the following reductions:\\
  $\redmls{\co\envr'_s\cup\bdr{Y_i}{\co\vm_i^s}_i}{\compile{s}{\sm}}
  {\envfrag_\bullet}{\envinj_\bullet}{\co\trace_s}$
  and
  $\redmlc{\co\envr_c\cup\bdr{Y_i}{\rf{R_i}}_i}{\compile{c}{\sm}}{\co\vm_c}
  {\envfrag_\bullet\cfrag\envfrag}{\envfrag}
  {\envinj_\bullet\cup\co\envg}{\co\envg'}{\co\trace_c}$
  \\
  with the usual invariants.
  We can now build the following executions:
  \begin{mathpar}
    \inferrule
    { \redmls{\co\envr'_s\cup\bdr{Y_i}{\co\vm_i^s}_i}
      {\fragm{\rf{F}}{(\getdyn{Y_i})_i}}{\rf{R}}
      {\envfrag_{\rf{R}}}
      {\emptym}{\niltr}\\
      \bdrin{F}{\co\vm_f}{\co\envr_s}\\
      \bdrin{X_i}{\co\vm_i^s}{\co\envr_s}\\
      \redmls{\co\envr'_s\cup\bdr{\dyn}{\rf{R}}\cup\bdr{Y_i}{\co\vm_i^s}_i}
      {\compile{s}{\sm}}
      {\envfrag_\bullet}{\envinj_\bullet}{\co\trace_s}
    }
    { \redmls{\co\envr_s}
      {\modulem[]{X}{F(X_1)\dots(X_n)};\pend}{\bdr{X}{\co\vm_s}}
      {\envfrag_{\rf{R}}\cfrag\envfrag_\bullet\cfrag\tokend}
      {\envinj_\bullet}{\co\trace_s}
    }
    \and
    \inferrule
    { \bdrin{\rf{F}}{\co\vm_{\rf{F}}}{\co\envg} \\
      \redmlc{\co\envr_{\rf{F}}\cup\bdr{Y_i}{\rf{R_i}}_i}
      {\compile{c}{\sm}}{\co\vm_c}
      {\envfrag_\bullet\cfrag\tokend\cfrag\envfrag}
      {\tokend\cfrag\envfrag}
      {\envinj_\bullet\cup\co\envg}{\co\envg'}
      {\co\trace_c}
    }
    { \redmlc{\co\envr_c}{\pexec}{\emptym}
      {\envfrag_{\rf{R}}\cfrag\envfrag_\bullet\cfrag\tokend\cfrag\envfrag}
      {\envfrag}
      {\envinj_\bullet\cup\co\envg}{\co\envg'\cup\bdr{\rf{R}}{\co\vm_c}}
      {\co\trace_c}
    }
  \end{mathpar}
  where $\envfrag_{\rf{R}} = \tokfrag{\rf{R}}{\rf{F}(\rf{R_1})\dots(\rf{R_n})}$.
  We respect the following invariants:
  \begin{align*}
    \co\envg' &\equivc{}\envg'&
    \co\vm_s&\equivs{}\vm_s&
    \co\trace_s&\equivs{}\trace_s\\
    \fragenv{&\co\envg_f',\vm_s}&
    \co\vm_c&\equivc{\co\envg'}\vm_c&
    \co\trace_c&\equivc{\co\envg'}\trace_c
  \end{align*}

  Let us now consider the client application.
  Since $\envr_c\equivc{\co\envg}\co\envr_c$, we
  have that the body of the functor $F$ is equivalent.
  We can thus build the following reduction:
  \begin{mathpar}
    \inferrule
    { \bdrin{F}
      {\functorenvi{\co\envr'_c}{Y}{\Mm}{\struct{\co\sm_c}}}
      {\co\envr_c} \\
      \bdrin{X_i}{\co\vm^c_i}{\co\envr_c} \\
      \redmlce{\co\envr'_c+\bdr{Y_i}{\co\vm_i^c}_i}{\co\sm_c}{\co\vm'_c}
      {\co\envg'\cup\bdr{\rf{R}}{\co\vm_c}}{\co\envg''}
      {\co\trace'_c}
    }
    { \redmlc{\co\envr_c}{\modulem[]{X}{F(X_1)\dots(X_n)}}{\bdr{X}{\co\vm'_c}}
      {\envfrag}{\envfrag}
      {\co\envg'\cup\bdr{\rf{R}}{\co\vm_c}}{\co\envg''}{\co\trace'_c} }
  \end{mathpar}
  By equivalence of $F$ and $X_i$ in $\envr_c$ and $\co\envr_c$, we have
  that $\co\envg'\equivc{}\envg'$, $\co\vm'_c\equivc{\co\envg'}\vm'_c$
  and $\co\trace_c\equivc{\co\envg'}\trace_c$.
  \\

  We already built the reduction for the compiled server program. We
  can now build the compiled client program:
  \begin{mathpar}
    \inferrule
    { \redmlc{\co\envr_c}{\pexec}{\emptym}
      {\tokfrag{\rf{R}}{\rf{F}(\rf{R_1})\dots(\rf{R_n})}
        \cfrag\envfrag_\bullet\cfrag\tokend\cfrag\envfrag}
      {\envfrag}
      {\envinj_\bullet\cup\co\envg}{\co\envg'\cup\bdr{\rf{R}}{\co\vm_c}}
      {\co\trace_c} \\
      \redmlc{\co\envr_c}{\modulem[]{X}{F(X_1)\dots(X_n)}}
      {\bdr{X}{\co\vm'_c}}
      {\envfrag}{\envfrag}
      {\co\envg'\cup\bdr{\rf{R}}{\co\vm_c}}{\co\envg''}{\co\trace'_c} }
    { \redmlc{\co\envr_c}
      {\compile{c}{\modulem[\sideX]{X}{F(X_1)\dots(X_n)}}}
      {\bdr{X}{\co\vm'_c}}
      {\envfrag_{\rf{R}}\cfrag\envfrag_\bullet\cfrag\tokend\cfrag\envfrag}
      {\envfrag}
      {\envinj_\bullet\cup\co\envg}{\co\envg''}
      {\co\trace_c\ctr\co\trace'_c}
    }
  \end{mathpar}
  where the invariants still hold. This concludes.
\end{subproof}
\end{proof}

\subsection{Proof of the main theorem}

Finally, we prove \cref{thm:simulation}. This is a direct consequence of
\cref{lemma:simulation:struct}.
\begin{proof}[Proof of \cref{thm:simulation}.]
  We have that $ \redp{\emptyr}{\Pm}{v}{\trace} $. By definition of an
  \etiny program execution, we can decompose this rule as following:
  \begin{mathpar}
    \inferrule
    { \redx{\emptyr}{\Pm}{()}{\clientprog}{\trace_s} \\
      \redc{\emptyr}{\clientprog}{v}{\trace_c}{\emptym}{\envg} }
    { \redp{\emptyr}{\Pm}{v}{\trace_s\ctr\trace_c} }
  \end{mathpar}

  We trivially have the following invariants:
  \begin{align*}
    \emptyr &\equivc{\emptyr}\emptyr&
    \emptyr &\equivs{}\emptyr&
    \co\envg &\equivc{}\envg&
    &\fragenv{\emptyr,\Pm}&
    &\fragenv{\emptyr,\emptyr}
  \end{align*}
  which allow us to apply \cref{lemma:simulation:struct} and conclude.
\end{proof}

%%% Local Variables:
%%% mode: latex
%%% TeX-master: "../main"
%%% End:

% \input{core/tyappendix}
% \input{module/appendix}

\end{document}
%%% Local Variables:
%%% mode: latex
%%% TeX-master: t
%%% End:

%  LocalWords:  Tierless IRILL Radanne Jérôme Vouillon Balat Cité UMR
%  LocalWords:  IRIF CNRS Inria BeSport composable specificities